\documentclass[apj, 8pt]{emulateapj-rtx4}
\usepackage{color}
\usepackage{amsmath}

\defcitealias{2015ApJS..219...15S}{Paper I}
\defcitealias{2015ApJ...800...39G}{G15}

\usepackage[dvipdfm]{hyperref}
\hypersetup{
	bookmarks=true,
	bookmarksnumbered=true,
	bookmarksopen=true,
	colorlinks   = true,
	citecolor    = blue,
	pdftitle = {Evolution of Clumpy Galaxies},
	pdfauthor = {Takatoshi Shibuya},
	pdfkeywords={cosmology: observations --- early universe --- galaxies: formation --- galaxies: high-redshift}
	pdfpagemode=UseThumbs
}

\renewcommand\thefootnote{\arabic{footnote}}

\slugcomment{Submitted to ApJ November 23, 2015; accepted February 2, 2016}

\shorttitle{Evolution of Clumpy Galaxies}
\shortauthors{T. Shibuya et al.}

\begin{document}

\title{Morphologies of $\sim 190,000$ Galaxies at $\lowercase{z}=0-10$ 
Revealed with HST Legacy Data \\II. Evolution of Clumpy Galaxies}

\author{Takatoshi Shibuya\altaffilmark{1}, Masami Ouchi\altaffilmark{1,2}, Mariko Kubo\altaffilmark{1}, and Yuichi Harikane\altaffilmark{1,3}}
\email{shibyatk\_at\_icrr.u-tokyo.ac.jp}

\altaffiltext{1}{Institute for Cosmic Ray Research, The University of Tokyo, 5-1-5 Kashiwanoha, Kashiwa, Chiba 277-8582, Japan}
\altaffiltext{2}{Kavli Institute for the Physics and Mathematics of the Universe (Kavli IPMU, WPI), University of Tokyo, Kashiwa, Chiba 277-8583, Japan}
\altaffiltext{3}{Department of Physics, Graduate School of Science, The University of Tokyo, 7-3-1 Hongo, Bunkyo, Tokyo, 113-0033, Japan}

\begin{abstract}

We investigate evolution of clumpy galaxies with the $\!${\it Hubble Space Telescope} ($\!${\it HST}) samples of $\sim17,000$ photo-$z$ and Lyman break galaxies at $z\simeq0-8$. We detect clumpy galaxies with off-center clumps in a self-consistent algorithm that is well tested with previous study results, and measure the number fraction of clumpy galaxies at the rest-frame UV, $f_{\rm clumpy}^{\rm UV}$. We identify an evolutionary trend of $f_{\rm clumpy}^{\rm UV}$ over $z\simeq0-8$ for the first time: $f_{\rm clumpy}^{\rm UV}$ increases from $z\simeq8$ to $z\simeq1-3$ and subsequently decreases from $z\simeq1$ to $z\simeq0$, which follows the trend of Madau-Lilly plot. A low average S\'ersic index of $n\simeq1$ is found in the underlining components of our clumpy galaxies at $z\simeq0-2$, indicating that typical clumpy galaxies have disk-like surface brightness profiles. Our $f_{\rm clumpy}^{\rm UV}$ values correlate with physical quantities related to star formation activities for star-forming galaxies at $z\simeq0-7$. We find that clump colors tend to be red at a small galactocentric distance for massive galaxies with $\log{M_*/M_\odot}\gtrsim11$. All of these results are consistent with a picture that a majority of clumps form in the violent disk instability and migrate into the galactic centers. 

\end{abstract}

\keywords{cosmology: observations --- early universe --- galaxies: formation --- galaxies: high-redshift}

\section{INTRODUCTION}\label{sec_intro}

Galaxy morphology offers invaluable insights into the galaxy formation and evolution. Nearby and low-$z$ galaxies are categorized mainly into three morphological types, i.e., spirals, spheroids, and irregular systems \citep[the Hubble sequence; ][]{1926ApJ....64..321H}. In contrast to this low-$z$ morphological classification, observations with $\!${\it Hubble Space Telescope} ($\!${\it HST}) have revealed that galaxies hosting several bright patchy structures, a.k.a. clumpy galaxies, are more abundant at $z\simeq1-3$ than the local Universe \citep[e.g., ][]{1995AJ....110.1576C,1996AJ....112..359V, 1996AJ....112..369G, 2004ApJ...604L..21E, 2005ApJ...627..632E, 2006ApJ...650..644E, 2005ApJ...623L..71E, 2008ApJ...688...67E, 2009ApJ...692...12E, 2009ApJ...701..306E, 2013ApJ...774...86E, 2013ApJ...778..170K, 2016MNRAS.455.3333K, 2013PASA...30...56G, 2014ApJ...780...77T,2014ApJ...786...15M,2015ApJ...800...39G,2015arXiv150603084H,2015ApJ...807..134G, 2016ASSL..418..355B}. 

Physical properties of the clumps have been intensively investigated by imaging observations with Advanced Camera for Survey (ACS) and Wide Field Camera 3 (WFC3)/IR-channel on board $\!${\it HST}. Spectral energy distribution (SED) analyses with $\!${\it HST} multi-wavebands suggest that typical clumps have a stellar mass of $\log{M_*/M_\odot}\simeq8-9$ and young stellar ages of $\lesssim0.5$ Gyr \citep[e.g., ][]{2012ApJ...757..120G, 2012ApJ...753..114W}. Combined with gravitational lensing effects, the high spatial resolving power of $\!${\it HST} reveals that the size of clumps ranges from $\simeq0.1$ kpc to $\simeq1$ kpc \citep[e.g., ][]{2009MNRAS.400.1121S, 2010MNRAS.404.1247J, 2012MNRAS.422.3339W, 2012MNRAS.427..688L,2015MNRAS.450.1812L}.

Even in substantial observational efforts, the clump formation mechanisms have not been fully understood. Two viable candidates have been proposed as a clump formation mechanism: 1) the violent disk instability \cite[VDI; ][]{2009ApJ...703..785D, 2013MNRAS.435..999D} and 2) galaxy mergers \citep[e.g., ][although cf. \citealt{2001ApJ...547..146T}]{2008A&A...492...31D}. The former and the latter are the ``{\it in-situ}" and ``{\it ex-situ}" origins for clumps, respectively. In the case of VDI, clumps are predicted to form in unstable regions where the Toomre $Q$ parameter \citep{1964ApJ...139.1217T} is below a critical value (an order of unity) in thick and gas-rich galaxy disks \citep[e.g., ][]{1998Natur.392..253N,2004A&A...413..547I,2004ApJ...611...20I,2007ApJ...670..237B, 2009ApJ...707L...1B, 2009ApJ...703..785D, 2010MNRAS.404.2151C, 2010Natur.463..781T, 2012MNRAS.425.1511H,2012ApJ...754...48F, 2010MNRAS.407.1223R, 2014MNRAS.442.1230R, 2009MNRAS.397L..64A, 2015MNRAS.449.2156A}.\footnote{One of the latest cosmological simulations suggests a possibility that clumps start to form at galactic regions with a high $Q$ value of $\simeq2-3$ \citep{2015arXiv151007695I}. } On the other hand, mergers of compact galaxies would also add clumpy components on galaxies, providing the similar clumpy morphology. 

The fate of clumps have also been a matter of debate. Theoretical studies and numerical simulations have predicted that clumps at disk regions migrate into the galactic centers due to dynamical frictions and/or clump-clump interactions \citep[e.g., ][]{2004A&A...413..547I,2004ApJ...611...20I,2009ApJ...703..785D, 2012MNRAS.422.1902I, 2012MNRAS.420.3490C,2014MNRAS.443.3675M, 2015arXiv151208791M}. The migrating clumps would subsequently contribute the formation of a central proto-bulge. In contrast, several numerical simulations have suggested that clumps are destroyed by their own and/or galactic feedback before the clump migration \citep[e.g., ][]{2012ApJ...745...11G, 2012MNRAS.427..968H, 2014MNRAS.444.1389M}. The disrupted clumps could become components of galaxy disks \citep[e.g. ][]{2014ApJ...780...57B}. These predictions indicate that clumps are important structures for characterizing the morphological evolution of galaxies.

A powerful observational technique is to construct spatial $Q$ and velocity ($v$) maps on clumpy galaxies for distinguishing the two clump formation mechanisms. During the last decade, spatially resolved kinematic properties for clumpy galaxies have been examined by exploiting adaptive optics systems and integral field spectrographs (IFS) \citep[e.g., ][]{2011ApJ...739...45F, 2012ApJ...752..111N, 2011ApJ...733..101G, 2014ApJ...796....7G}. According to these IFS studies, $Q$ values on clump regions tend to be below unity. The low $Q$ value is evidence that clumps have the in-situ origin and form through VDI. The velocity structure also provides important hints of whether clumps are dynamically bounded in host galaxies. The velocity of typical clumps follows global motions of host galaxy disks within $\Delta v\lesssim200$ km s$^{-1}$, which suggests the in-situ clump origin \citep[e.g., ][]{2008A&A...486..741B, 2011MNRAS.417.2601W,2012MNRAS.422.3339W,2013MNRAS.436..266W,2013ApJ...779..139S, 2015ApJ...802..101T,2015arXiv151003262B}. In contrast, statistical studies with IFS indicate that ex-situ clumps appear to be rare among $z\simeq2$ star-forming galaxies \citep[e.g., ][]{2008ApJ...682..231S, 2016ASSL..418..355B}. In the case of the ex-situ origin (i.e. mergers), clumps could show a large velocity offset ($\Delta v\gtrsim300$ km s$^{-1}$) with respect to host galaxies \citep[e.g. ][]{2013ApJ...767..151M}. 

These results support the picture that a majority of clumps have the in-situ origin for $z\simeq2$ massive galaxies. As a next step, we examine the major mechanism of clump formation over cosmic time. However, these kinematic analyses are expensive due to requirements of deep IFS spectroscopic observations. Recently, \citet{2015ApJ...800...39G} suggest that an abundance of clumpy galaxies is one of useful probes for investigating clump formation mechanisms. Using $\!${\it HST} imaging data, \citet{2015ApJ...800...39G} have systematically measured number fractions of clumpy galaxies in overall galaxy samples (the clumpy fraction; $f_{\rm clumpy}$) at $z\simeq0-3$. Comparisons between $f_{\rm clumpy}$ and theoretical predictions suggest that clumps are likely to form in VDI and galaxy mergers for galaxies with $\log{M_*/M_\odot}\gtrsim11$ and $\lesssim10$, respectively. However, $f_{\rm clumpy}$ has not still been unveiled beyond $z\simeq3$. Systematic $f_{\rm clumpy}$ measurements over a wide redshift range would provide useful hints for understanding the major clump formation mechanism and properties of clumpy host galaxies.

\begin{deluxetable*}{cccccccc}
\setlength{\tabcolsep}{0.35cm} 
\tabletypesize{\scriptsize}
\tablecaption{Numbers of Our Sample Galaxies for Clumpy Structure Analyses and Clumpy Galaxies}
\tablehead{\colhead{Population} & \multicolumn{6}{c}{$N_{\rm gal} (N_{\rm clumpy})$} & \colhead{$N_{\rm gal}^{\rm total} (N_{\rm clumpy}^{\rm total})$} \\
 & \colhead{$z=0-1$\tablenotemark{a}} & \colhead{$z=1-2$} & \colhead{$z=2-3$} & \colhead{$z=3-4$} & \colhead{$z=4-5$} & \colhead{$z=5-6$} \\
\colhead{(1)}& \colhead{(2)}& \colhead{(3)}& \colhead{(4)}& \colhead{(5)}& \colhead{(6)}& \colhead{(7)}& \colhead{(8)}} 

\startdata
\multicolumn{8}{c}{(Photo-$z$ galaxies)} \\
SFGs (UV) & 2187 (859) & 3236 (1842) & 1429 (672) & 94 (44) & 12 (6) & 4 (2) & 6962 (3425) \\
QGs (UV) & 1079 (195) & 786 (346) & 144 (50) & 11 (3) & 1 (1) & 0 (0) & 2021 (595) \\ 
SFGs (Opt) & 1353 (645) & 1624 (534) & \nodata & \nodata & \nodata & \nodata & 2977 (1179) \\ 
QGs (Opt) & 579 (137) & 523 (95) & \nodata & \nodata & \nodata & \nodata & 1102 (232) \\ \hline 
\multicolumn{8}{c}{} \\
 & $z\simeq4$ & $z\simeq5$ & $z\simeq6$ & $z\simeq7$ & $z\simeq8$ & ($z\simeq10$\tablenotemark{b}) & \\ \hline
LBGs (UV) & 2296 (391) & 1150 (158) & 277 (32) & 106 (11) & 19 (3) & 0 (0) & 3848 (595)  
\enddata

\tablecomments{Columns: (1) Galaxy population. The rest-frame wavelength for clumpy structure analyses is given in  parentheses. (2)-(7) Numbers of SFGs, QGs and LBGs used for our clumpy structure analyses in each redshift bin. The values in parentheses indicate the numbers of clumpy galaxies in each redshift bin. (8) Total numbers of sample galaxies. The values in parentheses indicate the total numbers of clumpy galaxies. }
\tablenotetext{a}{The redshift range is confined to $z=0.5-1$ because of the small galaxy number at $z\lesssim0.5$. }
\tablenotetext{b}{The LBGs at $z\simeq10$ are excluded in our clumpy structure analyses due to the small sample size. }
\label{tab_sample}
\end{deluxetable*}

In this paper, we investigate the redshift evolution of clumpy galaxies using $\!${\it HST} legacy data. We identify clumpy galaxies at $z\simeq0-8$ in a self-consistent clump detection algorithm using the large galaxy sample. This is the second paper in the series studying the galaxy morphology with the $\!${\it HST} samples. \footnote{The first paper presents a study on galaxy sizes at $z\simeq0-10$ \citep[][ hereafter \citetalias{2015ApJS..219...15S}]{2015ApJS..219...15S}.} The organization of this paper is as follows. In Section \ref{sec_data}, we describe the details of the $\!${\it HST} galaxy samples. Section \ref{sec_select} shows the $\!${\it HST} images and galaxies used for our clumpy structure analyses. We present methods to identify clumpy structures in Section \ref{sec_analysis}. In Section \ref{sec_complete}, we perform Monte Carlo simulations to evaluate the intrinsic clump luminosity and clump detection completeness. We show the redshift evolution of $f_{\rm clumpy}$, radial surface brightness (SB) profiles of clumpy galaxies, relations between clumps and physical quantities of host galaxies, and clump colors in Section \ref{sec_results}. In Section \ref{sec_discuss}, we discuss the implications for clump formation mechanisms. We summarize our findings in Section \ref{sec_conclusion}. 

Throughout this paper, we adopt the concordance cosmology with $(\Omega_m, \Omega_\Lambda, h)=(0.3, 0.7, 0.7)$, \citep{2011ApJS..192...18K}. All magnitudes are given in the AB system \citep{1983ApJ...266..713O}. We refer to the $\!${\it HST} F606W, F775W, F814W, F850LP, F098M, F105W, F125W, F140W, and F160W filters as $V_{606}, i_{775}, I_{814}, z_{850}, Y_{098}, Y_{105}, J_{125}, JH_{140}$, and $H_{160}$, respectively.

\section{Data and Samples}\label{sec_data}

We use the following two galaxy samples in this study. These galaxy samples are constructed from the deep optical and near-infrared imaging data taken by $\!${\it HST} deep extra-galactic legacy surveys. \citetalias{2015ApJS..219...15S} summarizes the limiting magnitudes and point spread function (PSF) sizes of the $\!${\it HST} images.

\subsection{Sample of Photo-$z$ Galaxies in 3D-HST+CANDELS}\label{sec_3dhst}

The first sample is made of $176,152$ $\!${\it HST}/WFC3-IR detected galaxies with photometric redshifts (hereafter photo-$z$ galaxies) at $z=0-6$ taken from \citet{2014ApJS..214...24S}. These galaxies are identified in five Cosmic Assembly Near-infrared Deep Extragalactic Legacy Survey (CANDELS) fields \citep{2011ApJS..197...35G, 2011ApJS..197...36K}, and detected in stacked WFC3-IR images. The stacked WFC3-IR images are created by combining three noise-equalized $\!${\it HST}/WFC3-IR bands, $J_{125}, JH_{140}$, and $H_{160}$ \citep[see ][for more details]{2014ApJS..214...24S}. The imaging of these IR-bands yields roughly a stellar mass-limited sample. The photometric properties and the results of SED fitting for all the sources are summarized in \citet{2014ApJS..214...24S}. The $\!${\it HST} images and catalogues are publicly released at the 3D-HST website. \footnote{http://3dhst.research.yale.edu/Home.html} We use galaxies whose physical quantities and photometric redshifts are well derived from SED fitting (specifically, sources with {\tt use\_phot}$=1$ in the public catalogues). In this paper, we assume \citet{1955ApJ...121..161S} initial mass function (IMF) for the fair comparison with previous studies. To obtain the Salpeter IMF values of stellar masses ($M_*$) and star formation rates (SFRs), we multiply the \citet{2003PASP..115..763C} IMF values from the \citet{2014ApJS..214...24S} catalogue by a factor of $1.8$. We divide the sample of photo-$z$ galaxies into two subsamples of star-forming galaxies (SFGs) and quiescent galaxies (QGs) by specific SFR (sSFR) values. Galaxies with sSFR $\geqslant0.1$ and $<0.1$ Gyr$^{-1}$ are classified as SFGs and QGs, respectively. 

Structural parameters such as semi-major axes ($R_{\rm e,major}$), position angle (P.A.), axial ratios ($q$), circularized effective radii ($r_{\rm e}$), and S\'ersic indices ($n$) have been obtained from our S\'ersic profile fitting \citep{1963BAAA....6...41S,1968adga.book.....S} with {\sc GALFIT} \citep{2002AJ....124..266P,2010AJ....139.2097P} \citepalias{2015ApJS..219...15S}. In the {\sc GALFIT} fitting, clumpy subcomponents of galaxies have been masked to acquire these structural parameters of major stellar components. By using {\sc SExtractor}, we have generated {\it mask images} that define regions of neighboring objects and clumpy subcomponents around the main galaxies. We have conducted the {\sc GALFIT} profile fitting with these mask images, omitting the masked regions \citep[see ][]{2002AJ....124..266P}. The SFR surface density (SFR SD), $\Sigma_{\rm SFR}$, has been calculated by an equation of $\Sigma_{\rm SFR} = {\rm SFR} / (2\pi r_{\rm e}^2)$. See \citetalias{2015ApJS..219...15S} for details.

\subsection{Sample of LBGs in CANDELS, HUDF09/12, and HFF}\label{sec_lbg}

The second sample consists of $10,454$ LBGs at $z\simeq4-10$ made by \citet{2015arXiv151107873H} in the CANDELS, the {\it Hubble} Ultra Deep Field 09+12 \citep[HUDF 09+12; ][]{2006AJ....132.1729B,2011ApJ...737...90B,2013ApJS..209....6I,2013ApJ...763L...7E} fields, \footnote{http://archive.stsci.edu/prepds/xdf/} and the parallel fields of Abell 2744 and MACS0416 in the {\it Hubble} Frontier Fields \citep[e.g., ][]{2015ApJ...800...84C, 2015ApJ...800...18A, 2015ApJ...808..104O,2015ApJ...799...12I}. These LBGs are selected with the color criteria, similar to those of \citet{2015ApJ...803...34B}. We perform source detections by {\sc SExtractor} \citep{1996A&AS..117..393B} in coadded images constructed from bands of $Y_{098}Y_{105}J_{125}H_{160}$, $J_{125}H_{160}$, and $H_{160}$ for the $z\simeq4-7$, $8$, and $10$ LBGs, respectively. The $JH_{140}$ band is included in the coadded image for the $z\simeq7-8$ LBGs in the HUDF09+12 field. The flux measurements are carried out in \cite{1980ApJS...43..305K}-type apertures with a \citeauthor{1980ApJS...43..305K} parameter of $1.6$ whose diameter is determined in the $H_{160}$ band. In two-color diagrams, we select objects with a Lyman break, no extreme-red stellar continuum, and no detection in passbands bluer than the spectral drop. See \citet{2015arXiv151107873H} for more details of the source detections and LBG selections. 

Similarly for the photo-$z$ galaxies, we have obtained $R_{\rm e,major}$, $q$, and $r_{\rm e}$ for the LBG sample. We note that the LBGs have no S\'ersic index measurements due to fixing $n$ values in the {\sc GALFIT} fitting. We have also derived $\Sigma_{\rm SFR}$ and UV slope $\beta$ for the LBGs. The $\beta$ value is defined by $f_{\lambda}\propto\lambda^\beta$, where $f_{\lambda}$ is the galaxy spectrum at the rest-frame wavelengths of $\simeq1500-3000$\,\AA. See \citetalias{2015ApJS..219...15S} for details.

\section{HST-Band Images and Galaxies for Clumpy Structure Analyses}\label{sec_select}

In this section, we describe {\it HST}-band images and galaxies used for our clumpy structure analyses. 

\subsection{HST-band images} 

To minimize the effect of morphological {\it K}-correction, we use images of $\!${\it HST} multi-wavebands. In this study, we investigate clumps at the rest-frame UV ($1500-3000$\,\AA) and optical ($4500-8000$\,\AA) wavelengths. These clumps are hereafter referred to as ``UV clumps" and ``optical clumps", respectively. For UV clumps, we use $V_{606}$ for $z=0-2$, $I_{814}$ for $z=2-3$ photo-$z$ galaxies, and coadded images of $Y_{098}Y_{105}J_{125}H_{160}$ and $J_{125}H_{160}$ for $z\gtrsim4$ LBGs. For optical clumps, we choose $J_{125}$ for $z=0-1$ and $H_{160}$ for $z=1-2$ photo-$z$ galaxies. \footnote{We make use of $z_{850}$ for GOODS-North that has not been observed with the $I_{814}$ band. } 

\citet{2015ApJ...800...39G} have analyzed a larger number of ACS bands, $B_{435}V_{606}i_{775}z_{850}$, for UV clumps in the GOODS-S field compared to ours. Here we restrict our bands to $V_{606}$ and $I_{814}$ for a homogeneous analysis in all the CANDELS fields due to the lack of $B_{435}, i_{775}$ and/or $z_{850}$ in AEGIS, COSMOS and UDS. The difference of these $\!${\it HST} band choices has provided no significant impacts on results of clumpy structure analyses \citep{2015ApJ...800...39G}.

\begin{figure*}[t!]
  \begin{center}
    \includegraphics[width=170mm]{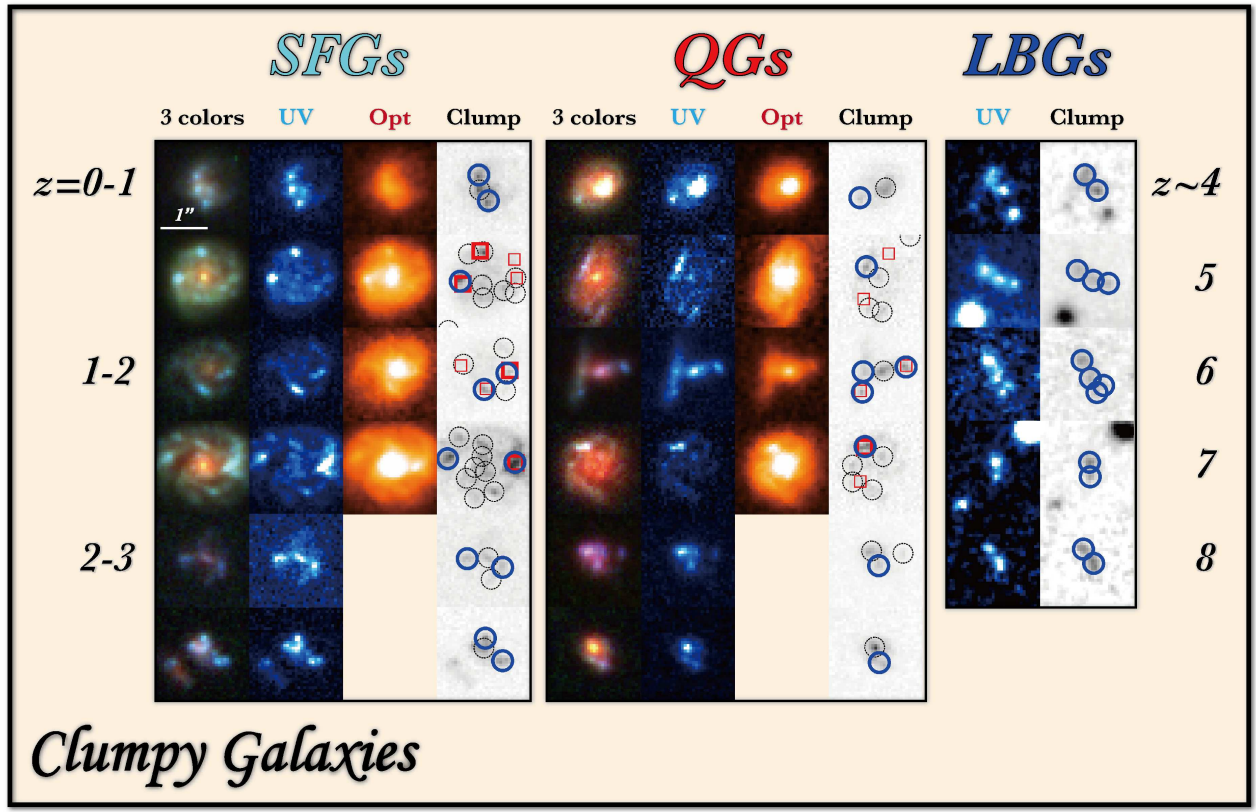}
  \end{center}
  \caption[]{{\footnotesize Examples of clumpy galaxies identified in our selections. The left, middle, and right panel-sets indicate SFGs, QGs, and LBGs, respectively. Each of the four-panel sets for the SFGs and the QGs presents the three-color composite, the rest-frame UV, the rest-frame optical images, and the clump position maps, from left to right. In the clump position maps, the black and blue circles denote UV clump candidates and selected UV clumps with $L_{\rm cl}/L_{\rm gal}\geqslant0.08$, respectively. The thick and thin red squares indicate optical clumps with $L_{\rm cl}/L_{\rm gal}\geqslant0.1$ and $L_{\rm cl}/L_{\rm gal}=0.08-0.1$, respectively. Each row, from top to bottom, represents example clumpy galaxies from $z\simeq0$ to $z\simeq3$. Each of the two-panel sets for the LBGs exhibits the coadded images (left) and the clump position maps (right). Each row, from top to bottom, denotes clumpy LBGs from $z\simeq4$ to $z\simeq8$. The white tick indicates the size of 1$^{\prime\prime}$. }}
  \label{fig_postage_clump}
\end{figure*}

\subsection{Sample galaxies} 

We select bright, large, and face-on galaxies from the two galaxy samples (Section \ref{sec_data}) for secure identifications of clumpy structures. We analyze photo-$z$ galaxies with $\log{M_*/M_\odot}\geqslant9$, $H_{160}<24.5$ mag, an $H_{160}$-band semi-major axis of $R_{\rm e,major}\geqslant0.\!\!^{\prime\prime}2$, and $q\geqslant0.5$ in the ACS band images. These selection criteria are the same as those of \citet{2015ApJ...800...39G}. We restrict analyses to galaxies with $R_{\rm e,major}\geqslant0.\!\!^{\prime\prime}4$ for the WFC3/IR band images due to a broader PSF size than that of ACS. In total, 6,962 (2,977) SFGs and 2,021 (1,102) QGs are selected for analyses of UV (optical) clumps. 

For the LBGs, we apply only a magnitude cut of $m_{\rm UV} < 27$ mag where we securely compare results of automated clump identification methods and visual inspections in typical \!{\it HST} deep fields \citep[e.g., ][]{2013ApJ...773..153J}. Moreover, with this $m_{\rm UV}$ threshold, we obtain at least $10$ sample galaxies at each redshift. This sample size enables us to perform statistical measurements of clumpy structures up to $z\simeq8$. The total number of LBGs at $z\simeq4-8$ is 3,848 for the clumpy structure analyses. Here we exclude the LBGs at $z\simeq10$ due to the small sample size. \footnote{Our visual inspections confirm that the LBGs at $z\simeq10$ show no clumpy structures \citep[see also, ][]{2015ApJ...808....6H}.} The effect of the selection criterion choices is discussed in Sections \ref{sec_fclumpy_evo} and \ref{sec_systematics}.  

Table \ref{tab_sample} summarizes the number of photo-$z$ galaxies and LBGs used for our clumpy structure analyses.

\section{Analysis}\label{sec_analysis}

In this section, we describe methods to identify clumpy structures in the $\!${\it HST} images. The clumpy structures of galaxies have been examined in several ways, e.g., by using {\sc SExtractor} \citep[e.g., ][]{2014ApJ...786...15M}, {\sc Clumpfind} \citep{1994ApJ...428..693W}, the non-parametric morphological indices \citep[e.g., ][]{2003ApJS..147....1C}, and based on the visual inspection \cite[e.g., ][]{2007ApJ...658..763E}. Recently, \citet{2015ApJ...800...39G} have have developed a method to investigate clumpy galaxies at $z\sim0-3$ and identified off-center UV clumps in large $\!${\it HST}/ACS data. In our study, we employ a method similar to that of \citet{2015ApJ...800...39G} to compare our measurements with those of \citeauthor{2015ApJ...800...39G}. We also newly apply this technique to both ACS and WFC3/IR images. The identification parameters for the WFC3/IR images are determined by our careful visual inspections. 

\begin{figure*}[t!]
  \begin{center}
    \includegraphics[height=60.5mm]{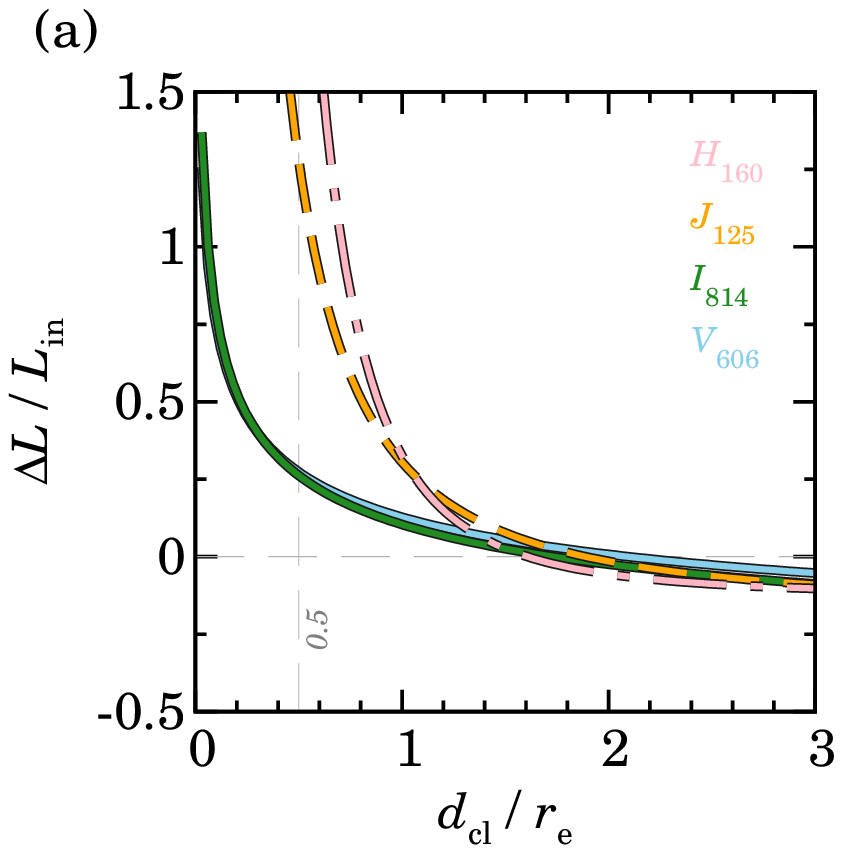}
    \includegraphics[height=60mm]{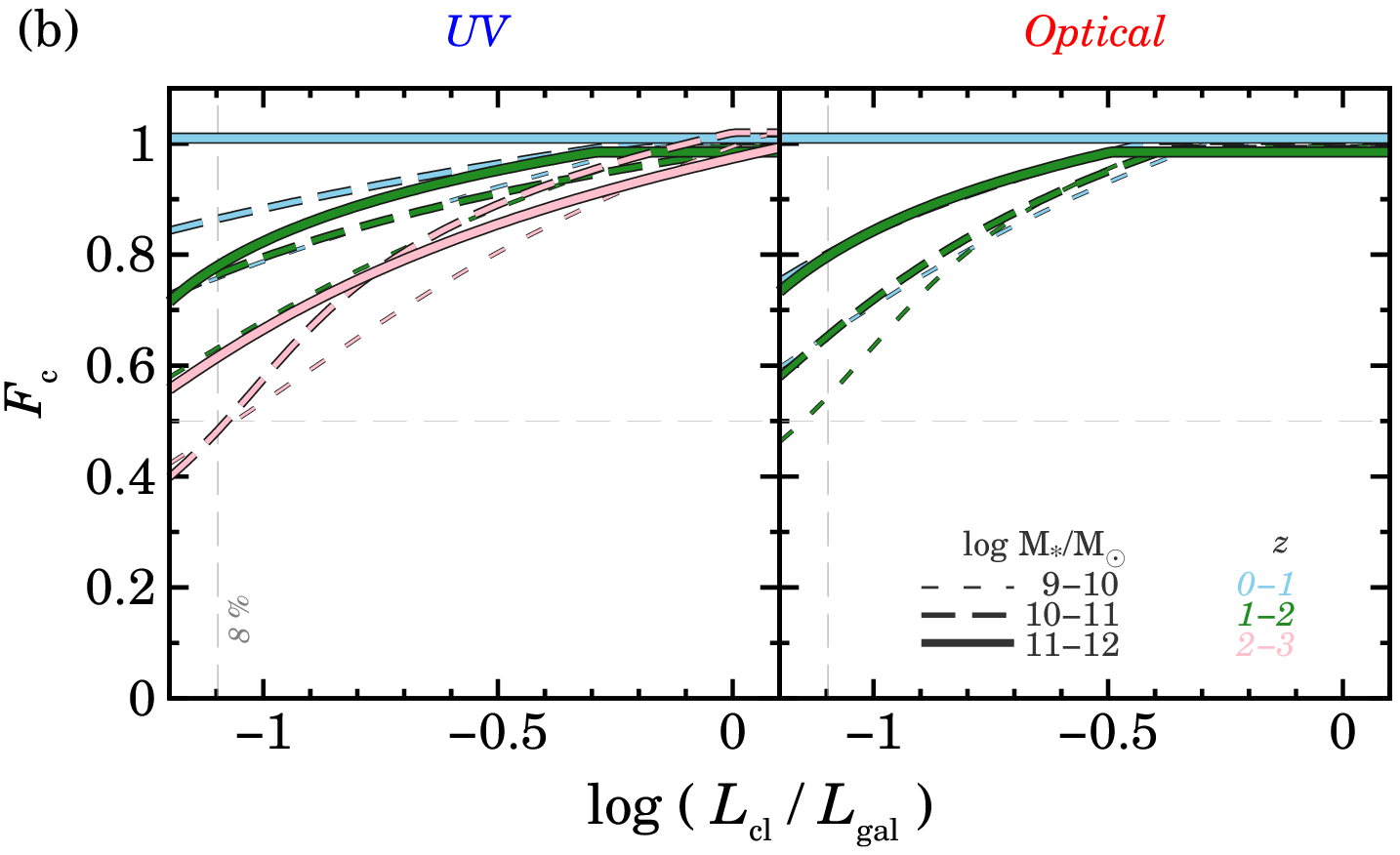}
  \end{center}
  \caption[]{{\footnotesize Results of the Monte-Carlo simulations for the clump identifications. (a) Luminosity overestimate ratios, $\Delta L/L_{\rm in}$, as a function of $d_{\rm cl}/r_{\rm e}$. The cyan solid, green solid, orange dashed, and magenta dot-dashed curves show $\Delta L/L_{\rm in}$ for the $V_{606}$, $I_{814}$, $J_{125}$, and $H_{160}$-band images, respectively. The vertical dashed line denotes the inner $d_{\rm cl} / r_{\rm e}$ threshold ($d_{\rm cl} / r_{\rm e}=0.5$; see Section \ref{sec_analysis} for details). The horizontal dashed line presents $\Delta L/L_{\rm in}=0$. (b) Detection completeness of clumps, $F_c$, as a function of $L_{\rm cl}/L_{\rm gal}$. The left and right panels indicate $F_c$ in the rest-frame UV and optical wavelengths, respectively. The cyan, green, and magenta curves denote $F_c$ at $z=0-1$, $1-2$, and $2-3$, respectively. The short-dashed, long-dashed, and solid curves present $\log{M_*/M_\odot}=9-10$, $10-11$, and $11-12$, respectively. The horizontal dashed lines denote $F_c = 0.5$. The vertical dashed lines indicate the threshold of fractional luminosity, $L_{\rm cl}/L_{\rm gal}=8\%$, for the clump identifications. }}
  \label{fig_comp}
\end{figure*}

\begin{deluxetable*}{lccccc}
\setlength{\tabcolsep}{0.35cm} 
\tabletypesize{\scriptsize}
\tablecaption{Summary of Previous Studies on Clumpy Galaxies}
\tablehead{\colhead{Reference} & \colhead{Sample ($N_{\rm gal}$)} & \colhead{$\log{M_*}, L_{\rm UV}$} & \colhead{Redshift} & \colhead{Method} & \colhead{Detection Band} \\
\colhead{} & \colhead{} & \colhead{($M_\odot, L_{z=3}^*$)} & \colhead{} & \colhead{} & \colhead{} \\
\colhead{(1)}& \colhead{(2)}& \colhead{(3)}& \colhead{(4)} &  \colhead{(5)}& \colhead{(6)}} 

\startdata
\citet{2007ApJ...658..763E} & Starbursts (1003) & N/A & $0 - 5$ & Visual & $i_{775}$ \\ 
\citet{2010MNRAS.406..535P} & Emission-line galaxies (63) & $\log{M_*}>10.3$ & $\simeq0.6$ & Visual & $B_{435}$ \\
\citet{2010ApJ...710..979O} & LBAs (30) & $\log{M_*}=9-11$ & $0.1\sim0.3$ & Visual & UV \\
\citet{2012ApJ...757..120G} & Star-forming galaxies (10) & $\log{M_*}>10$ & $1.5-2.5$ & Algorithm & $z_{850}$ \\
\citet{2012ApJ...753..114W} & Star-forming galaxies (649) & $\log{M_*}>10$ & $0.5-2.5$ & Algorithm & UV, opt \\
\citet{2014ApJ...780...77T} & HAEs (100) & $\log{M_*}9-11.5$ & $2-2.5$ & {\sc Clumpfind} & $V_{606}$, $H_{160}$ \\
\citet{2014ApJ...786...15M} & $I_{814}<22.5$ galaxies (24027) & $\log{M_*}>9.5$ & $0.2-1$ & Algorithm & $I_{814}$ \\
\citet{2015ApJ...800...39G} & Star-forming galaxies (3229) & $\log{M_*}\simeq9-11.5$ & $0\sim3$ & Algorithm & UV \\ \hline
\citet{2006ApJ...636..592L} & LBGs (82) & $L_{\rm UV}\gtrsim 0.4$ & $4$ & $CAGM_{20}$ & UV \\
\citet{2006ApJ...652..963R} & LBGs (1333) & $L_{\rm UV}\gtrsim 0.7$ & $3-5$ & S\'ersic & UV \\
\citet{2009MNRAS.397..208C} & LBGs (133) & $L_{\rm UV}\gtrsim 0.3$ & $4-6$\tablenotemark{a} & $CASGM_{20}$ & UV \\
\citet{2010ApJ...709L..21O} & LBGs (21) & $L_{\rm UV}\gtrsim 0.06$ & $7-8$ & Visual & UV \\
\citet{2012ApJ...759...29L} & LBGs (309) & $L_{\rm UV}\simeq0.3-1$\tablenotemark{b} & $1.5-3.6$ & $CAGM_{20}$ & UV \\
\citet{2013ApJ...773..153J}\tablenotemark{c} & LBGs (24) & $L_{\rm UV}\gtrsim 0.7$ & $5.7-7$ & Visual & UV \\
\cite{2014ApJ...785...64S}\tablenotemark{d} & LAEs (32) & $L_{\rm UV}\simeq 0.3-1$ & $2.2$ & $A$ & UV \\
\citet{2014arXiv1409.1832C} & LBGs (1318) & $L_{\rm UV}\gtrsim 0.3$ & $4-8$ & $A$ & UV \\ 
\cite{2015ApJ...804..103K} & LBGs (39) & $L_{\rm UV}\gtrsim 0.04$ & $6-8$ & Visual & UV \\ \hline
This work & Photo-$z$ galaxies (13062\tablenotemark{e}) & $\log{M_*}\simeq9-12$ & $0\sim6$ & \citetalias{2015ApJ...800...39G} & UV, opt \\ 
 & LBGs (3848) & $L_{\rm UV}\gtrsim 0.3$ & $4\sim8$ & \citetalias{2015ApJ...800...39G} & UV \\
 & ($N_{\rm gal}^{\rm total}=16910$) & \nodata & $0\sim8$ & \citetalias{2015ApJ...800...39G} & UV, opt

\enddata

\tablecomments{Columns: (1) Reference. (2) Galaxy population. The values in parentheses present the numbers of sample galaxies. (3) Stellar mass or UV luminosity ranges. (4) Redshift ranges. (5) Methods to identify clumpy structures (``{\tt Visual}": visual inspection; ``{\tt Algorithm}": original algorithm in the literature; ``{\sc Clumpfind}": the {\sc Clumpfind} software; ``$C$": Concentration; ``$A$": Asymmetry; ``$S$": Clumpiness; ``$G$": Gini coefficient; ``$M_{20}$": the second-order moment parameter; ``{\tt S\'ersic}": S\'ersic profile fitting; ``\citetalias{2015ApJ...800...39G}": the method of \citet{2015ApJ...800...39G}). See \citet{2014MNRAS.444.1125C} for the definition of the non-parametric indices. For comparison in Figure \ref{fig_z_fclumpy_lbg}, we choose $f_{\rm clumpy}^{\rm UV}$ measured with non-parametric indices (e.g., $GM_{20}$ and $A$) if the literature has several methods to identify clumps. This is because the non-parametric indices would select clumpy galaxies similar to ours compared to the other methods. (6) Detection bands for clump identifications. }
\tablenotetext{a}{\citet{2009MNRAS.397..208C} have also used galaxy samples at $z\lesssim3$, and have identified a possible peak of the galaxy merger fraction at $z\simeq2$ by fitting data points at $z\simeq0-5$.}
\tablenotetext{b}{The $L_{\rm UV}$ range is estimated from the $M_{\rm UV}-M_*$ relations in \citetalias{2015ApJS..219...15S}. }
\tablenotetext{c}{\citet{2013ApJ...773..153J} include a sample of Ly$\alpha$ emitters. }
\tablenotetext{d}{This LAE sample has $M_{\rm UV}$ and Ly$\alpha$ equivalent widths similar to those of typical LBGs. }
\tablenotetext{e}{The value is the total number of SFGs and QGs for the analyses of the rest-frame UV and optical wavelengths. See Table \ref{tab_sample}. }
\label{tab_previous}
\end{deluxetable*}

\begin{figure*}[t!]
  \begin{center}
   \includegraphics[width=130mm]{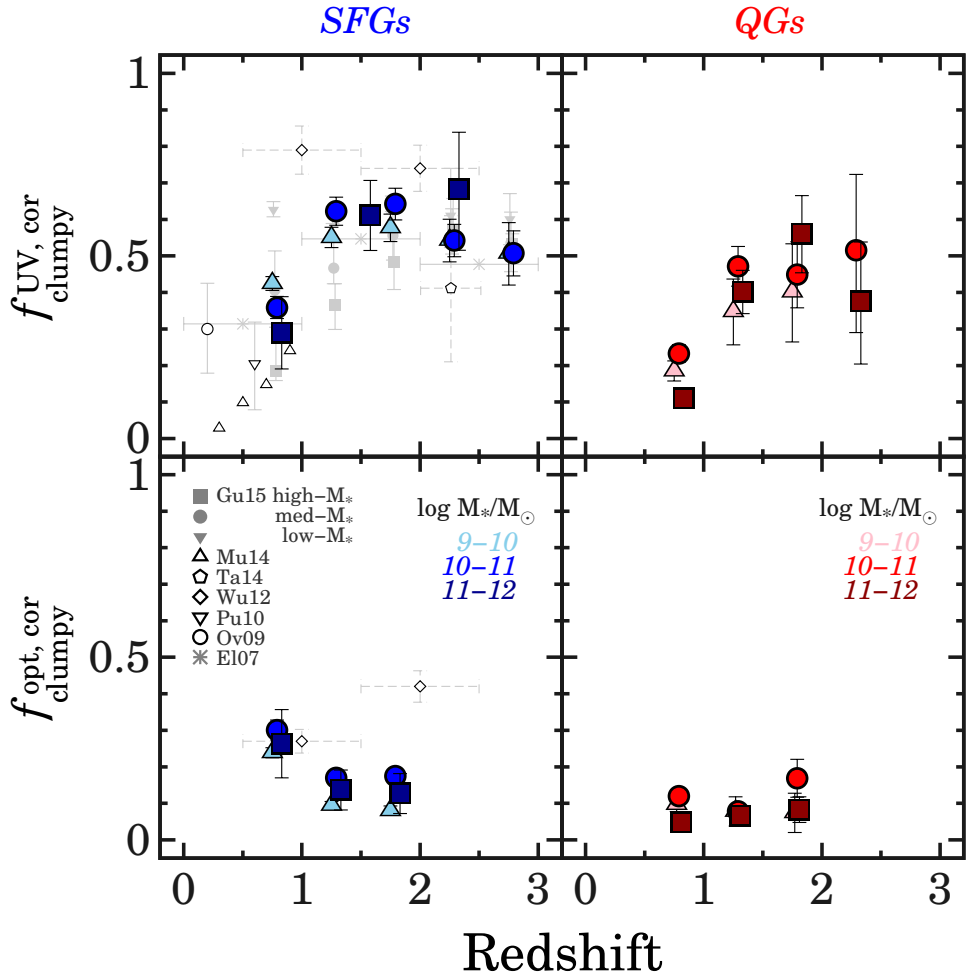}
  \end{center}
  \caption[]{{\footnotesize Redshift evolution of $f_{\rm clumpy}^{\rm UV, cor}$ (top) and $f_{\rm clumpy}^{\rm opt, cor}$ (bottom) for the photo-$z$ galaxies at $z\simeq0-3$. The left and right panels indicate $f_{\rm clumpy}^{\rm cor}$ for the SFGs and the QGs, respectively. The filled blue and red symbols denote our $f_{\rm clumpy}^{\rm cor}$ measurements for the SFGs and the QGs in each $M_*$ bin ($\log{M_*/M_\odot}=9-10$: triangles; $10-11$: circles; $11-12$: squares). The error bars of $f_{\rm clumpy}$ are calculated via analytical error propagation from Poisson errors, $\sqrt{N}$, where $N$ is the numbers of sample galaxies or selected clumpy galaxies. The gray symbols show $f_{\rm clumpy}$ in the literature (asterisks: \citealt{2007ApJ...658..763E}; open circle: \citealt{2010ApJ...710..979O}; inverse triangle:  \citealt{2010MNRAS.406..535P}; diamonds: \citealt{2012ApJ...753..114W}; pentagon: \citealt{2014ApJ...780...77T}; triangles: \citealt{2014ApJ...786...15M}; gray filled inverse triangles, circles, and squares: galaxies with $\log{M_*/M_\odot}=9-9.8$, $9.8-10.6$, and $10.6-11.4$, respectively, in \citealt{2015ApJ...800...39G}). The error bars are put by Poisson statistics from the galaxy number counts if the reference does not show uncertainties of $f_{\rm clumpy}$. Note that our SFGs and QGs are classified by the sSFR value, $\log{\rm sSFR}=0.1$ Gyr$^{-1}$. In this sSFR criterion, galaxies with a moderately high sSFR are included in our QG sample (see, Section \ref{sec_fclumpy_photoz_lbg}). }}
  \label{fig_z_fclumpy_tile}
\end{figure*}

\begin{enumerate}

\item {\it Object images} --- 
we extract $18^{\prime\prime}\times18^{\prime\prime}$ cutout images from the $\!${\it HST} data at the position of each photo-$z$ galaxy and LBG. The size of cutout images is sufficiently large to investigate entire galaxy structures even for extended objects at $z\simeq0-1$.

\item {\it Segmentation maps} --- 
we also cutout {\sc SExtractor} segmentation maps with the same size as for the object images to define areas searching for clumps. We use $H_{160}$ segmentation maps that are generated by our {\sc SExtractor} detections and publicly released in the 3D-HST website. {\sc SExtractor} frequently outputs segmentation areas smaller than intrinsic galaxy sizes for faint sources. We enlarge individual segmentation areas \citep[so-called ``dilation"; see, ][]{2013ApJS..207...24G,2013ApJS..206...10G} with a task of {\sc gauss} in {\sc IRAF} according to isophotal areas of each galaxy. Our dilation technique is slightly different from e.g. \citet{2013ApJS..206...10G} with the {\sc TFIT} software. We find that the difference of dilated areas provides negligible impacts on our results of our clumpy structure analyses. 

\item {\it Detection of clumps} --- 
we detect clumpy objects within galaxies. We first smooth the object images through a boxcar filter with a side of ten pixel ($=0.\!\!^{\prime\prime}6$). We find that the choice of the boxcar filter size does not significantly affect the results of clump identification (e.g., $f_{\rm clumpy}$). Next, we subtract the smoothed images from the original ones to stand out clumpy objects. These subtracted images are referred to {\it contrast} images. To avoid detections of sky noise, we mask all pixels with a flux less than $2\sigma$ of sky background fluctuation. Sky background levels are measured in the contrast images after a $3\sigma$ flux clipping. Finally, we perform {\sc SExtractor} detections within regions defined with the dilated segmentation maps. We require these clumpy objects to have an area larger than five (seven) contiguous pixels, ${\tt DETECT\_MINAREA}=5$ (7), for the ACS (WFC3/IR) images. We define ``clump candidates" as these detected objects.

\item {\it Photometry of clumps} --- 
we carry out photometry for the clump candidates. We measure an aperture flux, $f_{\rm aper}$, within a radius of $r_{\rm aper}=3$ (6) pixels, corresponding to $0.\!\!^{\prime\prime}18$ ($0.\!\!^{\prime\prime}36$), for clump candidates in the ACS (WFC3/IR) images. These photometric apertures would include background fluxes of host galaxies. We subtract the background fluxes of host galaxies from $f_{\rm aper}$. The average SB in a circular ring with at $r=4-10$ ($8-20$) pixels is considered to be the background fluxes for clump candidates in the ACS (WFC3/IR) images. The background-subtracted $f_{\rm aper}$ is converted to a total flux, $f_{\rm tot}$, by applying aperture corrections. The factors of aperture corrections range from $\simeq0.7$ to $\simeq0.8$. The aperture photometry with the aperture correction generally provides reliable measurements of the intrinsic clump magnitudes. This is because almost all of the clumps selected by our method have a point source-like radial profile \citep[see also, ][]{2015ApJ...800...39G}. We convert $f_{\rm tot}$ to a luminosity, $L_{\rm cl}$, under the assumption that these clump candidates are physically associated with host galaxies.

\citet{2015ApJ...800...39G} find that $L_{\rm cl}$ at a small galactocentric distance, $d_{\rm cl}$, tends to be overestimated due to bright central structures (e.g. bulges). We evaluate these ``{\it luminosity overestimate ratios}" in Monte-Carlo simulations (Section \ref{sec_complete}), and correct for $L_{\rm cl}$.

\item {\it Selection of clumps} --- 
we select off-center clumps among the clump candidates based on their luminosity and position. We employ the following criteria: 

\begin{equation}\label{eq_criteria}
\begin{aligned}
L_{\rm cl} / L_{\rm gal} \geqslant 0.08 \ {\rm and} & \\ 
0.5 \leqslant d_{\rm cl} / r_{\rm e} < 8
\end{aligned}
\end{equation}

where $L_{\rm gal}$ is a luminosity of a host galaxy. To measure $d_{\rm cl}$, we use flux-weighted centers determined by the {\sc SExtractor} detections of \citet{2014ApJS..214...24S} and \citet{2015arXiv151107873H}. The segmentation areas are typically more effective than the outer boundary of $d_{\rm cl} / r_{\rm e} = 8$. In this study, we define ``clumpy galaxies" as objects with at least one off-center clump that meets Equation \ref{eq_criteria}. 

For the LBG sample, we do not use the $d_{\rm cl} / r_{\rm e}$ criterion. This is because high-$z$ LBGs are unlikely to have an evolved bright cores at the centers. We define clumpy galaxies as objects with at least {\it two} clumps that meet only the $L_{\rm cl} / L_{\rm gal}$ criterion in Equation \ref{eq_criteria}.

\end{enumerate}

Figure \ref{fig_postage_clump} shows examples of clumpy galaxies. The results of clump identification appears to be in good agreement with visual inspections.

\begin{figure*}[t!]
  \begin{center}
    \includegraphics[width=160mm]{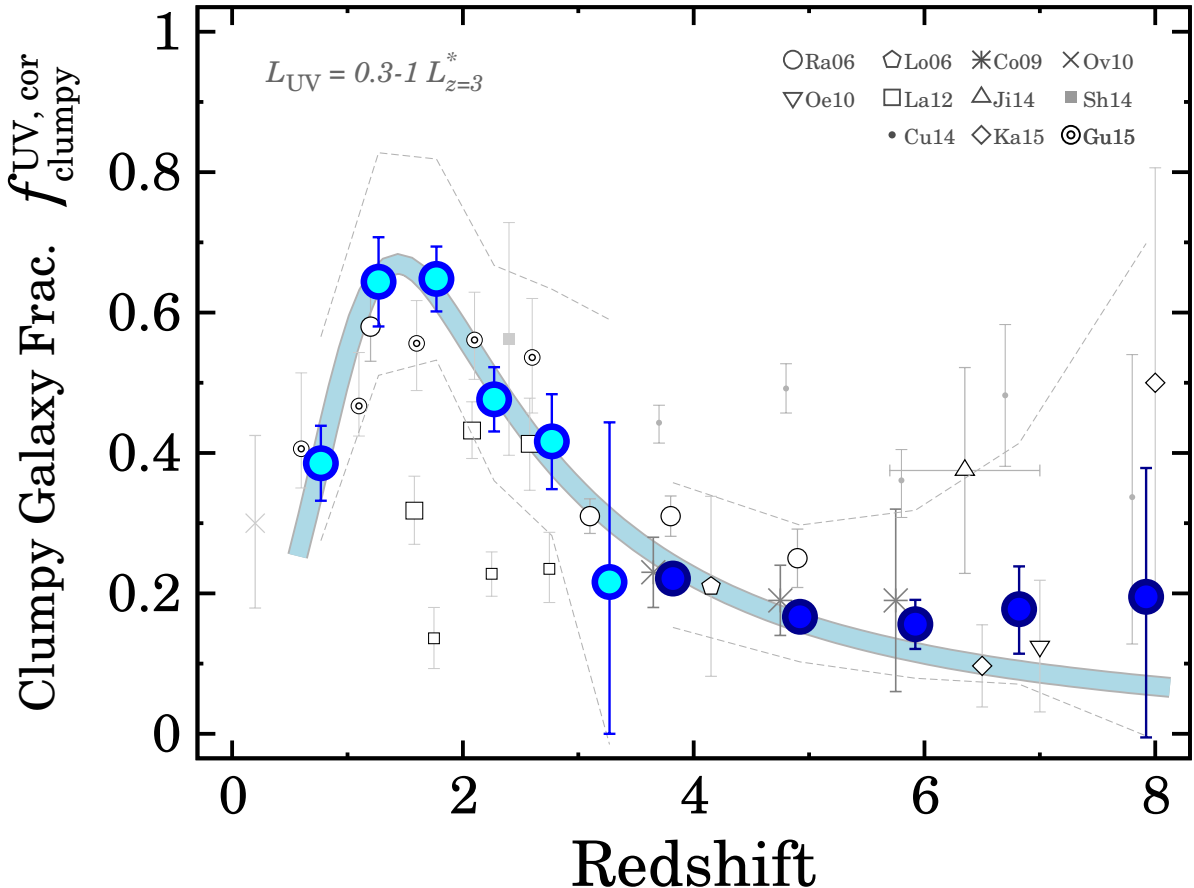}
  \end{center}
  \caption[]{{\footnotesize Redshift evolution of the clumpy galaxy fraction, $f_{\rm clumpy}^{\rm UV, cor}$, at $z\simeq0-8$ for the star-forming galaxies with $L_{\rm UV} = 0.3-1 L_{z=3}^*$. The filled cyan and blue circles denote the SFGs and the LBGs, respectively. The error  bars are given by Poisson statistics from the galaxy number counts. The cyan solid curve denotes the best-fit formula of Equation \ref{eq_madau} \citep{2014ARA&A..52..415M}. The gray dashed lines present the upper and lower limits of $f_{\rm clumpy}^{\rm UV, cor}$ derived by changing $F_c$ in a range of $0.5\times F_c^{\rm fid} - 2\times F_c^{\rm fid}$, where $F_c^{\rm fid}$ is the fiducial $F_c$ value for each redshift. The gray symbols show LBAs and LBGs with clumps or irregular morphologies in the literature (open circles: \citealt{2006ApJ...652..963R}; pentagon: \citealt{2006ApJ...636..592L}; asterisks: \citealt{2009MNRAS.397..208C}; inverse triangle: \citealt{2010ApJ...709L..21O}; cross: \citealt{2010ApJ...710..979O}; open squares: \citealt{2012ApJ...745...85L}; triangle: \citealt{2013ApJ...773..153J}; filled square: \citealt{2014ApJ...785...64S}; diamonds: \citealt{2015ApJ...804..103K}; dots: \citealt{2014arXiv1409.1832C}). The small and large open squares are based on the selection of clumpy galaxies with the $GM_{20}$ and $A$ indices, respectively. The double circles represent $f_{\rm clumpy}^{\rm UV, cor}$ for SFGs with $\log{M_*/M_\odot}=9.6-10.4$ in \citet{2015ApJ...800...39G}. The error bars are given by Poisson statistics from the number of sample galaxies, if no error bar has been presented in the literature. }}
  \label{fig_z_fclumpy_lbg}
\end{figure*}

\begin{figure}[t!]
  \begin{center}
    \includegraphics[width=84mm]{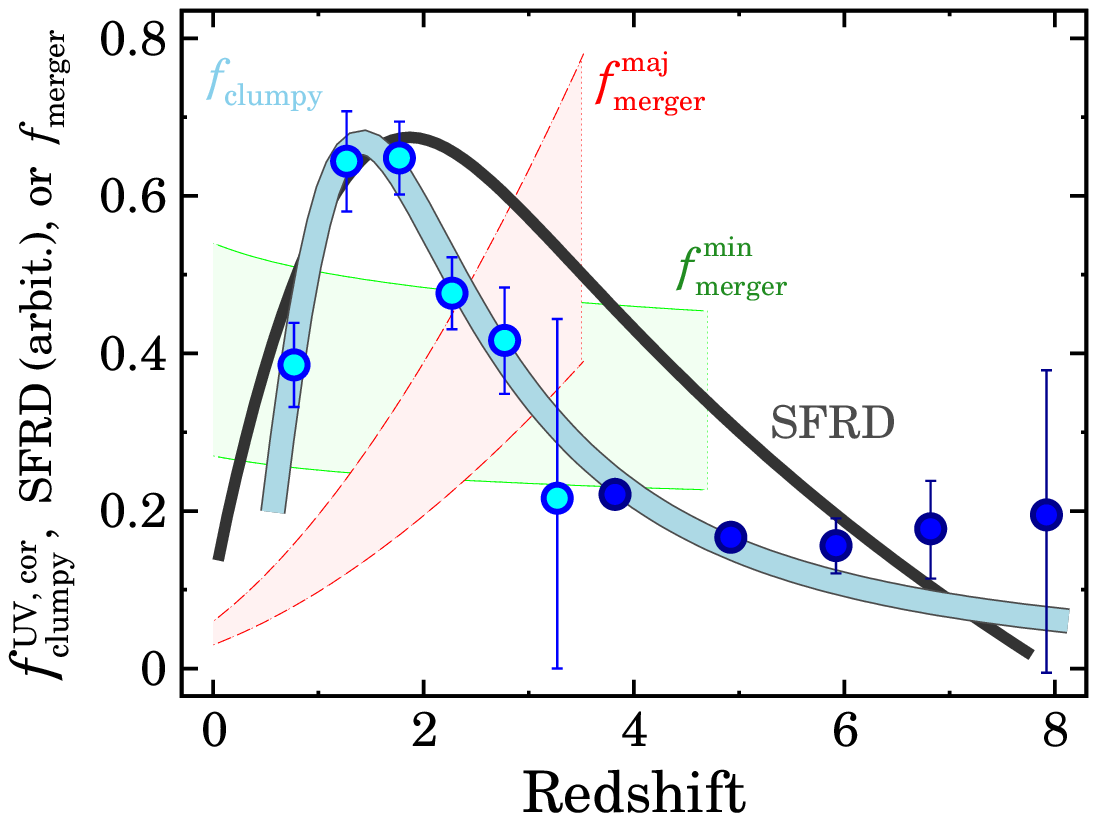}
  \end{center}
  \caption[]{{\footnotesize Same as Figure \ref{fig_z_fclumpy_lbg}, but for the comparisons with the cosmic SFR density \citep[the black solid line; ][]{2014ARA&A..52..415M} and merger fractions \cite[the shaded regions; ][]{2011ApJ...742..103L}. The red and green shaded regions present the galaxy major and minor merger fractions, $f_{\rm merger}^{\rm maj}$ and $f_{\rm merger}^{\rm min}$, respectively, assuming that the merger observability timescale ranges from $1$ to $2$ Gyr. The cosmic SFR density is arbitrarily shifted and scaled along the $y$ axis. }}
  \label{fig_z_fclumpy_lbg_model}
\end{figure}

\section{Luminosity Overestimate Ratios and Detection Completeness}\label{sec_complete}

We perform Monte-Carlo simulations to evaluate luminosity overestimate ratios ($\Delta L/L_{\rm in}$) and detection completeness ($F_c$) in the same manner as the analysis in \citet{2015ApJ...800...39G}. First, we create an artificial clump by using {\sc MKOBJECTS} in the {\sc IRAF} package. Second, the artificial clump is broadened to the PSF size of an $\!${\it HST} image. Third, the artificial clump is added to a segmentation area of a real galaxy. Here we randomly change luminosity $L_{\rm cl} / L_{\rm gal}$ and position $d_{\rm cl} / r_{\rm e}$ of the artificial clump in ranges of $-2.5\leqslant \log{(L_{\rm cl} / L_{\rm gal})} < 0$ and $0.1\leqslant d_{\rm cl} / r_{\rm e} < 3$, respectively. Finally, we detect the artificial clump and carry out photometry in the same manner as for the real galaxies (Section \ref{sec_analysis}). This procedure is repeated 50 times for each galaxy at a given $L_{\rm cl} / L_{\rm gal}$ and $d_{\rm cl} / r_{\rm e}$. We perform the simulation for $1,000$ galaxies  in each $\!${\it HST} band and field. 

The panel (a) of Figure \ref{fig_comp} shows the luminosity overestimate ratios, $\Delta L/L_{\rm in} \equiv (L_{\rm out} - L_{\rm in}) / L_{\rm in}$, as a function of $d_{\rm cl} / r_{\rm e}$, where $L_{\rm in}$ ($L_{\rm out}$) is the input (output) clump luminosity in the Monte-Carlo simulation. As shown in the panel (a), $\Delta L / L_{\rm in}$ gradually increases towards the galactic centers, and reach to $\Delta L / L_{\rm in}\simeq0.3$ at $d_{\rm cl} / r_{\rm e}=0.5$ for the $V_{606}I_{814}$ images. These trends are consistent with that of \citet{2015ApJ...800...39G}. In contrast, we find that the $J_{125}H_{160}$ images have a relatively high value of $\Delta L / L_{\rm in}\simeq1.5$ at $d_{\rm cl} / r_{\rm e}=0.5$. These high $\Delta L / L_{\rm in}$ values would result from the broad PSF sizes of $J_{125}H_{160}$ and/or a flux contribution of bright central bulge-like objects at the rest-frame optical wavelengths. 

The panel (b) of Figure \ref{fig_comp} represents $F_c$ as a function of $L_{\rm cl} / L_{\rm gal}$. We find that clumps are well identified even for low-$M_*$ and high-$z$ galaxies (i.e. $F_c\gtrsim0.5$ at $L_{\rm cl} / L_{\rm gal}\geqslant 0.08$). We correct for the detection incompleteness by using the following equation, 

\begin{equation}\label{eq_complete}
\begin{aligned}
f_{\rm clumpy}^{\rm cor} = f_{\rm clumpy} &+ \frac{1}{n_c} \Biggl( \frac{1}{F_c} - 1 \Biggr) f_{\rm clumpy} \\ 
&- \frac{1}{n_c} \Biggl( \frac{1}{F_c} - 1 \Biggr) (f_{\rm clumpy})^2, 
\end{aligned}
\end{equation}

where $f_{\rm clumpy}^{\rm cor}$ and $n_c$ are an incompleteness-corrected clumpy fraction and the number of clumps in a host galaxy, respectively. We assume $n_c=2$ for incompleteness correction similarly in \citet{2015ApJ...800...39G}. The correction with Equation \ref{eq_complete} typically makes $f_{\rm clumpy}$ to increase by a factor of $\simeq1.1-1.2$. We find that $F_c$ is most sensitive to $M_*$ ($\simeq L_{\rm gal}$) and $z$ of host galaxies. For this reason, we correct for the detection incompleteness for the redshift evolution of $f_{\rm clumpy}$ (Section \ref{sec_fclumpy}), but for analyses in a given $M_*$, $L_{\rm gal}$ and $z$ bin (e.g., Section \ref{sec_phys_photoz}).

\section{RESULTS}\label{sec_results}

\subsection{Evolution of Clumpy Fractions}\label{sec_fclumpy}

We present our $f_{\rm clumpy}$ measurements for the photo-$z$ galaxies and the LBGs in Section \ref{sec_fclumpy_photoz_lbg}. We compare our $f_{\rm clumpy}$ values for the photo-$z$ galaxies and the LBGs with results of previous studies in Sections \ref{sec_fclumpy_photoz_comp} and \ref{sec_fclumpy_lbg_comp}, respectively. In Section \ref{sec_fclumpy_evo}, we show the redshift evolution of $f_{\rm clumpy}$ at $z\simeq0-8$ in a compilation of the photo-$z$ galaxy and LBG samples. In Section \ref{sec_systematics}, we check whether our results of $f_{\rm clumpy}$ at $z\simeq0-8$ are reliable. 

\subsubsection{Clumpy Fractions for the Photo-$z$ Galaxies and the LBGs}\label{sec_fclumpy_photoz_lbg}

We present $f_{\rm clumpy}$ for the photo-$z$ galaxies at $z\simeq0-3$ and the LBGs at $z\gtrsim4$. Figure \ref{fig_z_fclumpy_tile} shows $F_c$-corrected clumpy fractions $f_{\rm clumpy}^{\rm cor}$ for the SFGs and the QGs. We measure the number fraction of galaxies with UV clumps, $f_{\rm clumpy}^{\rm UV, cor}$, and optical ones, $f_{\rm clumpy}^{\rm opt, cor}$. In Figure \ref{fig_z_fclumpy_tile}, galaxies at $z\geqslant3$ are not shown due to the small sample size (see Table \ref{tab_sample}). 

For the SFGs, we find that $f_{\rm clumpy}^{\rm UV, cor}$ basically decreases from $z\simeq3$ to $z\simeq0$ in all the $M_*$ bins. These $f_{\rm clumpy}^{\rm UV, cor}$ evolutions appear to slightly depend on $M_*$ in the sense that $f_{\rm clumpy}^{\rm UV, cor}$ in a high-$M_*$ bin drops quickly towards low-$z$. In a high-$M_*$ bin of $\log{M_*/M_\odot}=11-12$, $f_{\rm clumpy}^{\rm UV, cor}$ has $\simeq70\%$ at $z\gtrsim2$, and gradually decreases to $\simeq20\%$ at $z\simeq0$. In an intermediate-$M_*$ bin of $\log{M_*/M_\odot}=10-11$, $f_{\rm clumpy}^{\rm UV, cor}$ has $\simeq50\%$ at $z\simeq2-3$, and then drops to $\simeq40\%$ at $z\simeq0$. In a low-$M_*$ bin of $\log{M_*/M_\odot}=9-10$, $f_{\rm clumpy}^{\rm UV, cor}$ are an approximately constant value of $f_{\rm clumpy}^{\rm UV, cor}\simeq50-60\%$ at $z\simeq1-3$, and slightly decreases to $\simeq40\%$ at $z\simeq0$. In contrast, $f_{\rm clumpy}^{\rm opt, cor}$ shows small values, $\simeq10-30\%$, at $z\simeq0-3$ compared to $f_{\rm clumpy}^{\rm UV, cor}$. There is no significant dependence of $f_{\rm clumpy}^{\rm opt, cor}$ on $M_*$. 

For the QGs, the evolutional $f_{\rm clumpy}^{\rm UV, cor}$ trends are similar to those of the SFGs. Even in the similarity of the $f_{\rm clumpy}^{\rm UV, cor}$ evolution, $f_{\rm clumpy}^{\rm UV, cor}$ for the QGs are typically smaller than those for the SFGs by a factor of $\simeq1.4$. The $f_{\rm clumpy}^{\rm UV, cor}$ values for the QGs gradually decrease from $\simeq30-50\%$ at $z\simeq3$ to $\simeq10-30\%$ at $z\simeq0$. The relatively high $f_{\rm clumpy}^{\rm UV, cor}$ values could be explained by a high sSFR for our QGs (see also, Section \ref{sec_phys}). The average sSFR value is $\log{{\rm sSFR}}\simeq-1.8$ for our QGs at $z\simeq1-2$. The high sSFR indicates that our QGs have not been completely quenched. In contrast, $f_{\rm clumpy}^{\rm opt, cor}$ for the QGs has small values of $\simeq10-20\%$ at $z\simeq0-2$ in all the $M_*$ bins. We find no significant dependence of both $f_{\rm clumpy}^{\rm UV, cor}$ and $f_{\rm clumpy}^{\rm opt, cor}$ on $M_*$ for the QGs. Following the small sample size of the optical clumps, we mainly focus on UV clumps (e.g. $f_{\rm clumpy}^{\rm UV}$) in Sections \ref{sec_radprof}, \ref{sec_phys} and \ref{sec_color}. 

Figure \ref{fig_z_fclumpy_lbg} shows $f_{\rm clumpy}^{\rm UV, cor}$ for the $z\simeq4-8$ LBGs with a UV luminosity of $L_{\rm UV} = 0.3 -1 L_{z=3}^*$, where $L_{z=3}^*$ is the characteristic UV luminosity of LBGs at $z\simeq3$ ($M_{\rm UV}=-21$; \citealt{1999ApJ...519....1S}). Note that $f_{\rm clumpy}^{\rm opt}$ is not presented due to the lack of $\!${\it HST} filters covering the rest-frame optical wavelengths at $z\gtrsim3$. As shown in Figure \ref{fig_z_fclumpy_lbg}, we find that $f_{\rm clumpy}^{\rm UV, cor}$ remains roughly constant around $\simeq20 \%$ for the LBGs at $z\simeq4-8$ albeit with the large error bars at $z\gtrsim6$. 

\subsubsection{Comparisons of $f_{\rm clumpy}$ between our Photo-$z$ Galaxies and Previous Studies}\label{sec_fclumpy_photoz_comp}

We compare our $f_{\rm clumpy}$ for the photo-$z$ galaxies with those of previous studies at $z\simeq0-3$ in Figure \ref{fig_z_fclumpy_tile}. Table \ref{tab_previous} summarizes previous studies on clumpy galaxies at $z\simeq0-3$. The $f_{\rm clumpy}$ data points and the error bars of these previous studies are taken from \citet{2015ApJ...800...39G}. In most cases, our clump identification method is slightly different from those of these previous studies (see Table \ref{tab_previous}). We hereafter specify the methods in the following comparisons for several of these studies. 

The clumpy structures have widely been investigated for star-forming galaxies at the rest-frame UV wavelength. \citet{2007ApJ...658..763E} have reported that $f_{\rm clumpy}^{\rm UV}$ is $\simeq50-60\%$ at $z\simeq2-3$ and $\simeq30\%$ at $z\simeq0-1$ using a sample of 1,003 starburst galaxies. \citet{2010MNRAS.406..535P} have found $f_{\rm clumpy}^{\rm UV}\simeq20\%$ at $z\simeq0.6$ for 63 emission-line galaxies. \citet{2010ApJ...710..979O} have obtained $f_{\rm clumpy}^{\rm UV}\simeq20\%$ for 30 Lyman break analogues (LBAs) at $z\simeq0.2$ which is a low-$z$ UV bright galaxy population \citep[see also, ][]{2015A&A...576A..51G}. \citet{2014ApJ...780...77T} have examined clumpy structures for 100 H$\alpha$ emitters (HAEs) at $z\simeq 2.2-2.5$, and have obtained $f_{\rm clumpy}^{\rm UV}\simeq40\%$. \citet{2012ApJ...753..114W} have measured $f_{\rm clumpy}^{\rm UV}$ to be $\simeq70-80\%$ at $z\simeq0.5-2.5$ based on a sample of 649 star-forming galaxies. \citet{2014ApJ...786...15M} have identified clumpy structures among 24,027 galaxies at $z\simeq0-1$ in COSMOS, suggesting a gradual decrease from $f_{\rm clumpy}\simeq20\%$ at $z\simeq1$ to $\simeq0\%$ at $z\simeq0$. Recently, \citet{2015ApJ...800...39G} have investigated clumpy structures of 3,229 SFGs in the UDS and GOODS-S data for the evolution of $f_{\rm clumpy}^{\rm UV}$ at $z\simeq0-3$. In a $M_*$ bin of $\log{M_*/M_\odot}=9-9.8$, \citeauthor{2015ApJ...800...39G}'s $f_{\rm clumpy}^{\rm UV}$ show no evolution, $\simeq60\%$, at $\simeq0-3$. In contrast, $f_{\rm clumpy}^{\rm UV}$ gradually decreases from $\simeq60\%$ at $z\simeq3$ to $\simeq10-30\%$ at $z\simeq0$ in $M_*$ bins of $\log{M_*/M_\odot}=9.8-10.6$ and $10.6-11.4$. The evolutional $f_{\rm clumpy}^{\rm UV}$ trends of these studies are roughly consistent with our $f_{\rm clumpy}^{\rm UV, cor}$ measurements at $z\simeq0-3$.

On the other hand, a few of previous studies have derived $f_{\rm clumpy}^{\rm opt}$ for star-forming galaxies \citep[e.g. ][]{2011ApJ...739...45F, 2012ApJ...753..114W}. \citet{2012ApJ...753..114W} have measured $f_{\rm clumpy}^{\rm opt}$ for 649 star-forming galaxies at $z\simeq2-3$ in addition to $f_{\rm clumpy}^{\rm UV}$. Our measurement is remarkably consistent with $f_{\rm clumpy}^{\rm opt}$  of \citet{2012ApJ...753..114W} at $z\simeq1$. However, $f_{\rm clumpy}^{\rm opt}$ are smaller than theirs at $z\simeq2$ by a factor of three. This discrepancy might result from the difference of clump identification methods. 

These comparisons indicate that our $f_{\rm clumpy}$ measurements at $z\simeq0-3$ are consistent with the results of the previous studies. The $f_{\rm clumpy}$ agreements ensure the reliability of our clumpy structure analyses.

\subsubsection{Comparisons of $f_{\rm clumpy}$ between our LBGs and Previous Studies}\label{sec_fclumpy_lbg_comp}

We compare our $f_{\rm clumpy}^{\rm UV, cor}$ values for the LBGs with results of previous studies on UV-bright galaxies  at $z\gtrsim4$. In Figure \ref{fig_z_fclumpy_lbg}, we plot $f_{\rm clumpy}^{\rm UV}$ for UV-bright galaxies with a UV luminosity range of $L_{\rm UV}\simeq0.3-1 L_{z=3}^*$ similar to ours (see Table \ref{tab_previous}). All of these UV-bright galaxies have been selected in the dropout technique \citep{1999ApJ...519....1S}. Several of these studies aim to search for merger-like systems to derive the galaxy merger fractions. It is interesting to compare our ``clumpy" galaxies with ``merger-like" systems in these previous studies due to the morphological similarity. Here we refer all of these LBGs with irregular morphologies as ``clumpy" galaxies.

\citet{2006ApJ...636..592L} have derived a clumpy fraction of $f_{\rm clumpy}^{\rm UV}\simeq21\%$ for LBGs at $z\simeq4$ based on the $GM_{20}$ criterion (see the caption of Table \ref{tab_previous}). \citet{2006ApJ...652..963R} have obtained a roughly constant value of $f_{\rm clumpy}^{\rm UV}\simeq20-30\%$ for 1,333 LBGs at $z\simeq3-5$ in the S\'ersic profile fitting. \citet{2009MNRAS.397..208C} have measured $f_{\rm clumpy}^{\rm UV}$ using the $A$ index, and have obtained $f_{\rm clumpy}^{\rm UV}\simeq20\%$ at $z\simeq3-5$. \citet{2010ApJ...709L..21O} have found two clumpy objects among 21 LBGs at $z\simeq7$, corresponding to $f_{\rm clumpy}^{\rm UV}\simeq10\%$, identified by visual inspection. \citet{2013ApJ...773..153J} have suggested $f_{\rm clumpy}^{\rm UV}\simeq40\%$ given by visual inspections for LAEs and LBGs at $z\simeq5-7$. \citet{2015ApJ...804..103K} have found $f_{\rm clumpy}^{\rm UV}\simeq10\%$ at $z\simeq6-7$ and $\simeq50\%$ at $z\simeq8$, by visual inspection, for LBG samples selected in two HFF fields. These $f_{\rm clumpy}^{\rm UV}$ values are remarkably comparable to our results albeit with relatively large error bars at $z\gtrsim6$. The $f_{\rm clumpy}^{\rm UV}$ agreements with the results of the visual inspections encourage our automated method for faint sources at $z\gtrsim6$. 

On the other hand, we find that our $f_{\rm clumpy}^{\rm UV}$ values are smaller than those of LBGs at $z\simeq4-8$ in  \citet{2014arXiv1409.1832C}. As described in Section \ref{sec_phys_lbg}, $f_{\rm clumpy}^{\rm UV}$ depends on the UV luminosity. The sample galaxies of \citet{2014arXiv1409.1832C} would be slightly brighter than the range of $L_{\rm UV} = 0.3 -1 L_{z=3}^*$. Indeed, \citet{2014arXiv1409.1832C} have obtained an $r_{\rm e}$ evolution, which also depends on $L_{\rm UV}$, different from ours \citepalias[see ][]{2015ApJS..219...15S}. We thus attribute the $f_{\rm clumpy}^{\rm UV}$ difference to the $L_{\rm UV}$ range.

\begin{figure*}[t!]
  \begin{center}
    \includegraphics[width=140mm]{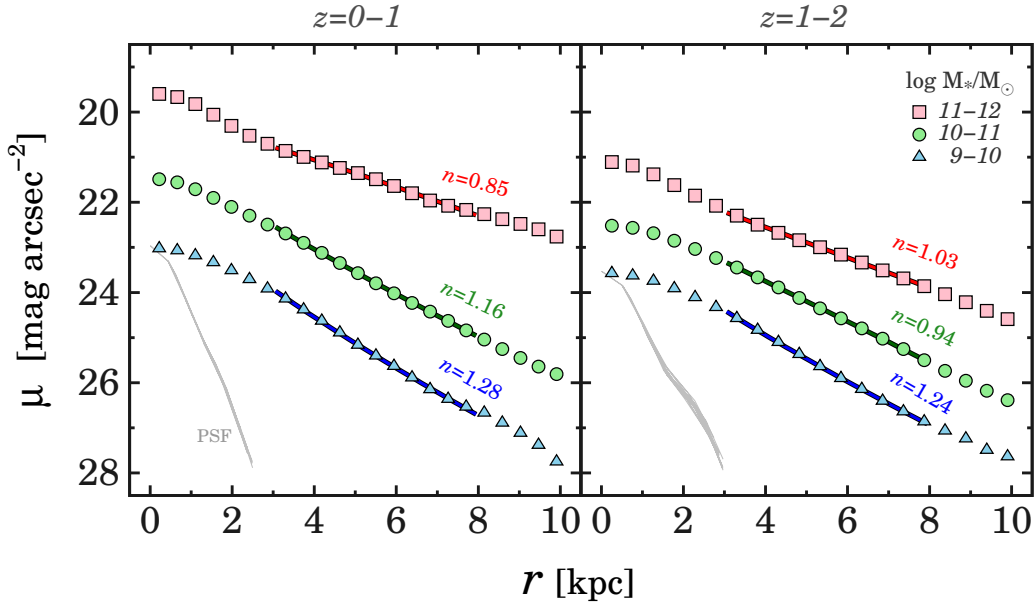}
  \end{center}
  \caption[]{{\footnotesize Radial SB profiles of clumpy galaxies at $z=0-1$ (left) and $z=1-2$ (right) in the median  stacked images at the rest-frame optical wavelength. The blue triangles, green circles, and red squares represent radial SB profiles in the bins of $\log{M_*/M_\odot}=9-10, 10-11$, and $11-12$, respectively. The error bars of each data point are smaller than the size of symbols. The colored curves indicate the best-fit S\'ersic functions to each radial SB profile at $r=3-8$ kpc with the color coding the same as for the symbols. The gray lines denote the PSFs of the $J_{125}$ (left) and $H_{160}$ (right) images. }}
  \label{fig_fd_radprof_tile}
\end{figure*}

\begin{deluxetable*}{cccc}
\setlength{\tabcolsep}{0.35cm} 
\tabletypesize{\scriptsize}
\tablecaption{The Best-fit S\'ersic Indices for the Radial SB Profiles of Clumpy Galaxies}
\tablehead{\colhead{Redshift} & \multicolumn{3}{c}{S\'ersic index $n$} \\
\colhead{} & \colhead{$\log{M_*/M_\odot}=9-10$}& \colhead{$\log{M_*/M_\odot}=10-11$} & \colhead{$\log{M_*/M_\odot}=11-12$} \\
\colhead{(1)}& \colhead{(2)}& \colhead{(3)}& \colhead{(4)}} 

\startdata

$0-1$ & $0.85\pm0.13$ & $1.16\pm0.06$ & $1.28\pm0.11$  \\
$1-2$ & $1.03\pm0.07$ & $0.94\pm0.03$ & $1.24\pm0.08$  

\enddata

\tablecomments{Columns: (1) Redshift. (2)-(4) Best-fit S\'ersic indices of the radial SB profiles for the photo-$z$ galaxies with $\log{M_*/M_\odot}=9-10$, $10-11$, and $11-12$, respectively. }
\label{tab_sersic}
\end{deluxetable*}

\subsubsection{Evolution of Clumpy Fractions at $z\simeq0-8$}\label{sec_fclumpy_evo}

We present the redshift evolution of $f_{\rm clumpy}^{\rm UV}$ at $z\simeq0-8$ in a compilation of the photo-$z$ galaxy and LBG samples. Figure \ref{fig_z_fclumpy_lbg} shows $f_{\rm clumpy}^{\rm UV, cor}$ for SFGs and LBGs with $L_{\rm UV} = 0.3 -1 L_{z=3}^*$. We find that the $f_{\rm clumpy}^{\rm UV, cor}$ value is in good agreement between the SFGs at $z\simeq3.5$ and the LBGs at $z\simeq4$. This $f_{\rm clumpy}^{\rm UV, cor}$ agreement suggests that the difference of galaxy selections (i.e. photo-$z$ or Lyman break) would minimally affect the $f_{\rm clumpy}^{\rm UV, cor}$ evolution. This argument is also supported by the evolution of galaxy sizes using the photo-$z$ and LBG samples same as those of this study \citepalias{2015ApJS..219...15S}. 

We identify an evolutional trend that $f_{\rm clumpy}^{\rm UV}$ increases from $z\simeq8$ to $\simeq1-3$ and subsequently decreases from $z\simeq1$ to $\simeq0$. This $f_{\rm clumpy}^{\rm UV, cor}$ trend is similar to the Madau-Lilly plot of the cosmic SFR density (SFRD) evolution \citep[e.g., ][]{1996MNRAS.283.1388M, 1996ApJ...460L...1L}. We fit the Madau-Lilly plot-type formula, 

\begin{equation}\label{eq_madau}
f_{\rm clumpy}^{\rm UV, cor}(z) = a\times\frac{(1+z)^b}{1 + [(1+z)/c]^d},
\end{equation}

\noindent where $a, b, c$, and $d$ are free parameters \citep{2014ARA&A..52..415M} to our $f_{\rm clumpy}^{\rm UV, cor}$ evolution. The best-fit parameters are $a=0.035$, $b=4.6$, $c=2.2$, and $d=6.7$. As shown in Figure \ref{fig_z_fclumpy_lbg}, the $f_{\rm clumpy}^{\rm UV, cor}$ evolution is well described by the Madau-Lilly plot-type formula. The best-fit function of $f_{\rm clumpy}^{\rm UV, cor}(z)$ indicates the presence of an $f_{\rm clumpy}^{\rm UV, cor}$ peak at $z\simeq1-2$. Figure \ref{fig_z_fclumpy_lbg_model} compares the best-fit $f_{\rm clumpy}^{\rm UV, cor}$ function with the cosmic SFRD in \citet{2014ARA&A..52..415M}. This comparison suggests the evolutional similarity between $f_{\rm clumpy}^{\rm UV, cor}$ and the cosmic SFRD albeit with a possible redshift difference of the peaks. Our self-consistent analyses for clumpy structures at $z\simeq0-8$ enable us to discover, for the first time, the evolutional trend and the peak of $f_{\rm clumpy}^{\rm UV}$.

For comparison, we plot $f_{\rm clumpy}^{\rm UV}$ in previous studies on UV-bright galaxies at $z\simeq1-3$ where the $f_{\rm clumpy}^{\rm UV}$ peak is shown in our sample in Figure \ref{fig_z_fclumpy_lbg}. \citet{2006ApJ...652..963R} have analyzed 153 starburst galaxies with $L_{\rm UV}\simeq0.3-1$ and have obtained $f_{\rm clumpy}^{\rm UV}\simeq60\%$ at $z\simeq1.2$. The $f_{\rm clumpy}^{\rm UV}$ value is in remarkably good agreement with our $f_{\rm clumpy}^{\rm UV}$ measurements. \citet{2012ApJ...745...85L} have selected clumpy objects among BX/BM galaxies and LBGs at $z\sim2-3$ using two selection criteria based on $GM_{20}$ and $A$ indices. We find that the $A$-based $f_{\rm clumpy}^{\rm UV}$ is consistent with ours at $z\simeq3$, but more quickly drops at $z\lesssim2$ than our $f_{\rm clumpy}^{\rm UV,cor}$ evolution. The $GM_{20}$-based $f_{\rm clumpy}^{\rm UV}$ measurements show a  lower values of $f_{\rm clumpy}^{\rm UV}\simeq20-30\%$ at $z\simeq2-3$ than ours. We do not reveal an exact cause for producing the $f_{\rm clumpy}^{\rm UV}$ differences. The $f_{\rm clumpy}^{\rm UV}$ difference may be caused by the combination of the $L_{\rm UV}$ range and clump detection methods. We also plot a morphological measurement for $z\simeq2.2$ Ly$\alpha$ emitters (LAEs) with $M_{\rm UV}$ and Ly$\alpha$ equivalent widths similar to those of LBGs \citep{2014ApJ...785...64S}. We find that the LAE sample has $f_{\rm clumpy}^{\rm UV}\simeq60\%$ that is consistent with our LBG results.

To enlarge the number of data points at $z\simeq1-3$, we re-plot $f_{\rm clumpy}^{\rm UV,cor}$ of the \citeauthor{2015ApJ...800...39G}'s SFGs with $\log{M_*/M_\odot}=9.8-10.6$ (see Figure \ref{fig_z_fclumpy_tile}). The \citeauthor{2015ApJ...800...39G}'s $f_{\rm clumpy}^{\rm UV,cor}$ has been measured in the $M_*$-limited sample. Despite the difference of the galaxy selection methods, this $M_*$ range roughly corresponds to $L_{\rm UV} \simeq 0.3 -1 L_{z=3}^*$ according to the $M_{\rm UV}$-$M_*$ relation \citepalias{2015ApJS..219...15S}. As shown in Figure \ref{fig_z_fclumpy_lbg}, the $f_{\rm clumpy}^{\rm UV}$ in \citet{2015ApJ...800...39G} follows our $f_{\rm clumpy}^{\rm UV,cor}$ evolution at $z\simeq0-3$. These previous studies of $f_{\rm clumpy}^{\rm UV}$ support that the peak of $f_{\rm clumpy}^{\rm UV}$ possibly exist at $z\simeq1-3$.

\begin{figure*}[t!]
  \begin{center}
    \includegraphics[width=150mm]{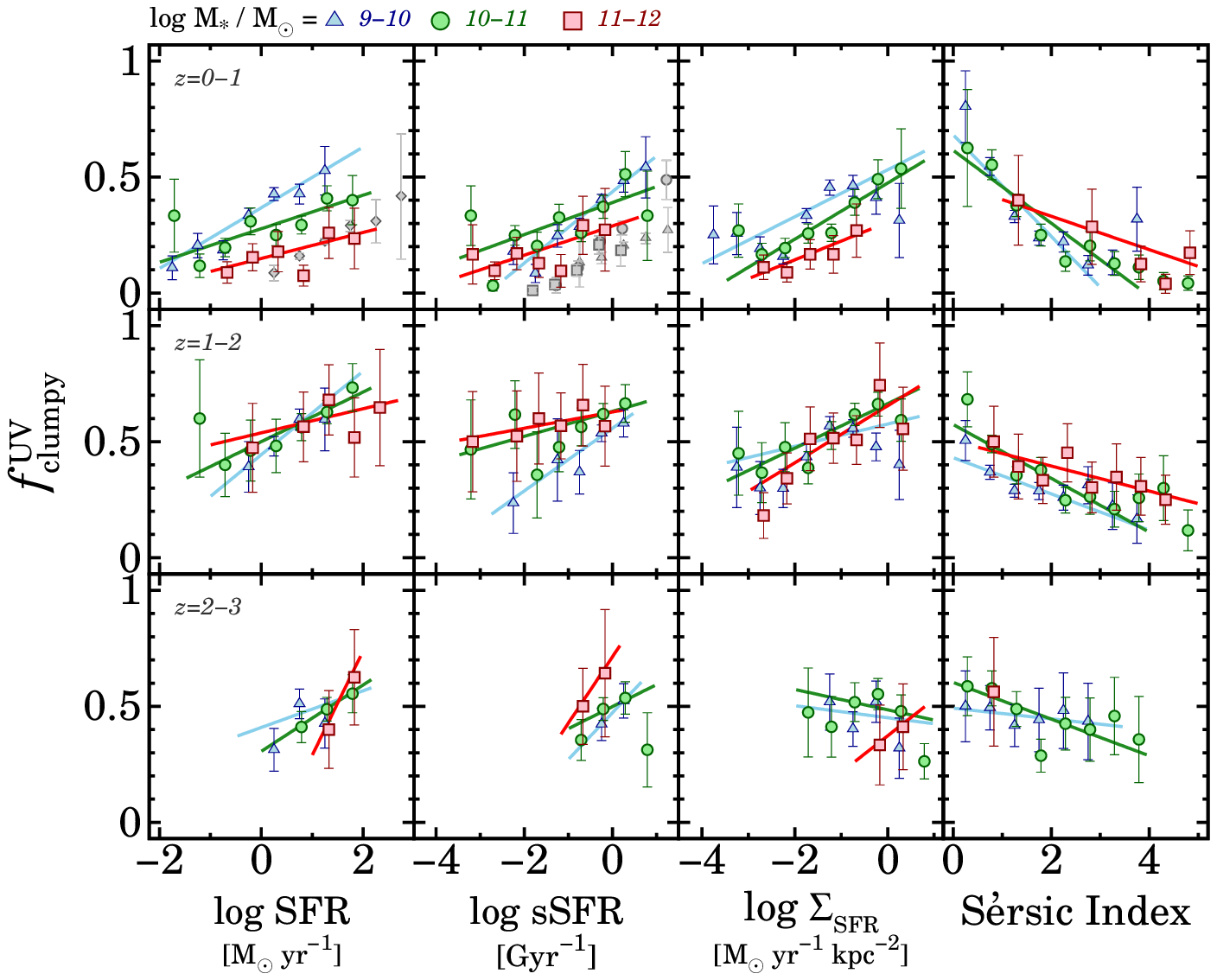}
  \end{center}
  \caption[]{{\footnotesize Dependences of $f_{\rm clumpy}^{\rm UV}$ on physical quantities (SFR, sSFR, $\Sigma_{\rm SFR}$, and $n$ from left to right) for the photo-$z$ galaxies. The top, middle, and bottom panels indicate the redshift bins of $z=0-1$, $1-2$, and $2-3$, respectively. The cyan triangles, green circles, and red squares denote $f_{\rm clumpy}^{\rm UV}$ for galaxies with $\log{M_*/M_\odot}=9-10$, $10-11$, and $11-12$, respectively. The colored lines indicate the best-fit linear function to $f_{\rm clumpy}^{\rm UV}$ in each $M_*$ bin with the color coding the same as for the symbols. The gray symbols are $f_{\rm clumpy}$ at $z=0.6-0.8$ in \citet{2014ApJ...786...15M} (diamonds: $\log{M_*/M_\odot}>9.5$; triangles: $\log{M_*/M_\odot}=9.5-10$; circles: $\log{M_*/M_\odot}=10-10.5$; squares: $\log{M_*/M_\odot}=10.5-11$). }}
  \label{fig_phys_fclump_tile}
\end{figure*}

\begin{figure*}[t!]
  \begin{center}
    \includegraphics[width=150mm]{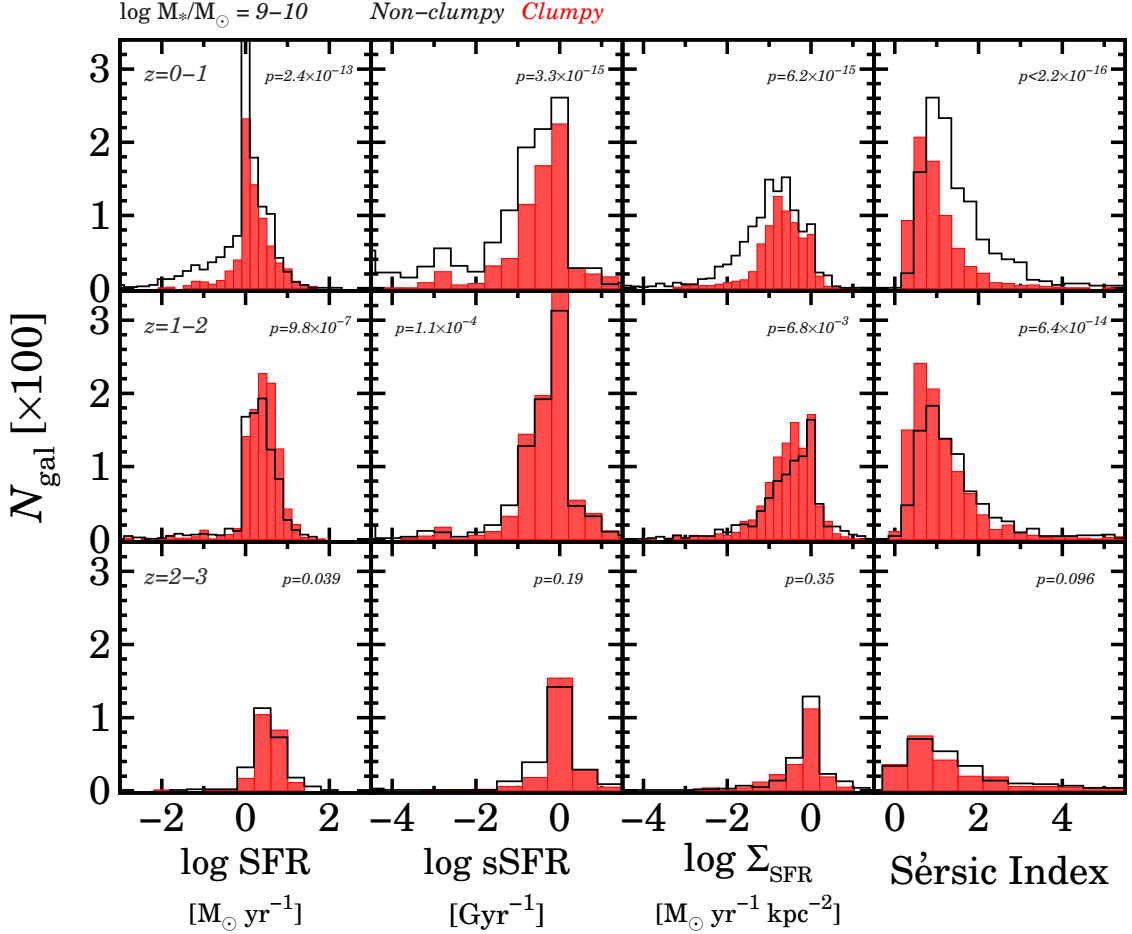}
  \end{center}
  \caption[]{{\footnotesize Number distributions of clumpy (filled red) and non-clumpy (open black) galaxies with respect to physical quantities (SFR, sSFR, $\Sigma_{\rm SFR}$, and $n$ from left to right) for the photo-$z$ galaxies with $\log{M_*/M_\odot}=9-10$. Each row, from top to bottom, denotes galaxies in a redshift bin of $z=0-1$, $2-3$, and $2-3$, respectively. The number in each panel represents $p$-values of K-S tests (see Section \ref{sec_phys} for details). }}
  \label{fig_hist_clump_tile}
\end{figure*}

\begin{figure*}[t!]
  \begin{center}
    \includegraphics[width=150mm]{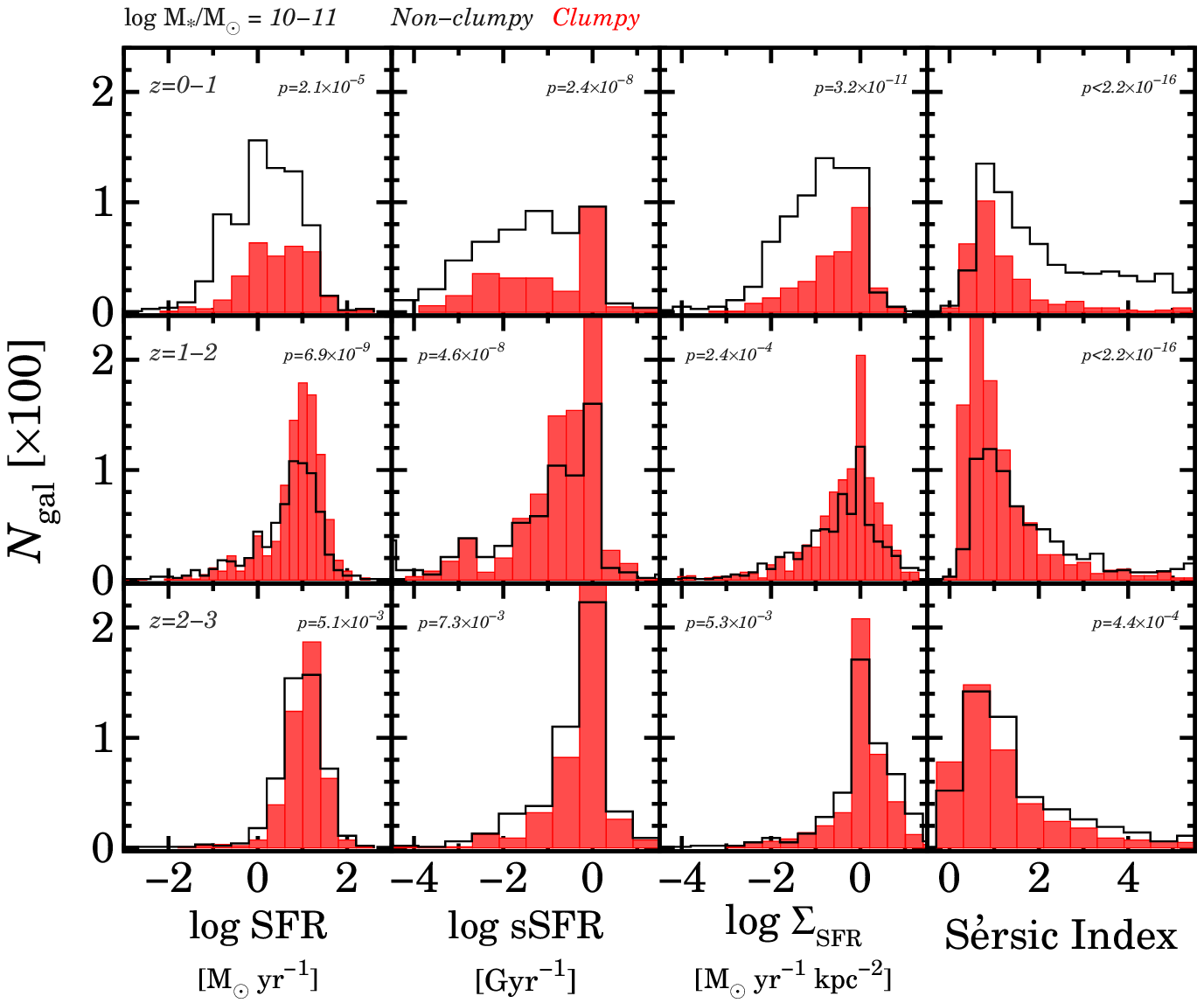}
  \end{center}
  \caption[]{{\footnotesize Same as Figure \ref{fig_hist_clump_tile}, but for the photo-$z$ galaxies with $\log{M_*/M_\odot}=10-11$. }}
  \label{fig_hist_clump_tile_midm}
\end{figure*}

\begin{figure*}[t!]
  \begin{center}
    \includegraphics[width=150mm]{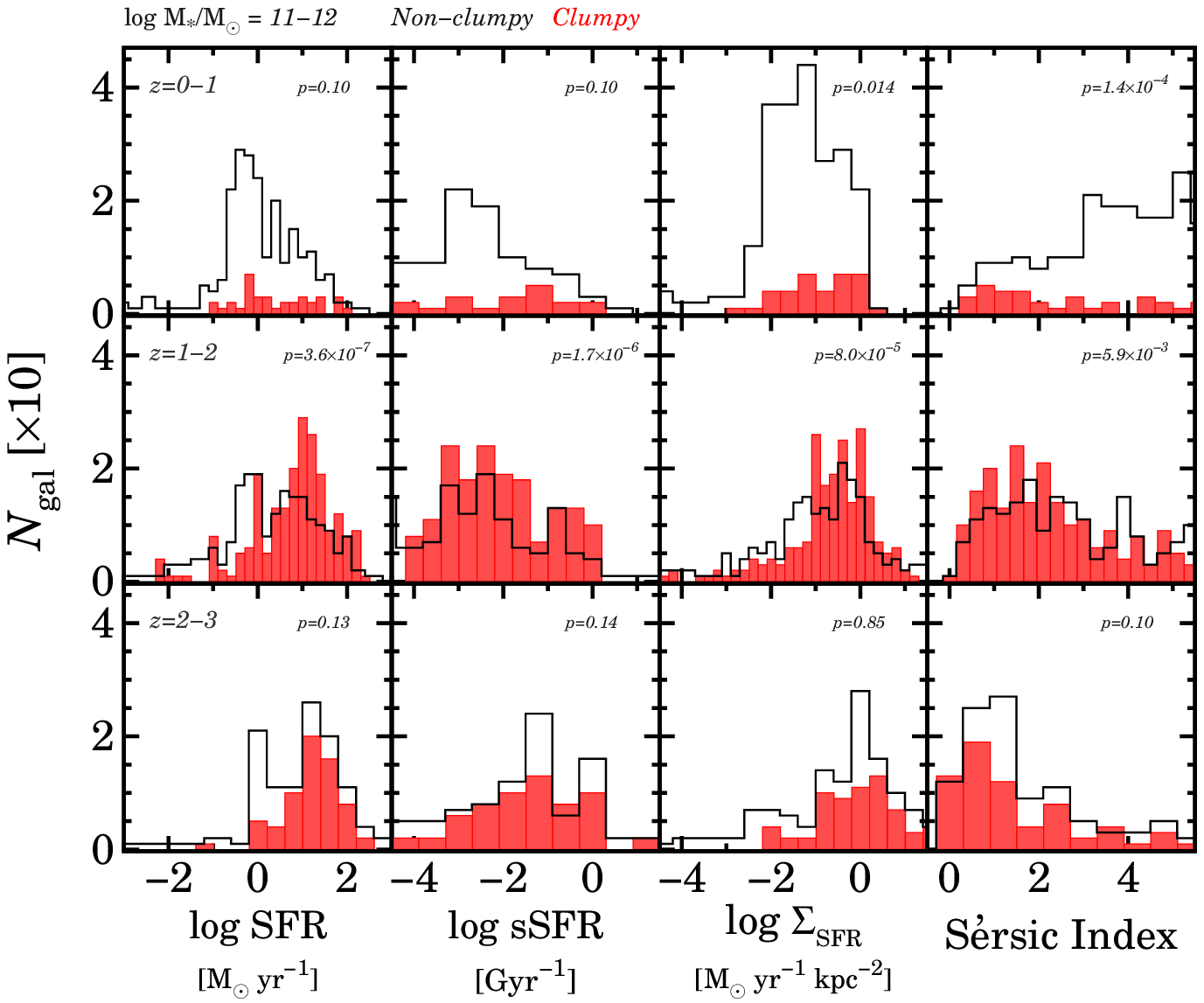}
  \end{center}
  \caption[]{{\footnotesize Same as Figure \ref{fig_hist_clump_tile}, but for the photo-$z$ galaxies with $\log{M_*/M_\odot}=11-12$. }}
  \label{fig_hist_clump_tile_highm}
\end{figure*}

\subsubsection{Is the $f_{\rm clumpy}$ Peak Real?}\label{sec_systematics}

We examine whether the possible $f_{\rm clumpy}^{\rm UV}$ peak at $z\simeq1-3$ is real. Basically, high spatial resolution and high sensitivity data are required to detect clumps in high-$z$ galaxies. In the limited spatial resolution and sensitivity of $\!${\it HST}, a low detection completeness for clumps at $z\gtrsim4$ might produce an $f_{\rm clumpy}^{\rm UV}$ peak at $z\simeq1-3$. Here we discuss five potential systematics that may produce an $f_{\rm clumpy}^{\rm UV}$ peak: the differences of 1) the spatial resolution, 2) clump search areas, 3) selection criteria of sample galaxies, 4) clump identification methods, and 5) the cosmological SB dimming effects between the SFGs at $z\lesssim3$ and the LBGs at $z\gtrsim4$.

\begin{enumerate}

\item {\it Spatial resolution} --- 
we check whether the difference in the spatial resolution changes $f_{\rm clumpy}^{\rm UV}$. In Section \ref{sec_analysis}, we have used the coadded images to select clumpy LBGs. Here we measure $f_{\rm clumpy}^{\rm UV}$ for the LBGs at $z\simeq4$ with the $I_{814}$ images whose PSF size is $\simeq2$ times smaller than that of the coadded images. The $I_{814}$ band covers the rest-frame wavelengths slightly bluer than that of the coadded images. Even in the wavelength shift, we find a good agreement in the $f_{\rm clumpy}^{\rm UV}$ values between the $I_{814}$ and coadded images (i.e. $f_{\rm clumpy}^{\rm UV}=0.146\pm0.012$ and $0.147\pm0.025$ for the $I_{814}$ and coadded images, respectively). This agreement indicates that the difference in the spatial resolution is unlikely to produce the $f_{\rm clumpy}^{\rm UV}$ peak.  

\item {\it Clump search areas} --- 
we test whether $f_{\rm clumpy}^{\rm UV}$ is affected by the difference in the clump search areas. In Section \ref{sec_analysis}, we have searched for clumps in areas defined by the segmentation maps. There is a possibility that clumps at outer galactic regions are not identified by small segmentation areas for high-$z$ compact sources. Here we enlarge the segmentation areas by a factor of $\simeq2$ and measure $f_{\rm clumpy}^{\rm UV}$ for the LBGs at $z\simeq4$. We find that $f_{\rm clumpy}^{\rm UV}$ changes within $\simeq\pm5$\% in the large clump search area. This test suggests that the difference in clump search areas does not significantly affect the $f_{\rm clumpy}^{\rm UV}$ values. 

\item {\it Selection criteria of sample galaxies} --- 
we examine whether $f_{\rm clumpy}^{\rm UV}$ is changed by the different selection criteria of sample galaxies. In Section \ref{sec_select}, the sample LBGs have been selected only by the $m_{\rm UV}$ criterion to improve the statistics at $z\gtrsim6$. Here we re-select LBGs in the selection criteria similar to those for the SFGs (i.e., the cuts of $R_{\rm e, major}$, $H_{160}$ magnitude, and $q$), and measure $f_{\rm clumpy}^{\rm UV}$ at $z\simeq4$. This sample selection significantly reduces the number of the LBGs at $z\simeq4$. Nevertheless, we find that $f_{\rm clumpy}^{\rm UV}$ does not significantly change. 

\item {\it Clump identification methods} --- 
we test whether the difference in the clump identification methods affects $f_{\rm clumpy}^{\rm UV}$. In Section \ref{sec_select}, we have not used the $d_{\rm cl}/r_{\rm e}$ criterion to identify clumpy LBGs. Here we include the $d_{\rm cl}/r_{\rm e}$ criterion in the selection for clumpy LBGs at $z\simeq4$, and measure $f_{\rm clumpy}^{\rm UV}$. We obtain $f_{\rm clumpy}^{\rm UV}=0.2\pm0.1$ that agrees with the old $f_{\rm clumpy}^{\rm UV, cor}$ value within a $1\sigma$ error. This agreement confirms that $f_{\rm clumpy}^{\rm UV}$ is not largely changed by the clump identification methods. 

\item {\it Cosmological SB dimming effects} --- 
we investigate how the cosmological SB dimming effects affect $f_{\rm clumpy}^{\rm UV}$. In the previous sections, we have corrected $F_c$ for the $f_{\rm clumpy}^{\rm UV}$ measurements. The $F_c$ correction takes into account the cosmological SB dimming effects. However, one might expect that the evolutional $f_{\rm clumpy}^{\rm UV}$ trends strongly depend on the $F_c$ values. Here we demonstrate the effect of the $F_c$ variation on $f_{\rm clumpy}^{\rm UV,cor}$ (dashed lines in Figure \ref{fig_z_fclumpy_lbg}; see the caption). We reproduce a broad peak of $f_{\rm clumpy}^{\rm UV, cor}$ at $z\simeq1-3$ even in the $F_c$ allowance. Moreover, we should note that typical clumps are reliably identified less affected by the cosmological SB dimming effect than low SB structures. 

\end{enumerate}

These tests indicate that the $f_{\rm clumpy}^{\rm UV}$ peak would not be produced by the difficulties of clump identifications at high $z$. We thus conclude that the $f_{\rm clumpy}^{\rm UV}$ peak is real.

\begin{figure*}[t!]
  \begin{center}
    \includegraphics[width=160mm]{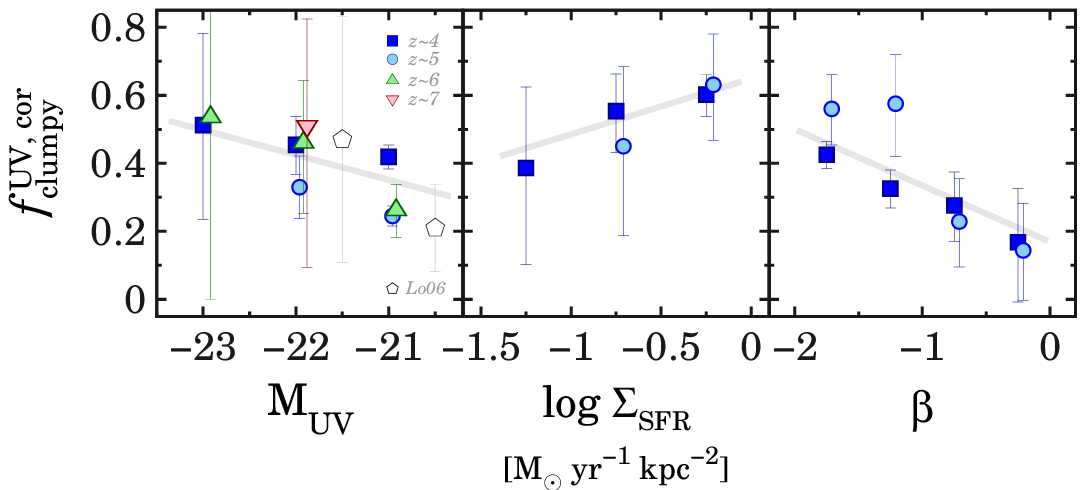}
  \end{center}
  \caption[]{{\footnotesize Dependence of $f_{\rm clumpy}^{\rm UV,cor}$ on the physical quantities, $M_{\rm UV}$ (left), $\Sigma_{\rm SFR}$ (middle), and $\beta$ (right), for the LBGs at $z\simeq4-7$. The relations for $\Sigma_{\rm SFR}$ and $\beta$ are derived in an $L_{\rm UV}$ range of $L_{\rm UV}=0.3-1L_{z=3}^*$. The blue squares, cyan circles, green triangles and magenta inverse triangle denote the LBGs at $z\simeq4, 5, 6$ and $7$, respectively. The gray lines indicate the best-fit linear functions of $f_{\rm clumpy}^{\rm UV,cor}(x)=ax+b$, where $a$ and $b$ are free parameters. The data points are not shown for LBGs with $F_c<0.2$ and/or $N_{\rm gal}<10$ in the bins of the physical quantities. The open pentagons present $f_{\rm clumpy}^{\rm UV, cor}$ for $z\simeq4$ LBGs in \cite{2006ApJ...636..592L}. }}
  \label{fig_muv_fclump}
\end{figure*}

\subsection{Radial SB Profiles of Clumpy Galaxies}\label{sec_radprof}

We investigate structural properties of our clumpy host galaxies. To obtain the underlying components of clumpy galaxies, we construct in the median stacked images of the clumpy galaxies. In general, clumps would be randomly distributed in a host galaxy. The median stacking analysis would allow us to examine the underlying galaxy components negligibly affected by the flux contributions of clumpy structures \citep[e.g., ][]{2005MNRAS.357..903C,2006ApJ...652..963R,2007ApJ...658..763E}. 

Figure \ref{fig_fd_radprof_tile} presents the radial SB profiles of galaxies with UV clumps at $z\simeq0-2$ in the median stacked images. Here we use the rest-frame optical images less affected by the clumpy structures than the rest-frame UV ones (see $f_{\rm clumpy}^{\rm opt}$ in Section \ref{sec_fclumpy}). The radial SB profiles of clumpy galaxies at $z\gtrsim2-3$ are not shown in Figure \ref{fig_fd_radprof_tile}. This is because the rest-frame optical wavelength at $z\gtrsim2-3$ are redshifted beyond the wavelength coverage of WFC3/IR. 

The radial SB profiles are obtained in the following processes. First, we align P.A. of individual clumpy galaxies in two-dimensional (2D) images. Next, we normalize the galaxy-light counts by the total magnitude of galaxies. We then stack the 2D images of clumpy galaxies in order to obtain the 2D median radial SB profiles. Finally, we extract one-dimensional (1D) radial SB profiles with a slice of three-pixel width along the major axis to avoid the effect of the $q$ variety. The $S/N$ ratios of the 1D radial SB profiles are high (i.e. $S/N\gtrsim30$ per pixel) even at an outer galactic region of $r\simeq8$ kpc in the low-$M_*$ and high-$z$ galaxy bins. Here we do not mask clumpy structures in the stacking. Even in no masking procedure, the {\it median} stacking analysis allows us to securely obtain the radial SB profiles of underlying galaxy components. The details of the stacking analysis are found in M. Kubo et al. in preparation. The number of stacked clumpy galaxies are 714, 300, and 36 (1,015, 953, and 206), at $z=0-1$ ($z=1-2$) in the bins of $\log{M_*/M_\odot}=9-10, 10-11$, and $11-12$, respectively.

We fit the S\'ersic function to the radial SB profiles at $r=3-8$ kpc where the PSF broadening effect would be negligible \citep[e.g., ][]{2015arXiv150703999N}. Table \ref{tab_sersic} summarizes the best-fit S\'ersic indices. The best-fit $n$ values are $n\simeq1$ for the radial SB profiles in all the $M_*$ and $z$ bins. These low $n$ values indicate that clumpy galaxies tend to have disk-like light profiles. 

In our stacking analysis, we have not matched $r_{\rm e}$ of individual galaxies. There is a possibility that the $n\simeq1$ results may be mimicked by the combination of {\it spheroidal} radial SB profiles of $n\simeq2-4$ and different $r_{\rm e}$ values. Here we test whether the $r_{\rm e}$ variation affects the results of our stacking analysis. First, we create artificial galaxies with {\tt GALLIST} and {\tt MKOBJECTS} of the {\sc IRAF} package. These artificial galaxies have the $r_{\rm e}$ and magnitude distributions similar to those of the real galaxies. These $r_{\rm e}$ values depend on the magnitudes following the size-luminosity relations \citepalias{2015ApJS..219...15S}. The axis ratio randomly varies in a range of $0\leqslant q\leqslant1$. To characterize the typical morphology of galaxies, we fix the input S\'ersic index $n_{\rm in}=1, 2, 3, $ or $4$ in each set of simulations. Next, we embed these artificial galaxies into the real images at random positions of the blank sky. Finally, we obtain stacked radial SB profiles and measure the S\'ersic index $n_{\rm out}$ for the artificial galaxies in the same manner as our analyses for the real galaxies. We do not find $n_{\rm out}\simeq1$ for the artificial galaxy sets of $n_{\rm in}=2, 3$, and $4$, but only for the one of $n_{\rm in}=1$. Thus, our test suggests that the $n\simeq1$ stacked radial SB profiles are truly made of galaxies with $n\simeq1$ radial SB profiles, and supports our conclusion that the clumpy galaxies have the disk-like morphology. 

In \citetalias{2015ApJS..219...15S}, the $n$ values have been derived in the profile fitting for the entire galaxy regions including the central bulge-like components. In general, the central components tend to have a large $n$ value of $n\gtrsim2$ for galaxies at $z\simeq0-2$. For this reason, we have obtained a slightly large value of $n\simeq1.5$ in \citetalias{2015ApJS..219...15S}. Here we re-measure the $n$ value including the central region for the radial SB profile fitting. We obtain the best-fit $n$ values of $n\simeq1$ similar to those found in this section.

\subsection{Relations between Clumps and Physical Properties of Host Galaxies}\label{sec_phys}

We investigate relations between clumps and physical properties of the photo-$z$ galaxies in Section \ref{sec_phys_photoz} and the LBGs in Section \ref{sec_phys_lbg}. 

\subsubsection{Results for the Photo-$z$ Galaxies}\label{sec_phys_photoz}

We first examine physical properties for clumpy photo-$z$ galaxies. Here we evaluate three quantities related to star formation (SF), i.e., SFR, sSFR, and $\Sigma_{\rm SFR}$, and a structural parameter, $n$. These quantities are derived from entire fluxes of host galaxies \citep[][\citetalias{2015ApJS..219...15S}, ]{2014ApJS..214...24S, 2015arXiv151107873H}. In Section \ref{sec_radprof}, we have measured $n$ of the radial SB profiles in the stacked images. In this section, we examine $n$ on the individual basis complimentary to the median stacking analysis. To test whether clumps are related to these physical quantities of host galaxies, we employ two approaches: i) to examine dependences of $f_{\rm clumpy}^{\rm UV}$ on physical quantities, and ii) to quantify differences of number distributions between clumpy and non-clumpy galaxies. Approach (i) enables us to clearly identify $f_{\rm clumpy}^{\rm UV}$ trends with respect to physical quantities. Approach (ii) is sensitive to differences of two distributions because of comparisons in a wide dynamic range of physical quantities. 

\paragraph{Approach {\rm (i)}} Figure \ref{fig_phys_fclump_tile} represents dependences of  $f_{\rm clumpy}^{\rm UV}$ on the physical quantities. Here we mainly use the $n$ values derived at the rest-frame optical wavelengths for $z\lesssim2$, but at the rest-frame UV ones for $z\gtrsim2$. We find trends that $f_{\rm clumpy}^{\rm UV}$ increases with increasing the three SF-related quantities (i.e., SFR, sSFR, and $\Sigma_{\rm SFR}$) for galaxies in most bins of $M_*$  and $z$. In contrast, $f_{\rm clumpy}^{\rm UV}$ appears to increase with decreasing $n$. This $f_{\rm clumpy}^{\rm UV}$-$n$ relation supports the results of the median stacking analysis in Section \ref{sec_radprof}. The correlations tend to be weaker at higher $z$. 

We compare our $f_{\rm clumpy}^{\rm UV}$ relations with those of a previous study. \citet{2014ApJ...786...15M} have reported increasing $f_{\rm clumpy}^{\rm UV}$ trends with SFR and sSFR for star-forming galaxies at $z\simeq0-1$. As shown in Figure \ref{fig_phys_fclump_tile}, the amplitudes of our $f_{\rm clumpy}$ values are slightly higher than those of \citet{2014ApJ...786...15M}. The difference of $f_{\rm clumpy}$ amplitudes would be caused by clump identification methods. Nevertheless, we find that our $f_{\rm clumpy}^{\rm UV}$ trends are comparable to the \citeauthor{2014ApJ...786...15M}'s measurements.

\paragraph{Approach {\rm (ii)}} Figures \ref{fig_hist_clump_tile}, \ref{fig_hist_clump_tile_midm}, and \ref{fig_hist_clump_tile_highm} compare the number distributions of clumpy and non-clumpy galaxies in the $M_*$ bins of $\log{M_*/M_\odot}=9-10$, $10-11$, and $11-12$, respectively. We perform Kolmogorov-Smirnov (K-S) tests to quantify the significance level of differences between the number distributions of clumpy and non-clumpy galaxies. The K-S tests determine probabilities ($p$-values) of accepting the null hypothesis that two samples are drawn from a statistically identical distribution. If the two distributions are statistically different, the $p$-values are significantly below $\simeq 5\%$. As shown in $p$-values in Figures \ref{fig_hist_clump_tile}, \ref{fig_hist_clump_tile_midm}, and \ref{fig_hist_clump_tile_highm}, the K-S tests suggest that the two number distributions of clumpy and non-clumpy galaxies are statistically different in almost all $M_*$- and $z$-bins with respect to most physical properties (i.e., $p\ll 5\%$). These results of the K-S tests are compatible with  the $f_{\rm clumpy}^{\rm UV}$ trends in Approach {\rm (i)}. However, we obtain a large $p$-value of  $\gtrsim5\%$ for galaxies in bins of ($\log{M_*/M_\odot}$, $z$) $=$ ($9-10$, $2-3$), ($11-12$, $0-1$), and ($11-12$, $2-3$). Such a high $p$-value might be attributed to the small sample size of high-$z$ clumpy galaxies.

\subsubsection{Results for the LBGs}\label{sec_phys_lbg}

We next examine physical properties for the clumpy LBGs. We have no measurements of $M_*$ and $n$ for the LBG sample. Instead of $M_*$ and $n$, we evaluate physical quantities of $M_{\rm UV}$, $\Sigma_{\rm SFR}$, and $\beta$ for the LBGs. Figure \ref{fig_muv_fclump} shows dependences of $f_{\rm clumpy}^{\rm UV,cor}$ on physical quantities, $M_{\rm UV}$, $\Sigma_{\rm SFR}$, and UV slope $\beta$. We perform the $F_c$ corrections to $f_{\rm clumpy}^{\rm UV}$ for the LBG sample. This is because the detection completeness is sensitive strongly to $M_{\rm UV}$ (see Section \ref{sec_complete}) and weakly to $\Sigma_{\rm SFR}$ and $\beta$ through weak correlations between $M_{\rm UV}$ and the two quantities of $\Sigma_{\rm SFR}$ and $\beta$ \citep[e.g., ][ \citetalias{2015ApJS..219...15S}]{2014ApJ...793..115B}. We exclude high SFR SD bins of $\log{\Sigma_{\rm SFR}}\gtrsim0$ M$_\odot$ yr$^{-1}$ kpc$^{-2}$ dominated by compact LBGs.

In Figure \ref{fig_muv_fclump}, we find a marginal trend that $f_{\rm clumpy}^{\rm UV,cor}$ increases for LBGs with a bright $M_{\rm UV}$, a high $\Sigma_{\rm SFR}$, and a small $\beta$ (i.e. blue UV slope). According to an empirical relation in \citet{1998ApJ...498..541K}, $M_{\rm UV}$ is an indicator of SFR. The $f_{\rm clumpy}^{\rm UV,cor}$-$M_{\rm UV}$ and $f_{\rm clumpy}^{\rm UV,cor}$-$\Sigma_{\rm SFR}$ trends for the LBGs are consistent with the correlations for the photo-$z$ galaxies in Section \ref{sec_phys_photoz}. On the other hand, $\beta$ is a coarse indicator of the stellar population and dust extinction of galaxies. The vigorous clump formations would yield young stellar populations on entire galactic regions, producing a blue UV slope. The $f_{\rm clumpy}^{\rm UV,cor}$-$\beta$ correlation could be another piece of evidence that clumps are related to the SF activity of host galaxies. 

For comparison, we plot LBGs at $z\simeq4$ in \citet{2006ApJ...636..592L} in the panel of $f_{\rm clumpy}^{\rm UV,cor}$-$M_{\rm UV}$ in Figure \ref{fig_muv_fclump}. The detection completeness has been corrected in \citet{2006ApJ...636..592L}. Our $f_{\rm clumpy}^{\rm UV,cor}$-$M_{\rm UV}$ relation is comparable to that of \citet{2006ApJ...636..592L}. 

We quantify the significance level of our $f_{\rm clumpy}^{\rm UV,cor}$ relations. We are not able to apply Approach {\rm (ii)} in Section \ref{sec_phys} to the LBG sample due to the small sample size and the differential $F_c$ values depending on $M_{\rm UV}$, $\Sigma_{\rm SFR}$, and $\beta$. Instead of Approach {\rm (ii)}, we carry out Spearman rank correlation tests for the $f_{\rm clumpy}^{\rm UV,cor}$ relations in Figure \ref{fig_muv_fclump}. In general, rank correlation tests do not adequately take into account fractional values (e.g. $f_{\rm clumpy}$ in the case of this study) and their error bars. We still expect that the test gives a rough estimate for significance levels of the correlations. We find that the significance levels of these $f_{\rm clumpy}^{\rm UV,cor}$ trends are $4.3\sigma$, $4.9\sigma$, and $4.5\sigma$ with respect to $M_{\rm UV}$, $\Sigma_{\rm SFR}$, and $\beta$, respectively. The correlation tests suggest that the $f_{\rm clumpy}^{\rm UV,cor}$ correlations are significant with these physical quantities for the LBGs at $z\gtrsim4$. We have extended the correlation study up to $z\lesssim7$, and have found the relation between clumps and the SF activity of host galaxies at $z\simeq0-7$.

\subsection{Relations between Clump Colors and Galactocentric Distance}\label{sec_color}

\begin{figure*}[t!]
  \begin{center}
    \includegraphics[width=160mm]{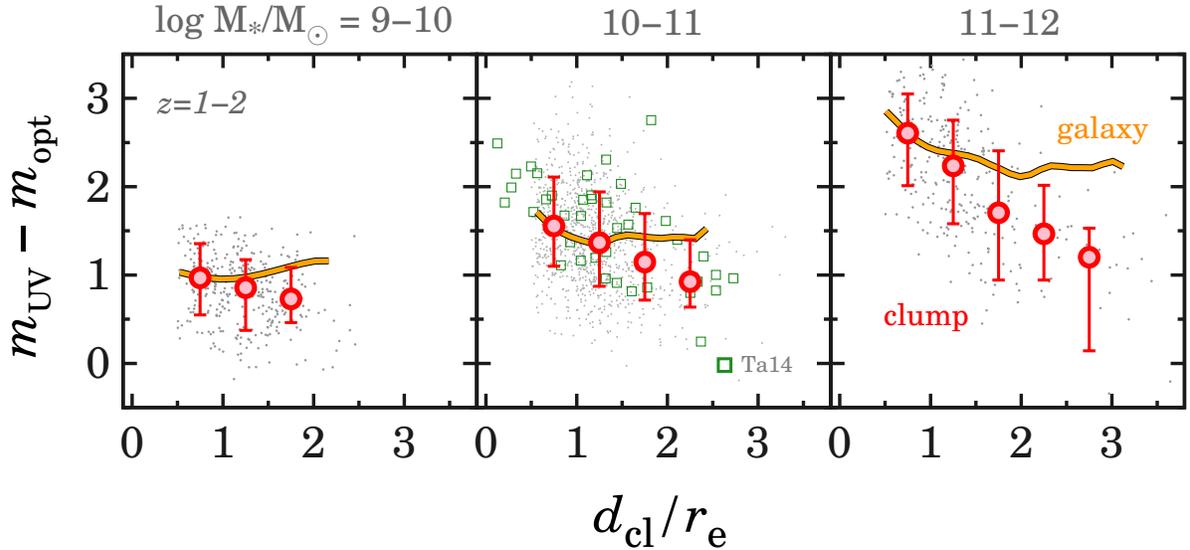}
  \end{center}
  \caption[]{{\footnotesize Clump $m_{\rm UV} - m_{\rm opt}$ colors as a function of $d_{\rm cl} / r_{\rm e}$ in the $M_*$ bins of $\log{M_*/M_\odot}=9-10$ (left), $10-11$ (middle), and $11-12$ (right). The gray dots indicate the color measurements for individual clumps. The red circles with error bars represent the median $m_{\rm UV} - m_{\rm opt}$ colors and the 16th and 84th percentiles of the distribution. The orange curves present average galaxy color gradients in each $M_*$ bin. For comparison, the galaxy color gradients are shifted along the $y$-axis to match the $m_{\rm UV} - m_{\rm opt}$ colors at the reddest median $m_{\rm UV} - m_{\rm opt}$ points. The green squares denote the clump color measurements for HAEs at $z\simeq2$ \citep{2014ApJ...780...77T} assuming that the galaxy size is $r_{\rm e}=2$ kpc. }}
  \label{fig_fd_uv_opt_tile}
\end{figure*}

We examine relations between colors and $d_{\rm cl}/r_{\rm e}$ for clumps. Figure \ref{fig_fd_uv_opt_tile} presents $m_{\rm UV} - m_{\rm opt}$ colors as a function of $d_{\rm cl}/r_{\rm e}$ for each clump, where $m_{\rm UV}$ and $m_{\rm opt}$ are the magnitudes of UV clumps at the rest-frame UV and optical wavelengths, respectively. Here we measure the $m_{\rm UV} - m_{\rm opt}$ colors in the double-detection mode of {\sc SExtractor} after the PSF homogenizations. We obtain $m_{\rm UV} - m_{\rm opt}$ colors at $z\simeq1-2$ where both $m_{\rm UV}$ and $m_{\rm opt}$ magnitudes are measurable with the $\!${\it HST} bands. As shown in Figure \ref{fig_fd_uv_opt_tile}, we find that the $m_{\rm UV} - m_{\rm opt}$ values tend to be red at a small $d_{\rm cl}/r_{\rm e}$. This color trend is clearly shown for high $M_*$ bins. To evaluate these trends, we perform Spearman rank correlation tests between $m_{\rm UV} - m_{\rm opt}$ and $d_{\rm cl}/r_{\rm e}$. The significance levels of these correlations are $5.1\sigma$, $9.7\sigma$, and $14\sigma$ for the mass bins of $\log{M_*/M_\odot}=9-10$, $10-11$, and $11-12$, respectively. 

For comparison, we plot $z\simeq2$ clumpy HAEs with $\log{M_*/M_\odot}\simeq10-11$ in \citet{2014ApJ...780...77T}. We assume that the galaxy size of these HAEs is $r_{\rm e}=2$ kpc (see the $r_{\rm e}$ evolution in \citetalias{2015ApJS..219...15S}). We find that our clumpy galaxies have a similar color-$d_{\rm cl}/r_{\rm e}$ correlations to those of these HAEs. 

There is a possible source of systematics in which we obtain the clump color-$d_{\rm cl}/r_{\rm e}$ correlations. The background flux of host galaxies is not subtracted in the measurements of the clump magnitudes in Figure \ref{fig_fd_uv_opt_tile} and \citet{2014ApJ...780...77T}. Evolved bright central structures (e.g. bulges) would make a similar color trend of host galaxies to those of clumps \citep[e.g., ][]{2010ApJ...709.1018V,2013ApJ...766...15P,2015ApJ...805...34M,2015arXiv150703999N}. To show this effect, we compare the typical $m_{\rm UV} - m_{\rm opt}$ color gradients of host galaxies with those of the clump colors in Figure \ref{fig_fd_uv_opt_tile}. The $m_{\rm UV} - m_{\rm opt}$ colors of host galaxies are derived from the radial SB profiles in the stacked images (Section \ref{sec_radprof}). The $m_{\rm UV} - m_{\rm opt}$ uncertainties resulting from the galaxy color variations are typically $\simeq\pm0.3$ mag \citep[e.g. ][]{2012ApJ...757..120G}. As indicated in Figure \ref{fig_fd_uv_opt_tile}, the host galaxies show similar color gradients to those of clumps. Nevertheless, we still identify the color-$d_{\rm cl}/r_{\rm e}$ correlations steeper than these galaxy color gradients for the massive bins of $\log{M_*/M_\odot}=11-12$ and marginally for $\log{M_*/M_\odot}=10-11$. These trends have already been reported in previous studies for massive galaxies \citep[e.g., ][]{2011ApJ...739...45F, 2012ApJ...757..120G}. In this study, we find, for the first time, the dependence of clump color correlations on $M_*$.

\section{DISCUSSION}\label{sec_discuss}

\subsection{Implications for Clump Formation Mechanisms; Galaxy Mergers or VDI?}\label{sec_discuss_mechanism}

We investigate whether our results are consistent with the clump formation mechanisms of galaxy mergers (Section \ref{sec_discuss_mergers}) or VDI (Section \ref{sec_discuss_vdi}). As shown in the following subsections, our results suggest that VDI would be favorable as a clump formation mechanism. In Section \ref{sec_discuss_color}, we discuss the possibility of the clump migration.

\subsubsection{Galaxy Mergers?}\label{sec_discuss_mergers}

We first examine the possibility that the galaxy merger is the major clump formation mechanism. In Section \ref{sec_fclumpy}, we identify the $f_{\rm clumpy}^{\rm UV}$ evolution at $z\simeq0-8$: $f_{\rm clumpy}^{\rm UV}$ increases from $z\simeq8$ to $z\simeq3$, and decreases from $z\simeq1$ to $z\simeq0$. In the case of the ex-situ clump origin, $f_{\rm clumpy}^{\rm UV}$ should follow the evolution of the merger fraction. Observational studies have measured the galaxy major and minor merger fractions, $f_{\rm merger}^{\rm maj}$ and $f_{\rm merger}^{\rm min}$, at $z\simeq0-3$, using galaxy close pairs which have a spatial scale of $\simeq20$ kpc different from the clump-clump separations \citep[e.g., ][]{2000MNRAS.311..565L,2009ApJ...697.1369B,2009MNRAS.394L..51B,2011ApJ...738L..25W, 2013A&A...553A..78L,2012ApJ...744...85M,2014arXiv1410.3479M}. In Figure \ref{fig_z_fclumpy_lbg_model}, we compare our $f_{\rm clumpy}^{\rm UV}$ evolution with the merger fractions of \citet{2011ApJ...742..103L} assuming that these $f_{\rm merger}^{\rm maj}$ and $f_{\rm merger}^{\rm min}$ values continuously evolve beyond $z\simeq3$. As shown in Figure \ref{fig_z_fclumpy_lbg_model}, the evolutional trend of the {\it major} merger fraction is probably similar to our $f_{\rm clumpy}^{\rm UV}$ evolution at $z\simeq0-1$, but is inconsistent beyond $z\simeq3$. In contrast, the evolutional trend of the {\it minor} merger fraction is comparable to our $f_{\rm clumpy}^{\rm UV}$ evolution at $z\gtrsim3$, but is incompatible at $z\simeq0-3$. These comparisons indicate that the merger fractions do not explain simultaneously both low-$z$ and high-$z$ trends of our $f_{\rm clumpy}^{\rm UV}$ evolution. \footnote{We have compared $f_{\rm clumpy}$ with the merger fractions, $f_{\rm merger}$, based on $M_*$ ratios of merging galaxies (i.e. gas-poor mergers). The detailed comparisons between $f_{\rm clumpy}$ and $f_{\rm merger}$ require comprehensive censuses of both gas-poor and gas-rich mergers at high-$z$ with merger ratios of $M_*$ and flux \citep[see e.g., ][]{2012ApJ...744...85M,2014arXiv1410.3479M}. } 

Moreover, numerical simulations predict that the halo-halo major and minor merger rates strongly increase from $z\simeq0$ to $z\simeq6$, $\propto(1+z)^{2\sim3}$ \citep[e.g., ][]{2008ApJ...688..789G, 2009ApJ...701.2002G, 2010MNRAS.406.2267F, 2010ApJ...724..915H, 2015MNRAS.449...49R}. Our decreasing $f_{\rm clumpy}^{\rm UV}$ evolution beyond $z\simeq3$ is also incompatible with the theoretical predictions. Thus, our evolutional $f_{\rm clumpy}^{\rm UV}$ trend is {\it inconsistent} with the scenario that the clump formation is mainly driven by galaxy mergers.

\subsubsection{VDI?}\label{sec_discuss_vdi}

We next examine the possibility that the VDI is the major clump formation mechanism. The S\'ersic index of clumpy host galaxies is one of key measurements for distinguishing the clump formation mechanisms. The VDI requires host galaxies to have disk-like underlying components. In Sections \ref{sec_radprof} and \ref{sec_phys}, we have obtained a low S\'ersic index of $n\simeq1$ in the SB profiles of clumpy galaxies. These low $n$ values support that clumpy galaxies tend to have disk-like structures. \footnote{The existence of galaxy disks at $z\gtrsim3$ is also supported by numerical simulations \citep[e.g., ][]{2011ApJ...738L..19R,2015MNRAS.451..418Y} and studies on galaxy sizes (e.g. \citetalias{2015ApJS..219...15S}). }

The $f_{\rm clumpy}^{\rm UV}$ is another key measurement. As described below, our $f_{\rm clumpy}^{\rm UV}$ evolution at $z\simeq0-8$ is consistent with the clump formation mechanism of VDI. The instability of galaxy disks could be sustained by the cold gas accretion as a result of continuous gas supply \cite[e.g., ][]{2003MNRAS.345..349B, 2005MNRAS.363....2K,2009MNRAS.395..160K,2012MNRAS.425.2027K,2006MNRAS.368....2D,2009ApJ...703..785D,2009Natur.457..451D, 2013MNRAS.429.3353N}. If clumps have the in-situ origin, $f_{\rm clumpy}^{\rm UV}$ is expected to evolve with the cold gas accretion rate. \citet{2005MNRAS.363....2K,2009MNRAS.395..160K} have calculated volume-averaged accretion rates of cold gas at $z\simeq0-7$ by performing cosmological hydrodynamics simulations. The simulations predict that the cold gas accretion rate shows an evolutional trend similar to the Madau-Lilly plot with a broad peak at $z\simeq3$. As shown in  Figure \ref{fig_z_fclumpy_lbg}, we have indeed found that our $f_{\rm clumpy}^{\rm UV}$ evolution at $z\simeq0-8$ is similar to such an evolutional trend of the cold gas accretion rate. Our $f_{\rm clumpy}^{\rm UV}$ peak is also remarkably consistent with the epoch of disk stabilization at $z\simeq1$ predicted by \citet{2012MNRAS.421..818C}. This evolutional similarity suggests that a majority of clumps form through the VDI. The $f_{\rm clumpy}^{\rm UV}$ measurements beyond $z\simeq3$ allow us to clearly identify the major clump formation mechanism over cosmic time.

The results of our clumpy structure analyses at $z\simeq0-3$ are also consistent with the in-situ clump formation mechanism. The following arguments have been found in previous studies on clumpy galaxies at $z\lesssim3$ \citep[e.g., ][]{2015ApJ...800...39G}. Our $z\simeq0-3$ $f_{\rm clumpy}^{\rm UV}$ and $n$ results would be explained by the evolution of unstable gas-rich galaxy disks. According to observations for gas fractions and kinematics \cite[e.g., ][]{2010Natur.463..781T}, $f_{\rm clumpy}^{\rm UV}$ is expected to decrease from $z\simeq3$ to $z\simeq0$ due to the disk stabilization towards low $z$. In Figure \ref{fig_z_fclumpy_tile}, we have indeed found that our $f_{\rm clumpy}^{\rm UV}$ shows such a decreasing trend at from $z\simeq3$ to $z\simeq0$. On the other hand, our $f_{\rm clumpy}^{\rm UV}$ correlates with the SF-related quantities of clumpy host galaxies (Section \ref{sec_phys}). These correlations could also be explained by the in-situ clump formation. The in-situ clump formation would enhance the global SF activity of clumpy host galaxies due to the vigorous star (clump) formation in galaxy disks. This enhancement of the SF activity would naturally result in a correlation between $f_{\rm clumpy}^{\rm UV}$ and the SF-related quantities. 

All of our $z\simeq0-3$ results also suggest the picture that a majority of clumps form in VDI. Combined with the discussion in Section \ref{sec_discuss_mergers}, the major clump formation mechanism at $z\simeq0-8$ is likely to be the VDI rather than the galaxy mergers.

\subsection{Clump Migration}\label{sec_discuss_color}

We discuss the evolution of clumps in galactic regions. We find that the clump color tends to be blue at a large $d_{\rm cl}/r_{\rm e}$, and tends to become red towards galactic centers for the clumpy galaxies with $\log{M/M_\odot} \gtrsim11$ (Figure \ref{fig_fd_uv_opt_tile} and Section \ref{sec_color}). This result suggests that clumps at outer galactic regions would be young if we assume that the blue $m_{\rm UV} - m_{\rm opt}$ color implies young stellar ages and/or low dust extinction. Recently, numerical simulations predict that these young clumps may form around the central compact sources (i.e. blue and red nuggets) after the wet compaction phase \citep[e.g., ][]{2014MNRAS.438.1870D, 2015MNRAS.450.2327Z, 2015arXiv150900017T, 2015arXiv150902529T, 2015Sci...348..314T}. The color-$d_{\rm cl}/r_{\rm e}$ correlation would be explained by a scenario that such a young clump is evolved in the stellar population during the clump migration to the galactic centers \citep[e.g., ][]{2009ApJ...703..785D, 2010MNRAS.406..112K, 2011ApJ...739...45F, 2012MNRAS.420.3490C, 2012MNRAS.422.1902I, 2013MNRAS.428..718O, 2014MNRAS.443.3675M, 2015arXiv151208791M}. In contrast, we find that no strong color-$d_{\rm cl}/r_{\rm e}$ correlation is shown for the low-$M_*$ and the intermediate-$M_*$ clumpy galaxies with $\log{M/M_\odot}=9-11$. The no correlation indicates that clumps may be destroyed by galaxy feedback effects before the clump migration in galaxies with $\log{M/M_\odot}\simeq9-11$.

\section{SUMMARY and CONCLUSION}\label{sec_conclusion}

We present the redshift evolution of $f_{\rm clumpy}$ and investigate the properties of clumpy host galaxies using the $\!${\it HST} samples of $\sim17,000$ galaxies at $z\simeq0-8$. Our {\it HST} samples consist of photo-$z$ galaxies at $z=0-6$ from the 3D-{\it HST}+CANDELS catalogue and LBGs at $z\simeq4-8$ identified in CANDELS, HUDF09/12, and HFF parallel fields. The large galaxy sample allows for the systematic search for clumpy galaxies over the wide-redshift range of $z\simeq0-8$. We identify clumpy galaxies with off-center clumps in a self-consistent clump detection algorithm for the $\!${\it HST} sample, and measure $f_{\rm clumpy}$ at $z\simeq0-8$. 
 
We summarize our major findings below. 
 
\begin{enumerate}

  \item We identify an evolutional trend of $f_{\rm clumpy}^{\rm UV}$ over $z\simeq0-8$ for the first time: $f_{\rm clumpy}^{\rm UV}$ increases from $z\simeq8$ to $z\simeq1-3$ and decreases from $z\simeq1$ to $z\simeq0$. This $f_{\rm clumpy}^{\rm UV}$ trend is similar to the Madau-Lilly plot of the cosmic SFR density evolution. We test whether the $f_{\rm clumpy}^{\rm UV}$ trend is caused by difficulties of the clump identification at high $z$. We confirm that the evolutional $f_{\rm clumpy}^{\rm UV}$ trend is real based on the tests.  
      
    \item We examine the underlying morphology of our clumpy galaxies. We obtain typical radial SB profiles by stacking the rest-frame optical images of our clumpy galaxies at $z\simeq0-2$. Our median stacking analysis enables us to examine underlying components of clumpy host galaxies, negligibly affected by flux contributions of individual clumps. The best-fit S\'ersic indices to these radial SB profiles show a low value of $n\simeq1$. These low $n$ values indicate that clumpy galaxies tend to have disk-like structures. 

  \item We investigate relations between $f_{\rm clumpy}^{\rm UV}$ and physical quantities of host galaxies. We find that $f_{\rm clumpy}^{\rm UV}$ tends to increase with increasing the SF-related quantities (i.e., SFR, sSFR, $\Sigma_{\rm SFR}$) for the photo-$z$ galaxies at $z\simeq0-2$. We also identify trends that $f_{\rm clumpy}^{\rm UV}$ increases for $z\simeq4-7$ LBGs with a bright $M_{\rm UV}$, a high $\Sigma_{\rm SFR}$, and a blue UV slope. These relations suggest that the clump formation would be intimately related to the entire SF activity of host galaxies.   

  \item We find a trend that the clump $m_{\rm UV} - m_{\rm opt}$ colors are red at a small galactocentric distance $d_{\rm cl}/r_{\rm e}$ for massive galaxies with $\log{M_*/M_\odot}=11-12$. The color-$d_{\rm cl}/r_{\rm e}$ correlations would be evidence that young clumps at galaxy disks migrate into the galactic centers. The clump color trends tend to be marginal for intermediate-$M_*$ and low-$M_*$ galaxies with $\log{M_*/M_\odot}=9-11$. The clumps may be destroyed by galaxy feedback effects before the clump migration for these galaxies with $\log{M_*/M_\odot}=9-11$. 

 \end{enumerate}

In Figure \ref{fig_z_fclumpy_lbg_model}, we find that $f_{\rm clumpy}^{\rm UV}$ does not show the evolutional trends similar to those of the galaxy major and minor merger fractions, but the one of the cold gas accretion rate. All of these results are consistent with a picture that a majority of clumps form in the violent disk instability and migrate into the galactic centers. 

The morphology of inner galactic regions remains unknown for compact sources with $r_{\rm e}\lesssim0.\!\!^{\prime\prime}2$ in the spatial resolution of PSF $\gtrsim0.\!\!^{\prime\prime}1-0.\!\!^{\prime\prime}2$ in FWHM of $\!${\it HST}. According to the evolution of galaxy size-luminosity relations, galaxies at $z\gtrsim3$ tend to be as small as $r_{\rm e}\lesssim 1\,{\rm kpc}$, corresponding to $r_{\rm e}\lesssim 0.\!\!^{\prime\prime}2$ (e.g. \citetalias{2015ApJS..219...15S}). The morphology of the inner regions for these compact galaxies will be revealed by future space missions and 30-meter class telescopes that resolve galaxy structures with a physical scale of $\gtrsim 0.05$ kpc at $z\simeq7-15$ \citep{2015RAA....15.1945S, 2015ApJ...808....6H}.

\acknowledgments

We would like to thank Avishai Dekel, Yicheng Guo, Takuya Hashimoto, Masao Hayashi, Masaru Kajisawa, Nobunari Kashikawa, Ryota Kawamata, Masakazu A. R. Kobayashi, Allison W. S. Man, Takahiro Morishita, Kentaro Motohara, Katsuhiro L. Murata, Kentaro Nagamine, Kouji Ohta, Takashi Okamoto, Masao Mori, Alessandro Romeo, Rhythm Shimakawa, Kazuhiro Shimasaku, Sandro Tacchella, Ken-ichi Tadaki, Masayuki Umemura, Kiyoto Yabe, and Toru Yamada for useful discussion and comments. We thank the anonymous referee for constructive comments and suggestions. This work is based on observations taken by the 3D-HST Treasury Program (GO 12177 and 12328) and CANDELS Multi-Cycle Treasury Program with the NASA/ESA HST, which is operated by the Association of Universities for Research in Astronomy, Inc., under NASA contract NAS5-26555. Support for this work was provided by NASA through an award issued by JPL/Caltech. This work was supported by World Premier International Research Center Initiative (WPI Initiative), MEXT, Japan, KAKENHI (23244025), (21244013), and (15H02064) Grant-in-Aid for Scientific Research (A) through Japan Society for the Promotion of Science (JSPS), and an Advanced Leading Graduate Course for Photon Science grant.

{\it Facilities:} \facility{HST (ACS, WFC3)}.

\bibliographystyle{apj}
\bibliography{reference}

\begin{thebibliography}{153}
\expandafter\ifx\csname natexlab\endcsname\relax\def\natexlab#1{#1}\fi

\bibitem[{{Agertz} {et~al.}(2015){Agertz}, {Romeo}, \&
  {Grisdale}}]{2015MNRAS.449.2156A}
{Agertz}, O., {Romeo}, A.~B., \& {Grisdale}, K. 2015, \mnras, 449, 2156

\bibitem[{{Agertz} {et~al.}(2009){Agertz}, {Teyssier}, \&
  {Moore}}]{2009MNRAS.397L..64A}
{Agertz}, O., {Teyssier}, R., \& {Moore}, B. 2009, \mnras, 397, L64

\bibitem[{{Atek} {et~al.}(2015){Atek}, {Richard}, {Kneib}, {Jauzac},
  {Schaerer}, {Clement}, {Limousin}, {Jullo}, {Natarajan}, {Egami}, \&
  {Ebeling}}]{2015ApJ...800...18A}
{Atek}, H., {et~al.} 2015, \apj, 800, 18

\bibitem[{{Beckwith} {et~al.}(2006){Beckwith}, {Stiavelli}, {Koekemoer},
  {Caldwell}, {Ferguson}, {Hook}, {Lucas}, {Bergeron}, {Corbin}, {Jogee},
  {Panagia}, {Robberto}, {Royle}, {Somerville}, \&
  {Sosey}}]{2006AJ....132.1729B}
{Beckwith}, S.~V.~W., {et~al.} 2006, \aj, 132, 1729

\bibitem[{{Bertin} \& {Arnouts}(1996)}]{1996A&AS..117..393B}
{Bertin}, E., \& {Arnouts}, S. 1996, \aaps, 117, 393

\bibitem[{{Birnboim} \& {Dekel}(2003)}]{2003MNRAS.345..349B}
{Birnboim}, Y., \& {Dekel}, A. 2003, \mnras, 345, 349

\bibitem[{{Bluck} {et~al.}(2009){Bluck}, {Conselice}, {Bouwens}, {Daddi},
  {Dickinson}, {Papovich}, \& {Yan}}]{2009MNRAS.394L..51B}
{Bluck}, A.~F.~L., {Conselice}, C.~J., {Bouwens}, R.~J., {Daddi}, E.,
  {Dickinson}, M., {Papovich}, C., \& {Yan}, H. 2009, \mnras, 394, L51

\bibitem[{{Bournaud}(2016)}]{2016ASSL..418..355B}
{Bournaud}, F. 2016, Galactic Bulges, 418, 355

\bibitem[{{Bournaud} {et~al.}(2007){Bournaud}, {Elmegreen}, \&
  {Elmegreen}}]{2007ApJ...670..237B}
{Bournaud}, F., {Elmegreen}, B.~G., \& {Elmegreen}, D.~M. 2007, \apj, 670, 237

\bibitem[{{Bournaud} {et~al.}(2009){Bournaud}, {Elmegreen}, \&
  {Martig}}]{2009ApJ...707L...1B}
{Bournaud}, F., {Elmegreen}, B.~G., \& {Martig}, M. 2009, \apjl, 707, L1

\bibitem[{{Bournaud} {et~al.}(2008){Bournaud}, {Daddi}, {Elmegreen},
  {Elmegreen}, {Nesvadba}, {Vanzella}, {Di Matteo}, {Le Tiran}, {Lehnert}, \&
  {Elbaz}}]{2008A&A...486..741B}
{Bournaud}, F., {et~al.} 2008, \aap, 486, 741

\bibitem[{{Bournaud} {et~al.}(2014){Bournaud}, {Perret}, {Renaud}, {Dekel},
  {Elmegreen}, {Elmegreen}, {Teyssier}, {Amram}, {Daddi}, {Duc}, {Elbaz},
  {Epinat}, {Gabor}, {Juneau}, {Kraljic}, \& {Le Floch'}}]{2014ApJ...780...57B}
---. 2014, \apj, 780, 57

\bibitem[{{Bouwens} {et~al.}(2011){Bouwens}, {Illingworth}, {Oesch},
  {Labb{\'e}}, {Trenti}, {van Dokkum}, {Franx}, {Stiavelli}, {Carollo},
  {Magee}, \& {Gonzalez}}]{2011ApJ...737...90B}
{Bouwens}, R.~J., {et~al.} 2011, \apj, 737, 90

\bibitem[{{Bouwens} {et~al.}(2014){Bouwens}, {Illingworth}, {Oesch},
  {Labb{\'e}}, {van Dokkum}, {Trenti}, {Franx}, {Smit}, {Gonzalez}, \&
  {Magee}}]{2014ApJ...793..115B}
---. 2014, \apj, 793, 115

\bibitem[{{Bouwens} {et~al.}(2015){Bouwens}, {Illingworth}, {Oesch}, {Trenti},
  {Labb{\'e}}, {Bradley}, {Carollo}, {van Dokkum}, {Gonzalez}, {Holwerda},
  {Franx}, {Spitler}, {Smit}, \& {Magee}}]{2015ApJ...803...34B}
---. 2015, \apj, 803, 34

\bibitem[{{Bundy} {et~al.}(2009){Bundy}, {Fukugita}, {Ellis}, {Targett},
  {Belli}, \& {Kodama}}]{2009ApJ...697.1369B}
{Bundy}, K., {Fukugita}, M., {Ellis}, R.~S., {Targett}, T.~A., {Belli}, S., \&
  {Kodama}, T. 2009, \apj, 697, 1369

\bibitem[{{Burkert} {et~al.}(2015){Burkert}, {F{\"o}rster Schreiber}, {Genzel},
  {Lang}, {Tacconi}, {Wisnioski}, {Wuyts}, {Bandara}, {Beifiori}, {Bender},
  {Brammer}, {Chan}, {Davies}, {Dekel}, {Fabricius}, {Fossati}, {Kulkarni},
  {Lutz}, {Mendel}, {Momcheva}, {Nelson}, {Naab}, {Renzini}, {Saglia},
  {Sharples}, {Sternberg}, {Wilman}, \& {Wuyts}}]{2015arXiv151003262B}
{Burkert}, A., {et~al.} 2015, ArXiv e-prints: 1510.03262

\bibitem[{{Cacciato} {et~al.}(2012){Cacciato}, {Dekel}, \&
  {Genel}}]{2012MNRAS.421..818C}
{Cacciato}, M., {Dekel}, A., \& {Genel}, S. 2012, \mnras, 421, 818

\bibitem[{{Cassata} {et~al.}(2005){Cassata}, {Cimatti}, {Franceschini},
  {Daddi}, {Pignatelli}, {Fasano}, {Rodighiero}, {Pozzetti}, {Mignoli}, \&
  {Renzini}}]{2005MNRAS.357..903C}
{Cassata}, P., {et~al.} 2005, \mnras, 357, 903

\bibitem[{{Ceverino} {et~al.}(2010){Ceverino}, {Dekel}, \&
  {Bournaud}}]{2010MNRAS.404.2151C}
{Ceverino}, D., {Dekel}, A., \& {Bournaud}, F. 2010, \mnras, 404, 2151

\bibitem[{{Ceverino} {et~al.}(2012){Ceverino}, {Dekel}, {Mandelker},
  {Bournaud}, {Burkert}, {Genzel}, \& {Primack}}]{2012MNRAS.420.3490C}
{Ceverino}, D., {Dekel}, A., {Mandelker}, N., {Bournaud}, F., {Burkert}, A.,
  {Genzel}, R., \& {Primack}, J. 2012, \mnras, 420, 3490

\bibitem[{{Chabrier}(2003)}]{2003PASP..115..763C}
{Chabrier}, G. 2003, \pasp, 115, 763

\bibitem[{{Coe} {et~al.}(2015){Coe}, {Bradley}, \&
  {Zitrin}}]{2015ApJ...800...84C}
{Coe}, D., {Bradley}, L., \& {Zitrin}, A. 2015, \apj, 800, 84

\bibitem[{{Conselice}(2003)}]{2003ApJS..147....1C}
{Conselice}, C.~J. 2003, \apjs, 147, 1

\bibitem[{{Conselice} \& {Arnold}(2009)}]{2009MNRAS.397..208C}
{Conselice}, C.~J., \& {Arnold}, J. 2009, \mnras, 397, 208

\bibitem[{{Conselice} {et~al.}(2014){Conselice}, {Bluck}, {Mortlock},
  {Palamara}, \& {Benson}}]{2014MNRAS.444.1125C}
{Conselice}, C.~J., {Bluck}, A.~F.~L., {Mortlock}, A., {Palamara}, D., \&
  {Benson}, A.~J. 2014, \mnras, 444, 1125

\bibitem[{{Cowie} {et~al.}(1995){Cowie}, {Hu}, \&
  {Songaila}}]{1995AJ....110.1576C}
{Cowie}, L.~L., {Hu}, E.~M., \& {Songaila}, A. 1995, \aj, 110, 1576

\bibitem[{{Curtis-Lake} {et~al.}(2014){Curtis-Lake}, {McLure}, {Dunlop},
  {Rogers}, {Targett}, {Dekel}, {Ellis}, {Faber}, {Ferguson}, {Grogin},
  {Huang}, {Kocevski}, {Koekemoer}, {Lai}, \&
  {Robertson}}]{2014arXiv1409.1832C}
{Curtis-Lake}, E., {et~al.} 2014, ArXiv e-prints: 1409.1832

\bibitem[{{Dekel} \& {Birnboim}(2006)}]{2006MNRAS.368....2D}
{Dekel}, A., \& {Birnboim}, Y. 2006, \mnras, 368, 2

\bibitem[{{Dekel} \& {Burkert}(2014)}]{2014MNRAS.438.1870D}
{Dekel}, A., \& {Burkert}, A. 2014, \mnras, 438, 1870

\bibitem[{{Dekel} {et~al.}(2009{\natexlab{a}}){Dekel}, {Sari}, \&
  {Ceverino}}]{2009ApJ...703..785D}
{Dekel}, A., {Sari}, R., \& {Ceverino}, D. 2009{\natexlab{a}}, \apj, 703, 785

\bibitem[{{Dekel} {et~al.}(2013){Dekel}, {Zolotov}, {Tweed}, {Cacciato},
  {Ceverino}, \& {Primack}}]{2013MNRAS.435..999D}
{Dekel}, A., {Zolotov}, A., {Tweed}, D., {Cacciato}, M., {Ceverino}, D., \&
  {Primack}, J.~R. 2013, \mnras, 435, 999

\bibitem[{{Dekel} {et~al.}(2009{\natexlab{b}}){Dekel}, {Birnboim}, {Engel},
  {Freundlich}, {Goerdt}, {Mumcuoglu}, {Neistein}, {Pichon}, {Teyssier}, \&
  {Zinger}}]{2009Natur.457..451D}
{Dekel}, A., {et~al.} 2009{\natexlab{b}}, \nat, 457, 451

\bibitem[{{Di Matteo} {et~al.}(2008){Di Matteo}, {Bournaud}, {Martig},
  {Combes}, {Melchior}, \& {Semelin}}]{2008A&A...492...31D}
{Di Matteo}, P., {Bournaud}, F., {Martig}, M., {Combes}, F., {Melchior}, A.-L.,
  \& {Semelin}, B. 2008, \aap, 492, 31

\bibitem[{{Ellis} {et~al.}(2013){Ellis}, {McLure}, {Dunlop}, {Robertson},
  {Ono}, {Schenker}, {Koekemoer}, {Bowler}, {Ouchi}, {Rogers}, {Curtis-Lake},
  {Schneider}, {Charlot}, {Stark}, {Furlanetto}, \&
  {Cirasuolo}}]{2013ApJ...763L...7E}
{Ellis}, R.~S., {et~al.} 2013, \apjl, 763, L7

\bibitem[{{Elmegreen} {et~al.}(2008){Elmegreen}, {Bournaud}, \&
  {Elmegreen}}]{2008ApJ...688...67E}
{Elmegreen}, B.~G., {Bournaud}, F., \& {Elmegreen}, D.~M. 2008, \apj, 688, 67

\bibitem[{{Elmegreen} \& {Elmegreen}(2005)}]{2005ApJ...627..632E}
{Elmegreen}, B.~G., \& {Elmegreen}, D.~M. 2005, \apj, 627, 632

\bibitem[{{Elmegreen} \& {Elmegreen}(2006)}]{2006ApJ...650..644E}
---. 2006, \apj, 650, 644

\bibitem[{{Elmegreen} {et~al.}(2009{\natexlab{a}}){Elmegreen}, {Elmegreen},
  {Fernandez}, \& {Lemonias}}]{2009ApJ...692...12E}
{Elmegreen}, B.~G., {Elmegreen}, D.~M., {Fernandez}, M.~X., \& {Lemonias},
  J.~J. 2009{\natexlab{a}}, \apj, 692, 12

\bibitem[{{Elmegreen} {et~al.}(2013){Elmegreen}, {Elmegreen}, {S{\'a}nchez
  Almeida}, {Mu{\~n}oz-Tu{\~n}{\'o}n}, {Dewberry}, {Putko}, {Teich}, \&
  {Popinchalk}}]{2013ApJ...774...86E}
{Elmegreen}, B.~G., {Elmegreen}, D.~M., {S{\'a}nchez Almeida}, J.,
  {Mu{\~n}oz-Tu{\~n}{\'o}n}, C., {Dewberry}, J., {Putko}, J., {Teich}, Y., \&
  {Popinchalk}, M. 2013, \apj, 774, 86

\bibitem[{{Elmegreen} {et~al.}(2005){Elmegreen}, {Elmegreen}, \&
  {Ferguson}}]{2005ApJ...623L..71E}
{Elmegreen}, D.~M., {Elmegreen}, B.~G., \& {Ferguson}, T.~E. 2005, \apjl, 623,
  L71

\bibitem[{{Elmegreen} {et~al.}(2004){Elmegreen}, {Elmegreen}, \&
  {Hirst}}]{2004ApJ...604L..21E}
{Elmegreen}, D.~M., {Elmegreen}, B.~G., \& {Hirst}, A.~C. 2004, \apjl, 604, L21

\bibitem[{{Elmegreen} {et~al.}(2009{\natexlab{b}}){Elmegreen}, {Elmegreen},
  {Marcus}, {Shahinyan}, {Yau}, \& {Petersen}}]{2009ApJ...701..306E}
{Elmegreen}, D.~M., {Elmegreen}, B.~G., {Marcus}, M.~T., {Shahinyan}, K.,
  {Yau}, A., \& {Petersen}, M. 2009{\natexlab{b}}, \apj, 701, 306

\bibitem[{{Elmegreen} {et~al.}(2007){Elmegreen}, {Elmegreen}, {Ravindranath},
  \& {Coe}}]{2007ApJ...658..763E}
{Elmegreen}, D.~M., {Elmegreen}, B.~G., {Ravindranath}, S., \& {Coe}, D.~A.
  2007, \apj, 658, 763

\bibitem[{{Fakhouri} {et~al.}(2010){Fakhouri}, {Ma}, \&
  {Boylan-Kolchin}}]{2010MNRAS.406.2267F}
{Fakhouri}, O., {Ma}, C.-P., \& {Boylan-Kolchin}, M. 2010, \mnras, 406, 2267

\bibitem[{{Forbes} {et~al.}(2012){Forbes}, {Krumholz}, \&
  {Burkert}}]{2012ApJ...754...48F}
{Forbes}, J., {Krumholz}, M., \& {Burkert}, A. 2012, \apj, 754, 48

\bibitem[{{F{\"o}rster Schreiber} {et~al.}(2011){F{\"o}rster Schreiber},
  {Shapley}, {Genzel}, {Bouch{\'e}}, {Cresci}, {Davies}, {Erb}, {Genel},
  {Lutz}, {Newman}, {Shapiro}, {Steidel}, {Sternberg}, \&
  {Tacconi}}]{2011ApJ...739...45F}
{F{\"o}rster Schreiber}, N.~M., {et~al.} 2011, \apj, 739, 45

\bibitem[{{Galametz} {et~al.}(2013){Galametz}, {Grazian}, {Fontana},
  {Ferguson}, {Ashby}, {Barro}, {Castellano}, {Dahlen}, {Donley}, {Faber},
  {Grogin}, {Guo}, {Huang}, {Kocevski}, {Koekemoer}, {Lee}, {McGrath}, {Peth},
  {Willner}, {Almaini}, {Cooper}, {Cooray}, {Conselice}, {Dickinson}, {Dunlop},
  {Fazio}, {Foucaud}, {Gardner}, {Giavalisco}, {Hathi}, {Hartley}, {Koo},
  {Lai}, {de Mello}, {McLure}, {Lucas}, {Paris}, {Pentericci}, {Santini},
  {Simpson}, {Sommariva}, {Targett}, {Weiner}, {Wuyts}, \& {the CANDELS
  Team}}]{2013ApJS..206...10G}
{Galametz}, A., {et~al.} 2013, \apjs, 206, 10

\bibitem[{{Garland} {et~al.}(2015){Garland}, {Pisano}, {Mac Low}, {Kreckel},
  {Rabidoux}, \& {Guzm{\'a}n}}]{2015ApJ...807..134G}
{Garland}, C.~A., {Pisano}, D.~J., {Mac Low}, M.-M., {Kreckel}, K., {Rabidoux},
  K., \& {Guzm{\'a}n}, R. 2015, \apj, 807, 134

\bibitem[{{Genel} {et~al.}(2009){Genel}, {Genzel}, {Bouch{\'e}}, {Naab}, \&
  {Sternberg}}]{2009ApJ...701.2002G}
{Genel}, S., {Genzel}, R., {Bouch{\'e}}, N., {Naab}, T., \& {Sternberg}, A.
  2009, \apj, 701, 2002

\bibitem[{{Genel} {et~al.}(2008){Genel}, {Genzel}, {Bouch{\'e}}, {Sternberg},
  {Naab}, {Schreiber}, {Shapiro}, {Tacconi}, {Lutz}, {Cresci}, {Buschkamp},
  {Davies}, \& {Hicks}}]{2008ApJ...688..789G}
{Genel}, S., {et~al.} 2008, \apj, 688, 789

\bibitem[{{Genel} {et~al.}(2012){Genel}, {Naab}, {Genzel}, {F{\"o}rster
  Schreiber}, {Sternberg}, {Oser}, {Johansson}, {Dav{\'e}}, {Oppenheimer}, \&
  {Burkert}}]{2012ApJ...745...11G}
---. 2012, \apj, 745, 11

\bibitem[{{Genzel} {et~al.}(2011){Genzel}, {Newman}, {Jones}, {F{\"o}rster
  Schreiber}, {Shapiro}, {Genel}, {Lilly}, {Renzini}, {Tacconi}, {Bouch{\'e}},
  {Burkert}, {Cresci}, {Buschkamp}, {Carollo}, {Ceverino}, {Davies}, {Dekel},
  {Eisenhauer}, {Hicks}, {Kurk}, {Lutz}, {Mancini}, {Naab}, {Peng},
  {Sternberg}, {Vergani}, \& {Zamorani}}]{2011ApJ...733..101G}
{Genzel}, R., {et~al.} 2011, \apj, 733, 101

\bibitem[{{Genzel} {et~al.}(2014){Genzel}, {F{\"o}rster Schreiber}, {Rosario},
  {Lang}, {Lutz}, {Wisnioski}, {Wuyts}, {Wuyts}, {Bandara}, {Bender}, {Berta},
  {Kurk}, {Mendel}, {Tacconi}, {Wilman}, {Beifiori}, {Brammer}, {Burkert},
  {Buschkamp}, {Chan}, {Carollo}, {Davies}, {Eisenhauer}, {Fabricius},
  {Fossati}, {Kriek}, {Kulkarni}, {Lilly}, {Mancini}, {Momcheva}, {Naab},
  {Nelson}, {Renzini}, {Saglia}, {Sharples}, {Sternberg}, {Tacchella}, \& {van
  Dokkum}}]{2014ApJ...796....7G}
---. 2014, \apj, 796, 7

\bibitem[{{Giavalisco} {et~al.}(1996){Giavalisco}, {Livio}, {Bohlin},
  {Macchetto}, \& {Stecher}}]{1996AJ....112..369G}
{Giavalisco}, M., {Livio}, M., {Bohlin}, R.~C., {Macchetto}, F.~D., \&
  {Stecher}, T.~P. 1996, \aj, 112, 369

\bibitem[{{Glazebrook}(2013)}]{2013PASA...30...56G}
{Glazebrook}, K. 2013, pasa, 30, 56

\bibitem[{{Grogin} {et~al.}(2011){Grogin}, {Kocevski}, {Faber}, {Ferguson},
  {Koekemoer}, {Riess}, {Acquaviva}, {Alexander}, {Almaini}, {Ashby}, {Barden},
  {Bell}, {Bournaud}, {Brown}, {Caputi}, {Casertano}, {Cassata}, {Castellano},
  {Challis}, {Chary}, {Cheung}, {Cirasuolo}, {Conselice}, {Roshan Cooray},
  {Croton}, {Daddi}, {Dahlen}, {Dav{\'e}}, {de Mello}, {Dekel}, {Dickinson},
  {Dolch}, {Donley}, {Dunlop}, {Dutton}, {Elbaz}, {Fazio}, {Filippenko},
  {Finkelstein}, {Fontana}, {Gardner}, {Garnavich}, {Gawiser}, {Giavalisco},
  {Grazian}, {Guo}, {Hathi}, {H{\"a}ussler}, {Hopkins}, {Huang}, {Huang},
  {Jha}, {Kartaltepe}, {Kirshner}, {Koo}, {Lai}, {Lee}, {Li}, {Lotz}, {Lucas},
  {Madau}, {McCarthy}, {McGrath}, {McIntosh}, {McLure}, {Mobasher},
  {Moustakas}, {Mozena}, {Nandra}, {Newman}, {Niemi}, {Noeske}, {Papovich},
  {Pentericci}, {Pope}, {Primack}, {Rajan}, {Ravindranath}, {Reddy}, {Renzini},
  {Rix}, {Robaina}, {Rodney}, {Rosario}, {Rosati}, {Salimbeni}, {Scarlata},
  {Siana}, {Simard}, {Smidt}, {Somerville}, {Spinrad}, {Straughn}, {Strolger},
  {Telford}, {Teplitz}, {Trump}, {van der Wel}, {Villforth}, {Wechsler},
  {Weiner}, {Wiklind}, {Wild}, {Wilson}, {Wuyts}, {Yan}, \&
  {Yun}}]{2011ApJS..197...35G}
{Grogin}, N.~A., {et~al.} 2011, \apjs, 197, 35

\bibitem[{{Guaita} {et~al.}(2015){Guaita}, {Melinder}, {Hayes}, {{\"O}stlin},
  {Gonzalez}, {Micheva}, {Adamo}, {Mas-Hesse}, {Sandberg},
  {Ot{\'{\i}}-Floranes}, {Schaerer}, {Verhamme}, {Freeland}, {Orlitov{\'a}},
  {Laursen}, {Cannon}, {Duval}, {Rivera-Thorsen}, {Herenz}, {Kunth}, {Atek},
  {Puschnig}, {Gruyters}, \& {Pardy}}]{2015A&A...576A..51G}
{Guaita}, L., {et~al.} 2015, \aap, 576, A51

\bibitem[{{Guo} {et~al.}(2012){Guo}, {Giavalisco}, {Ferguson}, {Cassata}, \&
  {Koekemoer}}]{2012ApJ...757..120G}
{Guo}, Y., {Giavalisco}, M., {Ferguson}, H.~C., {Cassata}, P., \& {Koekemoer},
  A.~M. 2012, \apj, 757, 120

\bibitem[{{Guo} {et~al.}(2013){Guo}, {Ferguson}, {Giavalisco}, {Barro},
  {Willner}, {Ashby}, {Dahlen}, {Donley}, {Faber}, {Fontana}, {Galametz},
  {Grazian}, {Huang}, {Kocevski}, {Koekemoer}, {Koo}, {McGrath}, {Peth},
  {Salvato}, {Wuyts}, {Castellano}, {Cooray}, {Dickinson}, {Dunlop}, {Fazio},
  {Gardner}, {Gawiser}, {Grogin}, {Hathi}, {Hsu}, {Lee}, {Lucas}, {Mobasher},
  {Nandra}, {Newman}, \& {van der Wel}}]{2013ApJS..207...24G}
{Guo}, Y., {et~al.} 2013, \apjs, 207, 24

\bibitem[{{Guo} {et~al.}(2015){Guo}, {Ferguson}, {Bell}, {Koo}, {Conselice},
  {Giavalisco}, {Kassin}, {Lu}, {Lucas}, {Mandelker}, {McIntosh}, {Primack},
  {Ravindranath}, {Barro}, {Ceverino}, {Dekel}, {Faber}, {Fang}, {Koekemoer},
  {Noeske}, {Rafelski}, \& {Straughn}}]{2015ApJ...800...39G}
---. 2015, \apj, 800, 39

\bibitem[{{Harikane} {et~al.}(2015){Harikane}, {Ouchi}, {Ono}, {More}, {Saito},
  {Lin}, {Coupon}, {Shimasaku}, {Shibuya}, {Price}, {Lin}, {Hsieh}, {Ishigaki},
  {Komiyama}, {Silverman}, {Takata}, {Tamazawa}, \&
  {Toshikawa}}]{2015arXiv151107873H}
{Harikane}, Y., {et~al.} 2015, ArXiv e-prints: 1511.07873

\bibitem[{{Hoffmann} \& {Romeo}(2012)}]{2012MNRAS.425.1511H}
{Hoffmann}, V., \& {Romeo}, A.~B. 2012, \mnras, 425, 1511

\bibitem[{{Holwerda} {et~al.}(2015){Holwerda}, {Bouwens}, {Oesch}, {Smit},
  {Illingworth}, \& {Labbe}}]{2015ApJ...808....6H}
{Holwerda}, B.~W., {Bouwens}, R., {Oesch}, P., {Smit}, R., {Illingworth}, G.,
  \& {Labbe}, I. 2015, \apj, 808, 6

\bibitem[{{Hopkins} {et~al.}(2012){Hopkins}, {Kere{\v s}}, {Murray},
  {Quataert}, \& {Hernquist}}]{2012MNRAS.427..968H}
{Hopkins}, P.~F., {Kere{\v s}}, D., {Murray}, N., {Quataert}, E., \&
  {Hernquist}, L. 2012, \mnras, 427, 968

\bibitem[{{Hopkins} {et~al.}(2010){Hopkins}, {Croton}, {Bundy}, {Khochfar},
  {van den Bosch}, {Somerville}, {Wetzel}, {Keres}, {Hernquist}, {Stewart},
  {Younger}, {Genel}, \& {Ma}}]{2010ApJ...724..915H}
{Hopkins}, P.~F., {et~al.} 2010, \apj, 724, 915

\bibitem[{{Hubble}(1926)}]{1926ApJ....64..321H}
{Hubble}, E.~P. 1926, \apj, 64, 321

\bibitem[{{Huertas-Company} {et~al.}(2015){Huertas-Company},
  {P{\'e}rez-Gonz{\'a}lez}, {Mei}, {Shankar}, {Bernardi}, {Daddi}, {Barro},
  {Cabrera-Vives}, {Cattaneo}, {Dimauro}, \& {Gravet}}]{2015arXiv150603084H}
{Huertas-Company}, M., {et~al.} 2015, ArXiv e-prints: 1506.03084

\bibitem[{{Illingworth} {et~al.}(2013){Illingworth}, {Magee}, {Oesch},
  {Bouwens}, {Labb{\'e}}, {Stiavelli}, {van Dokkum}, {Franx}, {Trenti},
  {Carollo}, \& {Gonzalez}}]{2013ApJS..209....6I}
{Illingworth}, G.~D., {et~al.} 2013, \apjs, 209, 6

\bibitem[{{Immeli} {et~al.}(2004{\natexlab{a}}){Immeli}, {Samland}, {Gerhard},
  \& {Westera}}]{2004A&A...413..547I}
{Immeli}, A., {Samland}, M., {Gerhard}, O., \& {Westera}, P.
  2004{\natexlab{a}}, \aap, 413, 547

\bibitem[{{Immeli} {et~al.}(2004{\natexlab{b}}){Immeli}, {Samland}, {Westera},
  \& {Gerhard}}]{2004ApJ...611...20I}
{Immeli}, A., {Samland}, M., {Westera}, P., \& {Gerhard}, O.
  2004{\natexlab{b}}, \apj, 611, 20

\bibitem[{{Inoue} {et~al.}(2015){Inoue}, {Dekel}, {Mandelker}, {Ceverino},
  {Bournaud}, \& {Primack}}]{2015arXiv151007695I}
{Inoue}, S., {Dekel}, A., {Mandelker}, N., {Ceverino}, D., {Bournaud}, F., \&
  {Primack}, J. 2015, ArXiv e-prints: 1510.07695

\bibitem[{{Inoue} \& {Saitoh}(2012)}]{2012MNRAS.422.1902I}
{Inoue}, S., \& {Saitoh}, T.~R. 2012, \mnras, 422, 1902

\bibitem[{{Ishigaki} {et~al.}(2015){Ishigaki}, {Kawamata}, {Ouchi}, {Oguri},
  {Shimasaku}, \& {Ono}}]{2015ApJ...799...12I}
{Ishigaki}, M., {Kawamata}, R., {Ouchi}, M., {Oguri}, M., {Shimasaku}, K., \&
  {Ono}, Y. 2015, \apj, 799, 12

\bibitem[{{Jiang} {et~al.}(2013){Jiang}, {Egami}, {Fan}, {Windhorst}, {Cohen},
  {Dav{\'e}}, {Finlator}, {Kashikawa}, {Mechtley}, {Ouchi}, \&
  {Shimasaku}}]{2013ApJ...773..153J}
{Jiang}, L., {et~al.} 2013, \apj, 773, 153

\bibitem[{{Jones} {et~al.}(2010){Jones}, {Swinbank}, {Ellis}, {Richard}, \&
  {Stark}}]{2010MNRAS.404.1247J}
{Jones}, T.~A., {Swinbank}, A.~M., {Ellis}, R.~S., {Richard}, J., \& {Stark},
  D.~P. 2010, \mnras, 404, 1247

\bibitem[{{Kawamata} {et~al.}(2015){Kawamata}, {Ishigaki}, {Shimasaku},
  {Oguri}, \& {Ouchi}}]{2015ApJ...804..103K}
{Kawamata}, R., {Ishigaki}, M., {Shimasaku}, K., {Oguri}, M., \& {Ouchi}, M.
  2015, \apj, 804, 103

\bibitem[{{Kennicutt}(1998)}]{1998ApJ...498..541K}
{Kennicutt}, Jr., R.~C. 1998, \apj, 498, 541

\bibitem[{{Kere{\v s}} {et~al.}(2009){Kere{\v s}}, {Katz}, {Fardal},
  {Dav{\'e}}, \& {Weinberg}}]{2009MNRAS.395..160K}
{Kere{\v s}}, D., {Katz}, N., {Fardal}, M., {Dav{\'e}}, R., \& {Weinberg},
  D.~H. 2009, \mnras, 395, 160

\bibitem[{{Kere{\v s}} {et~al.}(2005){Kere{\v s}}, {Katz}, {Weinberg}, \&
  {Dav{\'e}}}]{2005MNRAS.363....2K}
{Kere{\v s}}, D., {Katz}, N., {Weinberg}, D.~H., \& {Dav{\'e}}, R. 2005,
  \mnras, 363, 2

\bibitem[{{Kere{\v s}} {et~al.}(2012){Kere{\v s}}, {Vogelsberger}, {Sijacki},
  {Springel}, \& {Hernquist}}]{2012MNRAS.425.2027K}
{Kere{\v s}}, D., {Vogelsberger}, M., {Sijacki}, D., {Springel}, V., \&
  {Hernquist}, L. 2012, \mnras, 425, 2027

\bibitem[{{Koekemoer} {et~al.}(2011){Koekemoer}, {Faber}, {Ferguson}, {Grogin},
  {Kocevski}, {Koo}, {Lai}, {Lotz}, {Lucas}, {McGrath}, {Ogaz}, {Rajan},
  {Riess}, {Rodney}, {Strolger}, {Casertano}, {Castellano}, {Dahlen},
  {Dickinson}, {Dolch}, {Fontana}, {Giavalisco}, {Grazian}, {Guo}, {Hathi},
  {Huang}, {van der Wel}, {Yan}, {Acquaviva}, {Alexander}, {Almaini}, {Ashby},
  {Barden}, {Bell}, {Bournaud}, {Brown}, {Caputi}, {Cassata}, {Challis},
  {Chary}, {Cheung}, {Cirasuolo}, {Conselice}, {Roshan Cooray}, {Croton},
  {Daddi}, {Dav{\'e}}, {de Mello}, {de Ravel}, {Dekel}, {Donley}, {Dunlop},
  {Dutton}, {Elbaz}, {Fazio}, {Filippenko}, {Finkelstein}, {Frazer}, {Gardner},
  {Garnavich}, {Gawiser}, {Gruetzbauch}, {Hartley}, {H{\"a}ussler},
  {Herrington}, {Hopkins}, {Huang}, {Jha}, {Johnson}, {Kartaltepe},
  {Khostovan}, {Kirshner}, {Lani}, {Lee}, {Li}, {Madau}, {McCarthy},
  {McIntosh}, {McLure}, {McPartland}, {Mobasher}, {Moreira}, {Mortlock},
  {Moustakas}, {Mozena}, {Nandra}, {Newman}, {Nielsen}, {Niemi}, {Noeske},
  {Papovich}, {Pentericci}, {Pope}, {Primack}, {Ravindranath}, {Reddy},
  {Renzini}, {Rix}, {Robaina}, {Rosario}, {Rosati}, {Salimbeni}, {Scarlata},
  {Siana}, {Simard}, {Smidt}, {Snyder}, {Somerville}, {Spinrad}, {Straughn},
  {Telford}, {Teplitz}, {Trump}, {Vargas}, {Villforth}, {Wagner}, {Wandro},
  {Wechsler}, {Weiner}, {Wiklind}, {Wild}, {Wilson}, {Wuyts}, \&
  {Yun}}]{2011ApJS..197...36K}
{Koekemoer}, A.~M., {et~al.} 2011, \apjs, 197, 36

\bibitem[{{Komatsu} {et~al.}(2011){Komatsu}, {Smith}, {Dunkley}, {Bennett},
  {Gold}, {Hinshaw}, {Jarosik}, {Larson}, {Nolta}, {Page}, {Spergel},
  {Halpern}, {Hill}, {Kogut}, {Limon}, {Meyer}, {Odegard}, {Tucker}, {Weiland},
  {Wollack}, \& {Wright}}]{2011ApJS..192...18K}
{Komatsu}, E., {et~al.} 2011, \apjs, 192, 18

\bibitem[{{Kron}(1980)}]{1980ApJS...43..305K}
{Kron}, R.~G. 1980, \apjs, 43, 305

\bibitem[{{Krumholz} \& {Dekel}(2010)}]{2010MNRAS.406..112K}
{Krumholz}, M.~R., \& {Dekel}, A. 2010, \mnras, 406, 112

\bibitem[{{Kubo} {et~al.}(2016){Kubo}, {Yamada}, {Ichikawa}, {Kajisawa},
  {Matsuda}, {Tanaka}, \& {Umehata}}]{2016MNRAS.455.3333K}
{Kubo}, M., {Yamada}, T., {Ichikawa}, T., {Kajisawa}, M., {Matsuda}, Y.,
  {Tanaka}, I., \& {Umehata}, H. 2016, \mnras, 455, 3333

\bibitem[{{Kubo} {et~al.}(2013){Kubo}, {Uchimoto}, {Yamada}, {Kajisawa},
  {Ichikawa}, {Matsuda}, {Akiyama}, {Hayashino}, {Konishi}, {Nishimura},
  {Omata}, {Suzuki}, {Tanaka}, {Yoshikawa}, {Alexander}, {Fazio}, {Huang}, \&
  {Lehmer}}]{2013ApJ...778..170K}
{Kubo}, M., {et~al.} 2013, \apj, 778, 170

\bibitem[{{Law} {et~al.}(2012{\natexlab{a}}){Law}, {Steidel}, {Shapley},
  {Nagy}, {Reddy}, \& {Erb}}]{2012ApJ...759...29L}
{Law}, D.~R., {Steidel}, C.~C., {Shapley}, A.~E., {Nagy}, S.~R., {Reddy},
  N.~A., \& {Erb}, D.~K. 2012{\natexlab{a}}, \apj, 759, 29

\bibitem[{{Law} {et~al.}(2012{\natexlab{b}}){Law}, {Steidel}, {Shapley},
  {Nagy}, {Reddy}, \& {Erb}}]{2012ApJ...745...85L}
---. 2012{\natexlab{b}}, \apj, 745, 85

\bibitem[{{Le F{\`e}vre} {et~al.}(2000){Le F{\`e}vre}, {Abraham}, {Lilly},
  {Ellis}, {Brinchmann}, {Schade}, {Tresse}, {Colless}, {Crampton},
  {Glazebrook}, {Hammer}, \& {Broadhurst}}]{2000MNRAS.311..565L}
{Le F{\`e}vre}, O., {et~al.} 2000, \mnras, 311, 565

\bibitem[{{Lilly} {et~al.}(1996){Lilly}, {Le Fevre}, {Hammer}, \&
  {Crampton}}]{1996ApJ...460L...1L}
{Lilly}, S.~J., {Le Fevre}, O., {Hammer}, F., \& {Crampton}, D. 1996, \apjl,
  460, L1

\bibitem[{{Livermore} {et~al.}(2012){Livermore}, {Jones}, {Richard}, {Bower},
  {Ellis}, {Swinbank}, {Rigby}, {Smail}, {Arribas}, {Rodriguez Zaurin},
  {Colina}, {Ebeling}, \& {Crain}}]{2012MNRAS.427..688L}
{Livermore}, R.~C., {et~al.} 2012, \mnras, 427, 688

\bibitem[{{Livermore} {et~al.}(2015){Livermore}, {Jones}, {Richard}, {Bower},
  {Swinbank}, {Yuan}, {Edge}, {Ellis}, {Kewley}, {Smail}, {Coppin}, \&
  {Ebeling}}]{2015MNRAS.450.1812L}
---. 2015, \mnras, 450, 1812

\bibitem[{{L{\'o}pez-Sanjuan} {et~al.}(2013){L{\'o}pez-Sanjuan}, {Le
  F{\`e}vre}, {Tasca}, {Epinat}, {Amram}, {Contini}, {Garilli},
  {Kissler-Patig}, {Moultaka}, {Paioro}, {Perret}, {Queyrel}, {Tresse},
  {Vergani}, \& {Divoy}}]{2013A&A...553A..78L}
{L{\'o}pez-Sanjuan}, C., {et~al.} 2013, \aap, 553, A78

\bibitem[{{Lotz} {et~al.}(2011){Lotz}, {Jonsson}, {Cox}, {Croton}, {Primack},
  {Somerville}, \& {Stewart}}]{2011ApJ...742..103L}
{Lotz}, J.~M., {Jonsson}, P., {Cox}, T.~J., {Croton}, D., {Primack}, J.~R.,
  {Somerville}, R.~S., \& {Stewart}, K. 2011, \apj, 742, 103

\bibitem[{{Lotz} {et~al.}(2006){Lotz}, {Madau}, {Giavalisco}, {Primack}, \&
  {Ferguson}}]{2006ApJ...636..592L}
{Lotz}, J.~M., {Madau}, P., {Giavalisco}, M., {Primack}, J., \& {Ferguson},
  H.~C. 2006, \apj, 636, 592

\bibitem[{{Madau} \& {Dickinson}(2014)}]{2014ARA&A..52..415M}
{Madau}, P., \& {Dickinson}, M. 2014, \araa, 52, 415

\bibitem[{{Madau} {et~al.}(1996){Madau}, {Ferguson}, {Dickinson}, {Giavalisco},
  {Steidel}, \& {Fruchter}}]{1996MNRAS.283.1388M}
{Madau}, P., {Ferguson}, H.~C., {Dickinson}, M.~E., {Giavalisco}, M.,
  {Steidel}, C.~C., \& {Fruchter}, A. 1996, \mnras, 283, 1388

\bibitem[{{Man} {et~al.}(2012){Man}, {Toft}, {Zirm}, {Wuyts}, \& {van der
  Wel}}]{2012ApJ...744...85M}
{Man}, A.~W.~S., {Toft}, S., {Zirm}, A.~W., {Wuyts}, S., \& {van der Wel}, A.
  2012, \apj, 744, 85

\bibitem[{{Man} {et~al.}(2014){Man}, {Zirm}, \& {Toft}}]{2014arXiv1410.3479M}
{Man}, A.~W.~S., {Zirm}, A.~W., \& {Toft}, S. 2014, ArXiv e-prints:1411.2870

\bibitem[{{Mandelker} {et~al.}(2015){Mandelker}, {Dekel}, {Ceverino}, {DeGraf},
  {Guo}, \& {Primack}}]{2015arXiv151208791M}
{Mandelker}, N., {Dekel}, A., {Ceverino}, D., {DeGraf}, C., {Guo}, Y., \&
  {Primack}, J. 2015, ArXiv e-prints: 1512.08791

\bibitem[{{Mandelker} {et~al.}(2014){Mandelker}, {Dekel}, {Ceverino}, {Tweed},
  {Moody}, \& {Primack}}]{2014MNRAS.443.3675M}
{Mandelker}, N., {Dekel}, A., {Ceverino}, D., {Tweed}, D., {Moody}, C.~E., \&
  {Primack}, J. 2014, \mnras, 443, 3675

\bibitem[{{Men{\'e}ndez-Delmestre} {et~al.}(2013){Men{\'e}ndez-Delmestre},
  {Blain}, {Swinbank}, {Smail}, {Ivison}, {Chapman}, \& {Gon{\c
  c}alves}}]{2013ApJ...767..151M}
{Men{\'e}ndez-Delmestre}, K., {Blain}, A.~W., {Swinbank}, M., {Smail}, I.,
  {Ivison}, R.~J., {Chapman}, S.~C., \& {Gon{\c c}alves}, T.~S. 2013, \apj,
  767, 151

\bibitem[{{Moody} {et~al.}(2014){Moody}, {Guo}, {Mandelker}, {Ceverino},
  {Mozena}, {Koo}, {Dekel}, \& {Primack}}]{2014MNRAS.444.1389M}
{Moody}, C.~E., {Guo}, Y., {Mandelker}, N., {Ceverino}, D., {Mozena}, M.,
  {Koo}, D.~C., {Dekel}, A., \& {Primack}, J. 2014, \mnras, 444, 1389

\bibitem[{{Morishita} {et~al.}(2015){Morishita}, {Ichikawa}, {Noguchi},
  {Akiyama}, {Patel}, {Kajisawa}, \& {Obata}}]{2015ApJ...805...34M}
{Morishita}, T., {Ichikawa}, T., {Noguchi}, M., {Akiyama}, M., {Patel}, S.~G.,
  {Kajisawa}, M., \& {Obata}, T. 2015, \apj, 805, 34

\bibitem[{{Murata} {et~al.}(2014){Murata}, {Kajisawa}, {Taniguchi},
  {Kobayashi}, {Shioya}, {Capak}, {Ilbert}, {Koekemoer}, {Salvato}, \&
  {Scoville}}]{2014ApJ...786...15M}
{Murata}, K.~L., {et~al.} 2014, \apj, 786, 15

\bibitem[{{Nelson} {et~al.}(2013){Nelson}, {Vogelsberger}, {Genel}, {Sijacki},
  {Kere{\v s}}, {Springel}, \& {Hernquist}}]{2013MNRAS.429.3353N}
{Nelson}, D., {Vogelsberger}, M., {Genel}, S., {Sijacki}, D., {Kere{\v s}}, D.,
  {Springel}, V., \& {Hernquist}, L. 2013, \mnras, 429, 3353

\bibitem[{{Nelson} {et~al.}(2015){Nelson}, {van Dokkum}, {F{\"o}rster
  Schreiber}, {Franx}, {Brammer}, {Momcheva}, {Wuyts}, {Whitaker}, {Skelton},
  {Fumagalli}, {Kriek}, {Labb{\'e}}, {Leja}, {Rix}, {Tacconi}, {van der Wel},
  {van den Bosch}, {Oesch}, {Dickey}, \& {Ulf Lange}}]{2015arXiv150703999N}
{Nelson}, E.~J., {et~al.} 2015, ArXiv e-prints: 1507.03999

\bibitem[{{Newman} {et~al.}(2012){Newman}, {Shapiro Griffin}, {Genzel},
  {Davies}, {F{\"o}rster-Schreiber}, {Tacconi}, {Kurk}, {Wuyts}, {Genel},
  {Lilly}, {Renzini}, {Bouch{\'e}}, {Burkert}, {Cresci}, {Buschkamp},
  {Carollo}, {Eisenhauer}, {Hicks}, {Lutz}, {Mancini}, {Naab}, {Peng}, \&
  {Vergani}}]{2012ApJ...752..111N}
{Newman}, S.~F., {et~al.} 2012, \apj, 752, 111

\bibitem[{{Noguchi}(1998)}]{1998Natur.392..253N}
{Noguchi}, M. 1998, \nat, 392, 253

\bibitem[{{Oesch} {et~al.}(2015){Oesch}, {Bouwens}, {Illingworth}, {Franx},
  {Ammons}, {van Dokkum}, {Trenti}, \& {Labb{\'e}}}]{2015ApJ...808..104O}
{Oesch}, P.~A., {Bouwens}, R.~J., {Illingworth}, G.~D., {Franx}, M., {Ammons},
  S.~M., {van Dokkum}, P.~G., {Trenti}, M., \& {Labb{\'e}}, I. 2015, \apj, 808,
  104

\bibitem[{{Oesch} {et~al.}(2010){Oesch}, {Bouwens}, {Carollo}, {Illingworth},
  {Trenti}, {Stiavelli}, {Magee}, {Labb{\'e}}, \&
  {Franx}}]{2010ApJ...709L..21O}
{Oesch}, P.~A., {et~al.} 2010, \apjl, 709, L21

\bibitem[{{Okamoto}(2013)}]{2013MNRAS.428..718O}
{Okamoto}, T. 2013, \mnras, 428, 718

\bibitem[{{Oke} \& {Gunn}(1983)}]{1983ApJ...266..713O}
{Oke}, J.~B., \& {Gunn}, J.~E. 1983, \apj, 266, 713

\bibitem[{{Overzier} {et~al.}(2010){Overzier}, {Heckman}, {Schiminovich},
  {Basu-Zych}, {Gon{\c c}alves}, {Martin}, \& {Rich}}]{2010ApJ...710..979O}
{Overzier}, R.~A., {Heckman}, T.~M., {Schiminovich}, D., {Basu-Zych}, A.,
  {Gon{\c c}alves}, T., {Martin}, D.~C., \& {Rich}, R.~M. 2010, \apj, 710, 979

\bibitem[{{Patel} {et~al.}(2013){Patel}, {van Dokkum}, {Franx}, {Quadri},
  {Muzzin}, {Marchesini}, {Williams}, {Holden}, \&
  {Stefanon}}]{2013ApJ...766...15P}
{Patel}, S.~G., {et~al.} 2013, \apj, 766, 15

\bibitem[{{Peng} {et~al.}(2002){Peng}, {Ho}, {Impey}, \&
  {Rix}}]{2002AJ....124..266P}
{Peng}, C.~Y., {Ho}, L.~C., {Impey}, C.~D., \& {Rix}, H.-W. 2002, \aj, 124, 266

\bibitem[{{Peng} {et~al.}(2010){Peng}, {Ho}, {Impey}, \&
  {Rix}}]{2010AJ....139.2097P}
---. 2010, \aj, 139, 2097

\bibitem[{{Puech}(2010)}]{2010MNRAS.406..535P}
{Puech}, M. 2010, \mnras, 406, 535

\bibitem[{{Ravindranath} {et~al.}(2006){Ravindranath}, {Giavalisco},
  {Ferguson}, {Conselice}, {Katz}, {Weinberg}, {Lotz}, {Dickinson}, {Fall},
  {Mobasher}, \& {Papovich}}]{2006ApJ...652..963R}
{Ravindranath}, S., {et~al.} 2006, \apj, 652, 963

\bibitem[{{Rodriguez-Gomez} {et~al.}(2015){Rodriguez-Gomez}, {Genel},
  {Vogelsberger}, {Sijacki}, {Pillepich}, {Sales}, {Torrey}, {Snyder},
  {Nelson}, {Springel}, {Ma}, \& {Hernquist}}]{2015MNRAS.449...49R}
{Rodriguez-Gomez}, V., {et~al.} 2015, \mnras, 449, 49

\bibitem[{{Romano-D{\'{\i}}az} {et~al.}(2011){Romano-D{\'{\i}}az}, {Choi},
  {Shlosman}, \& {Trenti}}]{2011ApJ...738L..19R}
{Romano-D{\'{\i}}az}, E., {Choi}, J.-H., {Shlosman}, I., \& {Trenti}, M. 2011,
  \apjl, 738, L19

\bibitem[{{Romeo} \& {Agertz}(2014)}]{2014MNRAS.442.1230R}
{Romeo}, A.~B., \& {Agertz}, O. 2014, \mnras, 442, 1230

\bibitem[{{Romeo} {et~al.}(2010){Romeo}, {Burkert}, \&
  {Agertz}}]{2010MNRAS.407.1223R}
{Romeo}, A.~B., {Burkert}, A., \& {Agertz}, O. 2010, \mnras, 407, 1223

\bibitem[{{Salpeter}(1955)}]{1955ApJ...121..161S}
{Salpeter}, E.~E. 1955, \apj, 121, 161

\bibitem[{{S{\'e}rsic}(1963)}]{1963BAAA....6...41S}
{S{\'e}rsic}, J.~L. 1963, Boletin de la Asociacion Argentina de Astronomia La
  Plata Argentina, 6, 41

\bibitem[{{S{\'e}rsic}(1968)}]{1968adga.book.....S}
---. 1968, {Atlas de galaxias australes} ({S{\'e}rsic}, J.~L.)

\bibitem[{{Shapiro} {et~al.}(2008){Shapiro}, {Genzel}, {F{\"o}rster Schreiber},
  {Tacconi}, {Bouch{\'e}}, {Cresci}, {Davies}, {Eisenhauer}, {Johansson},
  {Krajnovi{\'c}}, {Lutz}, {Naab}, {Arimoto}, {Arribas}, {Cimatti}, {Colina},
  {Daddi}, {Daigle}, {Erb}, {Hernandez}, {Kong}, {Mignoli}, {Onodera},
  {Renzini}, {Shapley}, \& {Steidel}}]{2008ApJ...682..231S}
{Shapiro}, K.~L., {et~al.} 2008, \apj, 682, 231

\bibitem[{{Shibuya} {et~al.}(2015){Shibuya}, {Ouchi}, \&
  {Harikane}}]{2015ApJS..219...15S}
{Shibuya}, T., {Ouchi}, M., \& {Harikane}, Y. 2015, \apjs, 219, 15

\bibitem[{{Shibuya} {et~al.}(2014){Shibuya}, {Ouchi}, {Nakajima}, {Yuma},
  {Hashimoto}, {Shimasaku}, {Mori}, \& {Umemura}}]{2014ApJ...785...64S}
{Shibuya}, T., {Ouchi}, M., {Nakajima}, K., {Yuma}, S., {Hashimoto}, T.,
  {Shimasaku}, K., {Mori}, M., \& {Umemura}, M. 2014, \apj, 785, 64

\bibitem[{{Skelton} {et~al.}(2014){Skelton}, {Whitaker}, {Momcheva}, {Brammer},
  {van Dokkum}, {Labb{\'e}}, {Franx}, {van der Wel}, {Bezanson}, {Da Cunha},
  {Fumagalli}, {F{\"o}rster Schreiber}, {Kriek}, {Leja}, {Lundgren}, {Magee},
  {Marchesini}, {Maseda}, {Nelson}, {Oesch}, {Pacifici}, {Patel}, {Price},
  {Rix}, {Tal}, {Wake}, \& {Wuyts}}]{2014ApJS..214...24S}
{Skelton}, R.~E., {et~al.} 2014, \apjs, 214, 24

\bibitem[{{Skidmore} {et~al.}(2015){Skidmore}, {TMT International Science
  Development Teams}, \& {Science Advisory Committee}}]{2015RAA....15.1945S}
{Skidmore}, W., {TMT International Science Development Teams}, \& {Science
  Advisory Committee}, T. 2015, Research in Astronomy and Astrophysics, 15,
  1945

\bibitem[{{Sobral} {et~al.}(2013){Sobral}, {Swinbank}, {Stott}, {Matthee},
  {Bower}, {Smail}, {Best}, {Geach}, \& {Sharples}}]{2013ApJ...779..139S}
{Sobral}, D., {et~al.} 2013, \apj, 779, 139

\bibitem[{{Steidel} {et~al.}(1999){Steidel}, {Adelberger}, {Giavalisco},
  {Dickinson}, \& {Pettini}}]{1999ApJ...519....1S}
{Steidel}, C.~C., {Adelberger}, K.~L., {Giavalisco}, M., {Dickinson}, M., \&
  {Pettini}, M. 1999, \apj, 519, 1

\bibitem[{{Swinbank} {et~al.}(2009){Swinbank}, {Webb}, {Richard}, {Bower},
  {Ellis}, {Illingworth}, {Jones}, {Kriek}, {Smail}, {Stark}, \& {van
  Dokkum}}]{2009MNRAS.400.1121S}
{Swinbank}, A.~M., {et~al.} 2009, \mnras, 400, 1121

\bibitem[{{Tacchella} {et~al.}(2015{\natexlab{a}}){Tacchella}, {Dekel},
  {Carollo}, {Ceverino}, {DeGraf}, {Lapiner}, {Mandelker}, \&
  {Primack}}]{2015arXiv150900017T}
{Tacchella}, S., {Dekel}, A., {Carollo}, C.~M., {Ceverino}, D., {DeGraf}, C.,
  {Lapiner}, S., {Mandelker}, N., \& {Primack}, J.~R. 2015{\natexlab{a}}, ArXiv
  e-prints: 1509.00017

\bibitem[{{Tacchella} {et~al.}(2015{\natexlab{b}}){Tacchella}, {Dekel},
  {Carollo}, {Ceverino}, {DeGraf}, {Lapiner}, {Mandelker}, \&
  {Primack}}]{2015arXiv150902529T}
---. 2015{\natexlab{b}}, ArXiv e-prints: 1509.02529

\bibitem[{{Tacchella} {et~al.}(2015{\natexlab{c}}){Tacchella}, {Carollo},
  {Renzini}, {Schreiber}, {Lang}, {Wuyts}, {Cresci}, {Dekel}, {Genzel},
  {Lilly}, {Mancini}, {Newman}, {Onodera}, {Shapley}, {Tacconi}, {Woo}, \&
  {Zamorani}}]{2015Sci...348..314T}
{Tacchella}, S., {et~al.} 2015{\natexlab{c}}, Science, 348, 314

\bibitem[{{Tacchella} {et~al.}(2015{\natexlab{d}}){Tacchella}, {Lang},
  {Carollo}, {F{\"o}rster Schreiber}, {Renzini}, {Shapley}, {Wuyts}, {Cresci},
  {Genzel}, {Lilly}, {Mancini}, {Newman}, {Tacconi}, {Zamorani}, {Davies},
  {Kurk}, \& {Pozzetti}}]{2015ApJ...802..101T}
---. 2015{\natexlab{d}}, \apj, 802, 101

\bibitem[{{Tacconi} {et~al.}(2010){Tacconi}, {Genzel}, {Neri}, {Cox}, {Cooper},
  {Shapiro}, {Bolatto}, {Bouch{\'e}}, {Bournaud}, {Burkert}, {Combes},
  {Comerford}, {Davis}, {Schreiber}, {Garcia-Burillo}, {Gracia-Carpio}, {Lutz},
  {Naab}, {Omont}, {Shapley}, {Sternberg}, \& {Weiner}}]{2010Natur.463..781T}
{Tacconi}, L.~J., {et~al.} 2010, \nat, 463, 781

\bibitem[{{Tadaki} {et~al.}(2014){Tadaki}, {Kodama}, {Tanaka}, {Hayashi},
  {Koyama}, \& {Shimakawa}}]{2014ApJ...780...77T}
{Tadaki}, K.-i., {Kodama}, T., {Tanaka}, I., {Hayashi}, M., {Koyama}, Y., \&
  {Shimakawa}, R. 2014, \apj, 780, 77

\bibitem[{{Taniguchi} \& {Shioya}(2001)}]{2001ApJ...547..146T}
{Taniguchi}, Y., \& {Shioya}, Y. 2001, \apj, 547, 146

\bibitem[{{Toomre}(1964)}]{1964ApJ...139.1217T}
{Toomre}, A. 1964, \apj, 139, 1217

\bibitem[{{van den Bergh} {et~al.}(1996){van den Bergh}, {Abraham}, {Ellis},
  {Tanvir}, {Santiago}, \& {Glazebrook}}]{1996AJ....112..359V}
{van den Bergh}, S., {Abraham}, R.~G., {Ellis}, R.~S., {Tanvir}, N.~R.,
  {Santiago}, B.~X., \& {Glazebrook}, K.~G. 1996, \aj, 112, 359

\bibitem[{{van Dokkum} {et~al.}(2010){van Dokkum}, {Whitaker}, {Brammer},
  {Franx}, {Kriek}, {Labb{\'e}}, {Marchesini}, {Quadri}, {Bezanson},
  {Illingworth}, {Muzzin}, {Rudnick}, {Tal}, \& {Wake}}]{2010ApJ...709.1018V}
{van Dokkum}, P.~G., {et~al.} 2010, \apj, 709, 1018

\bibitem[{{Williams} {et~al.}(1994){Williams}, {de Geus}, \&
  {Blitz}}]{1994ApJ...428..693W}
{Williams}, J.~P., {de Geus}, E.~J., \& {Blitz}, L. 1994, \apj, 428, 693

\bibitem[{{Williams} {et~al.}(2011){Williams}, {Quadri}, \&
  {Franx}}]{2011ApJ...738L..25W}
{Williams}, R.~J., {Quadri}, R.~F., \& {Franx}, M. 2011, \apjl, 738, L25

\bibitem[{{Wisnioski} {et~al.}(2012){Wisnioski}, {Glazebrook}, {Blake},
  {Poole}, {Green}, {Wyder}, \& {Martin}}]{2012MNRAS.422.3339W}
{Wisnioski}, E., {Glazebrook}, K., {Blake}, C., {Poole}, G.~B., {Green}, A.~W.,
  {Wyder}, T., \& {Martin}, C. 2012, \mnras, 422, 3339

\bibitem[{{Wisnioski} {et~al.}(2013){Wisnioski}, {Glazebrook}, {Blake}, \&
  {Swinbank}}]{2013MNRAS.436..266W}
{Wisnioski}, E., {Glazebrook}, K., {Blake}, C., \& {Swinbank}, A.~M. 2013,
  \mnras, 436, 266

\bibitem[{{Wisnioski} {et~al.}(2011){Wisnioski}, {Glazebrook}, {Blake},
  {Wyder}, {Martin}, {Poole}, {Sharp}, {Couch}, {Kacprzak}, {Brough},
  {Colless}, {Contreras}, {Croom}, {Croton}, {Davis}, {Drinkwater}, {Forster},
  {Gilbank}, {Gladders}, {Jelliffe}, {Jurek}, {Li}, {Madore}, {Pimbblet},
  {Pracy}, {Woods}, \& {Yee}}]{2011MNRAS.417.2601W}
{Wisnioski}, E., {et~al.} 2011, \mnras, 417, 2601

\bibitem[{{Wuyts} {et~al.}(2012){Wuyts}, {F{\"o}rster Schreiber}, {Genzel},
  {Guo}, {Barro}, {Bell}, {Dekel}, {Faber}, {Ferguson}, {Giavalisco}, {Grogin},
  {Hathi}, {Huang}, {Kocevski}, {Koekemoer}, {Koo}, {Lotz}, {Lutz}, {McGrath},
  {Newman}, {Rosario}, {Saintonge}, {Tacconi}, {Weiner}, \& {van der
  Wel}}]{2012ApJ...753..114W}
{Wuyts}, S., {et~al.} 2012, \apj, 753, 114

\bibitem[{{Yajima} {et~al.}(2015){Yajima}, {Shlosman}, {Romano-D{\'{\i}}az}, \&
  {Nagamine}}]{2015MNRAS.451..418Y}
{Yajima}, H., {Shlosman}, I., {Romano-D{\'{\i}}az}, E., \& {Nagamine}, K. 2015,
  \mnras, 451, 418

\bibitem[{{Zolotov} {et~al.}(2015){Zolotov}, {Dekel}, {Mandelker}, {Tweed},
  {Inoue}, {DeGraf}, {Ceverino}, {Primack}, {Barro}, \&
  {Faber}}]{2015MNRAS.450.2327Z}
{Zolotov}, A., {et~al.} 2015, \mnras, 450, 2327

\end{thebibliography}

\clearpage

\end{document}